\documentclass[twocolumn]{aa}  
\usepackage{natbib}
\bibpunct{(}{)}{;}{a}{}{,} 
\usepackage{orcidlink}
\usepackage{multirow}
\usepackage{hyperref}
\usepackage{xcolor}
\usepackage[utf8]{inputenc}
\DeclareUnicodeCharacter{2212}{\ensuremath{-}}
\usepackage{ulem}
\usepackage{changes}
\usepackage{siunitx}
\usepackage{multirow}
\usepackage{ulem}
\usepackage{pifont}
\usepackage[document]{}
\usepackage{graphicx}
\usepackage{cancel}
\usepackage{booktabs}
\usepackage{txfonts}
\usepackage{ulem}
\usepackage{siunitx}
\usepackage[document]{}
\usepackage{graphicx}

\usepackage{pifont}
\begin{document}


    

   
 \title{Gas Phase Distribution in the Neutral ISM: A Comparison between Observation and Numerical Simulation}

\author{
A. Koley\inst{}\thanks{Fondecyt Postdoctoral fellow; email:atanuphysics15@gmail.com}
\orcidlink{0000-0003-2713-0211}
}

\authorrunning{Koley, A.}

\institute{
Departamento de Astronom\'{i}a, Universidad de Concepci\'{o}n,
Casilla 160-C, Concepci\'{o}n, Chile
}

\date{Received xxxxxxx; accepted xxxxxx}
   

 \abstract
    {The neutral hydrogen (H{\sc i}) 21-cm line serves as a powerful tracer of the neutral interstellar medium (ISM). Thermal stability analysis suggests that the neutral ISM is bistable in nature, consisting of the cold neutral medium (CNM) embedded within the warm neutral medium (WNM), both in approximate thermal pressure equilibrium. When turbulence is incorporated into the numerical simulations, a third thermally unstable medium (UNM) emerges between the CNM and the WNM. Although observational studies support the existence of this intermediate phase, a clear empirical correlation between the fraction of the UNM gas and the strength of the turbulence remains elusive. In this study, we investigate the various phases of neutral ISM using H{\sc i} 21-cm emission-absorption spectra from the publicly available GWA and LAB surveys. We compare our results with several numerical simulations, including TIGRESS-NCR and TIGRESS-CLASSIC, and find that our results are more closely consistent with TIGRESS-NCR than with TIGRESS-CLASSIC. From our observational modeling, we find that  19.8\% of the gas reside in the CNM phase, 32.5\% in the UNM phase, and 47.8\% in the WNM phase, assuming phase boundaries defined by spin temperature: $T_{\text{s}}$ $<$ 250 K for the CNM, 250 K $<$ $T_{\text{s}}$ $<$ 4000 K for the UNM, and $T_{\text{s}}$ $>$ 4000 K for the WNM. We further expect that deep, sensitive absorption studies with the Square Kilometre Array (SKA) or the Next Generation Very Large Array (ngVLA), capable of robustly detecting WNM clouds in absorption will place more tighter observational constraints on the fraction of the gas in three different phases of the neutral ISM.\\ 
}
   

   \maketitle
%

\section{Introduction}\label{section_1}
Thermal stability analysis indicates that the neutral interstellar medium (ISM) is bistable in nature and consists of two major phases: the clumpy cold neutral medium (CNM) and the more diffuse, widespread warm neutral medium (WNM) \citep{1965ApJ...142..531F,1969ApJ...155L.149F, 2003ApJ...587..278W}. However, both observational and numerical studies provide evidence for an intermediate phase between these two, referred to as the thermally unstable medium (UNM) \citep{2003ApJ...586.1067H,2013MNRAS.436.2366R,2014ApJ...793..132S}. Although these three phases are primarily distinguished by their temperatures, the precise temperature ranges remain uncertain. Recent observational studies suggest that the spin temperature ($T_{\text{s}}$) of the CNM is up to 250 K, followed by the UNM phase extending up to approximately 4000 K, and the WNM phase above this \citep[and references therein]{2023ARA&A..61...19M}. The UNM, which lies between the CNM and the WNM, is thought to arise due to turbulent effects; where turbulence prevents the complete condensation of WNM gas into the CNM phase, leaving a substantial fraction of gas in this intermediate regime. This implies that the fraction of thermally unstable gas is strongly correlated with the level of turbulence \citep{2005AA...433....1A,2024arXiv240714199H}.
To study gas clouds across these three phases (CNM, UNM, and WNM), the most commonly used tracer is neutral hydrogen (H{\sc i}). The standard approach involves comparing H{\sc i} emission and absorption spectra observed along nearly the same lines of sight. The absorption spectra are obtained toward background quasars, while the emission spectra are taken from nearby directions. Through decomposition of these spectra, one can derive various physical and kinematical parameters of the gas clouds, such as column density ($N$[H{\sc i}]), spin temperature ($T_{\text{s}}$), upper limit of kinetic temperature ($T_{\text{k,max}}$), peak optical depth ($\tau_{\text{peak}}$), line width ($\Delta V$), and central velocity ($v_{\text{c}}$). While this technique has its limitations, it is currently considered as the standard method for probing the kinematic and physical properties of gas clouds in the neutral ISM and has been widely used for several decades \citep{2003ApJS..145..329H,2003ApJ...586.1067H,2014ApJ...793..132S,2018ApJS..238...14M,2019ApJ...880..141N}.\\

However, it may not always be possible to successfully decompose the emission and absorption spectra into multiple Gaussian components. Several complications may arise that hinder the reliable identification of components, particularly those that are narrow and associated with small spatial scales \citep{2023PASA...40...46K,2025AJ....169..284C,2026arXiv260821115C}. Cloud properties may also change between the emission and absorption lines of sight, even when the spectra are observed toward nearby lines of sight. Moreover, joint decomposition of the emission-absorption spectral line does not directly provide the kinetic temperature of the gas cloud; instead, it yields the spin temperature, which in certain cases only serves as a lower limit to the kinetic temperature \citep{2003ApJ...586.1067H,2014ApJ...793..132S,2018ApJS..238...14M,2019ApJ...880..141N}. 
This raises the question whether the physical properties of H{\sc i} gas clouds can be inferred using only absorption or emission spectra; thereby overcoming these limitations. To address this, we developed a simple iterative method based solely on absorption spectra to derive various physical properties of the H{\sc i} gas clouds \citep{2019MNRAS.483..593K}. Current sensitive radio telescopes such as Australia Telescope Compact Array (ATCA), Karl G. Jansky Very Large Array (JVLA), Westerbork Synthesis Radio Telescope (WSRT), and Giant Metrewave Radio Telescope (GMRT) are still unable to detect WNM clouds easily in absorption \citep{2003ApJS..145..329H,2003ApJ...586.1067H,2013MNRAS.436.2352R,2013MNRAS.436.2366R,2018ApJS..238...14M}. Therefore, the resultant emission spectra toward nearby lines of sight are still required to constrain the fraction of WNM gas clouds. However, this method can be reliably applied solely on absorption spectra with future highly sensitive facilities such as the square kilometer array (SKA) and the Next Generation Very Large Array (ngVLA), where WNM clouds are expected to be detected more frequently in absorption within a reasonable integration of time.  Reliable application of this method requires accurate knowledge of the thermal pressure ($P_{\text{th}}$), the turbulent scaling relation, and the conversion between spin temperature ($T_{\text{s}}$) and kinetic temperature ($T_{\text{k}}$) in neutral ISM \citep{2001A&A...371..698L, 2011ApJ...734...65J, 2023PASA...40...46K}. Without proper information on these, the results derived may be misleading, leading to incorrect interpretations of the gas properties and phase fractions (CNM, UNM, and WNM). In this study, we investigate the gas phase distribution in the neutral ISM using this iterative method, with a particular focus on the thermally unstable medium (UNM), incorporating new observational data from recent work \citep{2024MNRAS.529.4037P} in addition to previous observational data \citep{2013MNRAS.436.2366R}. A larger data set would allow for significantly tighter constraints on the gas fraction in the different phases of the neutral interstellar medium.\\

Earlier, \citet{2019MNRAS.483..593K} performed the analysis using the \cite{1979MNRAS.186..479L} linewidth--size relation. Although this study provided important evidence for the existence of a linewidth--size relation in the interstellar medium, their result was based on a relatively limited number of data points. Moreover, the adopted relation does not directly represent the non-thermal velocity dispersion as a function of the physical scale.  The line width--size relation adopted by \citet{1979MNRAS.186..479L} leads to an overestimation of the $T_\text{s}$, and consequently also results in an overestimation of the inferred H{\sc i} column density. This effect is clearly seen and mentioned in fig.~7 of \citet{2024MNRAS.527.8475B}, where the use of the \cite{1979MNRAS.186..479L} linewidth--size relation leads to systematically higher estimates of these quantities, and most of the gas is artificially accumulated in the UNM phase. In this work, we instead adopt the linewidth--size relation derived by \citet{2023PASA...40...46K}, which provides a more direct characterization of the non-thermal velocity dispersion and is broadly consistent with the relations reported in several other studies \citep{2003ApJ...587..278W, 2025A&A...694L..11K}.\\

Furthermore, we compare our derived results with those from numerical simulations, particularly the Three-phase ISM in Galaxies Resolving Evolution with Star formation and Supernova feedback \texttt{(TIGRESS)-} (Non-equilibrium Cooling and Radiation) \texttt{NCR} simulations of \citet{2023ApJ...946....3K}, to assess the consistency between the observationally inferred properties and numerical predictions. An important consideration in such a comparison is the trade-off between physical realism and numerical resolution. Simulations that adequately incorporate large-scale processes, such as supernova explosions and stellar feedback, can be limited by their spatial resolution and may therefore not fully resolve the small-scale structures relevant to our observations \citep{2021MNRAS.504.1039R,2023ApJ...946....3K}. Conversely, very high-resolution simulations can resolve small-scale structures more effectively but often do not self-consistently capture the full range of large-scale processes, such as clustered stellar feedback and supernova-driven turbulence, and instead commonly adopt simplified setups, such as large-scale converging flows, to generate dense structures \citep{2005AA...433....1A,2022MNRAS.514..957S}. Thus, differences between the observational and numerical results may partly reflect the different spatial scales, physical processes, and numerical resolutions represented in the respective approaches.\\

\S~\ref{section_2} describes the iterative method in detail. \S~\ref{section_3} presents the H{\sc i} data adopted from previous studies. \S~\ref{section_4} presents the results derived using the iterative method. The main results are discussed in \S~\ref{section_5}. Finally, \S~\ref{section_6} summarizes the main conclusions of this study.\\

\section{Methodology}\label{section_2}

This iterative method critically relies on the thermal pressure profile, the proper turbulent scaling law, and the relation between spin and kinetic temperatures in neutral ISM. We first outline how these quantities are adopted from previous studies before describing the iterative method.\\

\subsection{Thermal pressure}\label{subsection_2_1}

The value of the thermal pressure is obtained from the work of \cite{2011ApJ...734...65J}. They studied the ionization balance of [C{\sc i}] and [C{\sc ii}] (spectra taken towards 89 bright stars) using the rotational temperature ($T_{\text{rot}}$) obtained from the H$_{2}$ molecule, and obtained the log-normal thermal pressure  ($P_{\text{th}}$) profile with a median value of $\sim$ 3800 K cm$^{-3}$. In some cases where no direct measurements of [C{\sc ii}] were available, alternative tracers such as [O{\sc i}] and [Si{\sc ii}] were used. Since the detection rate of the H$_{2}$ molecule is biased towards the CNM clouds, this pressure profile is biased towards the CNM cloud. Although the thermal pressure ranges from $\sim$ 10$^{2.5}$ to $\sim$ 10$^{4.5}$ K cm$^{-3}$, the lower and upper extremes were excluded at the end due to the theoretical lower limit for the CNM and the exceptionally high pressures near the H{\sc ii} regions \citep{2003ApJ...587..278W}. Excluding these regions reduces the lognormal tail and the deviation from the Gaussian profile.
It is evident from the works of \cite{2003ApJ...586.1067H} and \cite{2018ApJS..238...14M} that most of the components that are observed both in emission and absorption lie below 500 K, implying that the majority of the components belong to CNM gas clouds. The reason for this is that because of the low optical depth ($\tau$) of the line, diffuse WNM gas clouds are hardly detected in absorption. Consequently, it is appropriate to consider the value of $P_{\text{th}}$ derived from the aforementioned work.\\

\subsection{Turbulent scaling relation}\label{subsection_2_2}

The scaling relation of turbulence is equally critical for this iterative process. Isotropic hydrodynamic (HD) turbulence follows the Kolmogorov law of turbulence, based on which the non-thermal velocity dispersion ($\sigma_{\text{nth}}$) varies with the length scale ($L$) with a power law index ($\alpha$) of 0.33 \citep{1941DoSSR..32...16K,1995tlan.book.....F}. However, ISM, particularly the CNM phase, is compressible as well as magnetized, which means that the power law may be different from the Kolmogorov law of turbulence. Theoretical and simulation studies predict that, in a compressible magnetohydrodynamic system (in neutral ISM mainly in CNM phase), turbulence perpendicular to the magnetic field where energy is carried by Alfven waves; follows a power-law spectrum ($\alpha$) similar to the Kolmogorov turbulence (0.33). In addition, along the magnetic field, shock compression associated with slow modes leads to a steeper power-law spectrum, closer to that of Burger law of turbulence (0.50) \citep{BURGERS1948171, 2007ApJ...665..416K,2007ApJ...666L..69K,2010A&A...512A..81F}. When researchers study observationally the turbulent velocity dispersion along the line of sight, contributions from both parallel and perpendicular components are combined. In addition, the observed spectra are density-weighted. As a result of these effects the line of sight mixes the contributions from different turbulence modes.\\


Observational studies of the compressible interstellar medium particularly in the cold neutral medium and molecular clouds, generally found a power-law index close to 0.5, consistent with Burger-like turbulence \citep{1987ApJ...319..730S,2004ApJ...615L..45H,2023PASA...40...46K,2025A&A...702A.133K}. For example, \cite{2023PASA...40...46K} using a simple analytical method (after taking the observed H{\sc i} emission-absorption gas components of H{\sc i} from various large-scale surveys) showed that the intensity ($A$) and the power-law index ($\alpha$) in the CNM phase are 1.14 km s$^{-1}$ and 0.55, respectively. Likewise,  \cite{2025A&A...694L..11K}  fitted the velocity dispersion of high-latitude CNM gas clouds with the values of $A$ = 1.2 km s$^{-1}$ and $\alpha$ = 0.37 and 0.50 obtained from \cite{2003ApJ...587..278W} and noticed that for $\alpha$ = 0.50, the data set is fitted more accurately than $\alpha$ = 0.37. Based on all these results, it is likely that in the CNM phase, $\alpha$ is close to 0.50. For this reason, we have decided to use the values of $A$ and $\alpha$ from the work of \cite{2023PASA...40...46K} for this iterative method. This is because current absorption studies primarily detect the CNM, as well as the lower-temperature portion of the UNM. We note that in \cite{2023PASA...40...46K}, a clear turbulence correlation is found above 0.4 pc, whereas no significant correlation is observed below this scale. In Appendix \ref{Appendix: appendix1}, we discuss the plausible reasons for this behavior. In our analysis, we adopt a single turbulence scaling relation for the entire data set where the value of $A$ is 1.14 km s$^{-1}$ and $\alpha$ is 0.55.\\

\begin{figure*}
\centering 
 \includegraphics[width=6.8in,height=8.8in,angle=0]{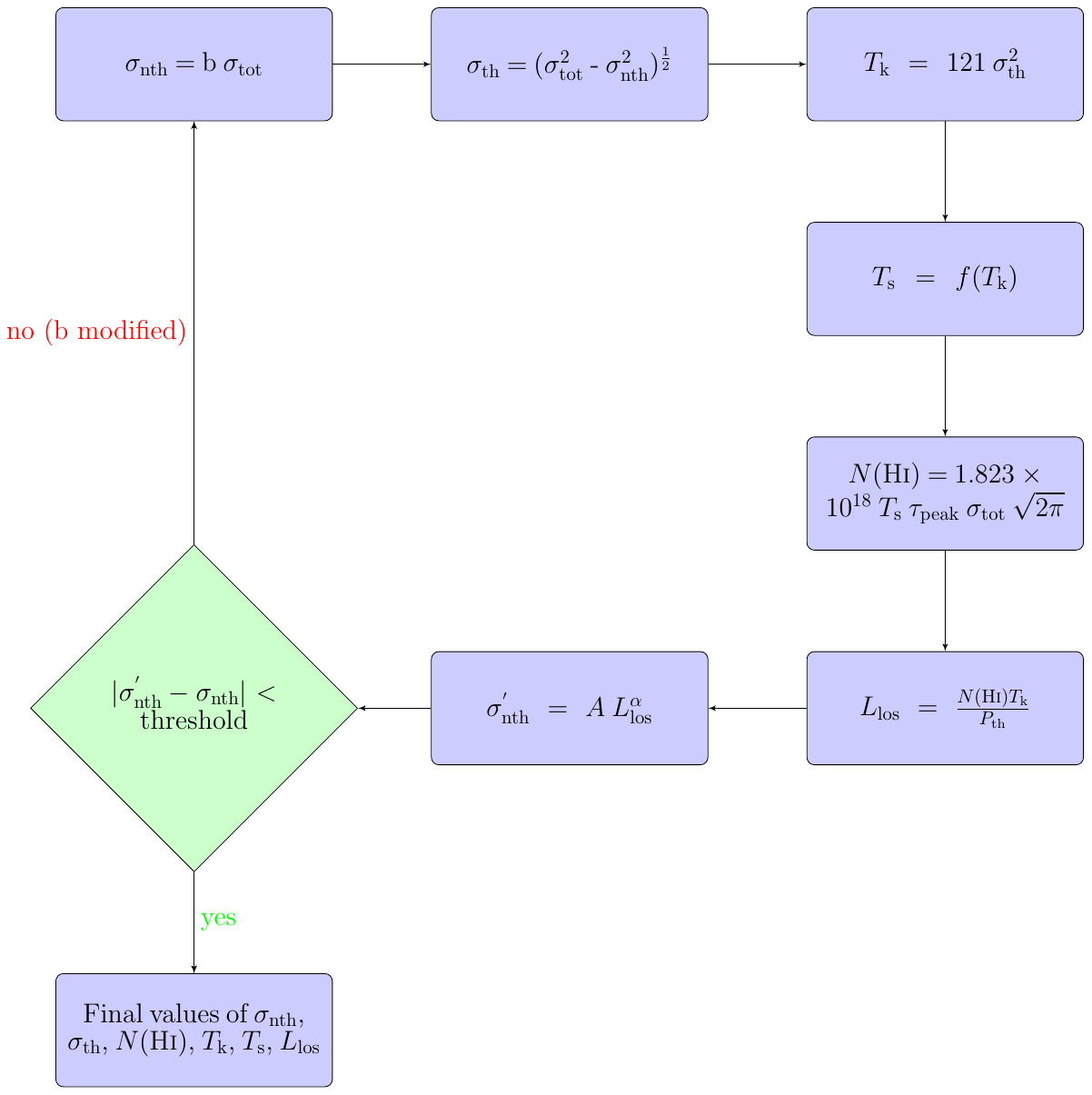}\\
\caption{Flowchart of the iterative method. Here the input parameters are total velocity dispersion ($\sigma_{\text{tot}}$), peak optical depth ($\tau_{\text{peak}}$), observed thermal pressure ($P_{\text{th}}$) \citep{2011ApJ...734...65J}, spin temperature ($T_{\text{s}}$) vs. kinetic temperature ($T_{\text{k}}$) relation \citep{2001A&A...371..698L}. Final parameters obtained from this iterative method are column density ($N$[H{\sc i}]), spin temperature ($T_{\text{s}}$), kinetic temperature ($T_{\text{k}}$), line-of-sight length scale ($L_{\text{los}}$) for each of the gas cloud component.}
  \label{fig:fig1}
\end{figure*}

\subsection{Relation between spin and kinetic temperatures}\label{subsection_2_3}

Along with the thermal pressure and the turbulent scaling law, it is also essential to establish the relationship between the spin temperature and the kinetic temperature of the gas clouds. The spin temperature $T_{\text{s}}$ determines the population ratio between the two hyperfine levels of the ground state of the H{\sc i} atom. Several mechanisms influence the population of these levels, including the impact of Lyman-$\alpha$ radiation, collisions between H{\sc i} atoms, and collisions with electrons and ions. Using empirical formulations for these processes, \citet{2001A&A...371..698L} derived the correlation between $T_{\text{s}}$ and $T_{\text{k}}$ of H{\sc i} gas under typical interstellar medium (ISM) conditions. It is important to note that these correlations depend on the thermal pressure ($P_{\text{th}}$) of the neutral ISM and the strength of the Lyman-$\alpha$ radiation field (the Wouthuysen–Field effect, hereafter WF effect). In the CNM phase, collisions are sufficiently frequent that $T_{\text{s}} \approx T_{\text{k}}$, and the effect of Lyman-$\alpha$ radiation is minimal. In contrast, within the WNM phase, $T_{\text{s}}$ is significantly lower than $T_{\text{k}}$ in the absence of this effect, but strong Lyman-$\alpha$ radiation can raise $T_{\text{s}}$ to values comparable to  $T_{\text{k}}$ \citep{1958PIRE...46..240F,2020ApJS..250....9S}. Consequently, \citet{2001A&A...371..698L} computed the $T_{\text{s}}$–$T_{\text{k}}$ relationship for different scenarios  based on varying strengths of the Lyman-$\alpha$ field. However, in most H{\sc i} absorption surveys, the majority of gas clouds belong to the CNM or UNM phase, where the Lyman-$\alpha$ radiation has little influence \citep{2003ApJ...586.1067H, 2013MNRAS.432.3074C,2018ApJS..238...14M}. Therefore, by combining the appropriate thermal pressure, the turbulent scaling law, and the conversion factor between $T_{\text{s}}$ and $T_{\text{k}}$, it is possible to determine the physical properties of H{\sc i} gas clouds solely from the absorption spectra. In the following, we describe the outline of this iterative method.\\

\subsection{Iterative Method}\label{subsection_2_4}

The absorption spectra observed toward background quasars need to first decompose in order to identify multiple gas components along the line of sight. From the decomposed spectra, the peak optical depth ($\tau_{\text{peak}}$), central velocity ($v_{\text{c}}$) and total velocity dispersion ($\sigma_{\text{tot}}$) can be obtained for each component. These parameters serve as initial input for the iterative process. In the first step, we assume that a small fraction $\text{b}$ (close to zero) of the total velocity dispersion ($\sigma_{\text{tot}}$) arises from non-thermal motions ($\sigma_{\text{nth}}$):\\

\begin{eqnarray}
\sigma_{\text{nth}} = \text{b}~\sigma_{\text{tot}}.
\end{eqnarray}

\hspace{-5mm}The remaining part is attributed to thermal broadening ($\sigma_{\text{th}}$), from which $T_{\text{k}}$ is calculated using the following relation:\\

\begin{eqnarray}
T_{\text{k}}= 121~\sigma^{2}_{\text{th}}. 
\end{eqnarray}\\

\hspace{-5mm}Now, from $T_{\text{k}}$, we obtain $T_{\text{s}}$ for each gas cloud from the numerical model of \citet{2001A&A...371..698L}. This then allows us to calculate the column density for each component by:\\

\begin{equation}
N({\text{H\sc i}}) = 1.823 \times 10^{18} \sqrt{2 \pi}~T_{\text{s}}~\tau_{\text{peak}}~\sigma_{\text{tot}}. 
\end{equation}\\

\hspace{-5mm}Consequently, we calculate the line-of-sight length scale ($L_{\text{los}}$) for the gas cloud using the formula:

\begin{equation}
L_{\text{los}} = \frac{N({\text{H\sc i}})~T_{\text{k}}}{P_{\text{th}}}.
\end{equation}\\

\hspace{-5mm}Here, $P_{\text{th}}$ denotes the thermal pressure in the neutral ISM. In molecular clouds, when the distance to a cloud is known from different observational methods such as trigonometric parallax, stellar extinction method, the projected angular distance on the plane of the sky can be converted into a physical (linear) distance. However, in the present case, the distances to individual H{\sc i} clouds are unknown, and multiple gas components are present within each velocity channel. As a result, it is not possible to convert the angular separation on the sky into a physical cloud size using distance estimates. Nevertheless, if the gas cloud is assumed to be isothermal (which is the case we have assumed in our analysis), the $L_{\text{los}}$ can be estimated from the observed $P_{\text{th}}$.\\

\hspace{-5mm}After obtaining $L_{\text{los}}$, we calculate the non-thermal velocity dispersion using the following formula:\\

\begin{equation}
\sigma{'}_{\text{nth}} = A L^{\alpha}_{\text{los}}.
\end{equation}\\

\hspace{-5mm}We have taken the normalization factor $A$ = 1.14 km s$^{-1}$ and the power law index $\alpha$ is 0.55 adopted from \cite{2023PASA...40...46K}. We then compare the newly obtained $\sigma^{'}_{\text{nth}}$ with the previous value of $\sigma_{\text{nth}}$. If the difference exceeds a defined threshold, the fraction $\text{b}$ is automatically updated to a new value, and the process continues iteratively until convergence is achieved. Once convergence is reached, we derive the column density (N[{\text{H\sc i}}]), length scale ($L_{\text{los}}$), spin temperature ($T_{\text{s}}$), and kinetic temperature ($T_{\text{k}}$) of the individual gas components. A flow chart illustrating this iterative process is shown in Fig.~\ref{fig:fig1}. We note that, in the bisection method to find a root of $f(x)$ within an interval [a,b], we first calculate $f(a)$ and  $f(b)$. If these values have opposite signs, it confirms the existence of a root within that interval. We then compute 
$c=(a+b)/2$ and evaluate  $f(a)f(c)$  and $f(b)f(c)$ to determine in which sub-interval the sign changes, thereby identifying the new range containing the root. This process is repeated iteratively until the solution converges. However, in the present iterative process, such a sign-change criterion does not exist. Therefore, if at any step we assume that the modified fraction $\text{b}$ is obtained from the midpoint of ($\sigma_{\text{nth}}$ + $\sigma^{'}_{\text{nth}}$)/2 and proceed in the same manner, convergence may not be achieved. To ensure convergence, the value of $\text{b}$ must instead be gradually incremented by a small amount in successive iterations. We implemented this iterative process on data from the ongoing Galactic H{\sc i} absorption surveys \citep{2013MNRAS.436.2366R,2024MNRAS.529.4037P}. The details of the data acquisition are discussed in the following section.\\


\section{Data Acquisition}\label{section_3}
\begin{figure*}
\centering 
 \includegraphics[width=3.2in,height=2.4in,angle=0]{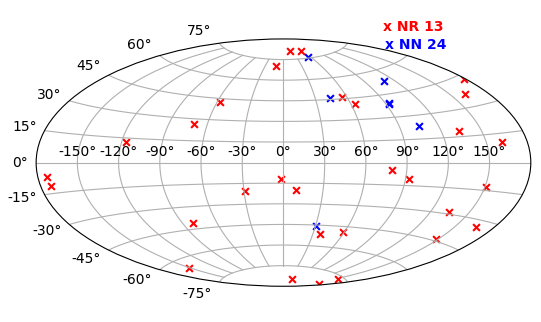}\includegraphics[width=3.4in,height=2.4in,angle=0]{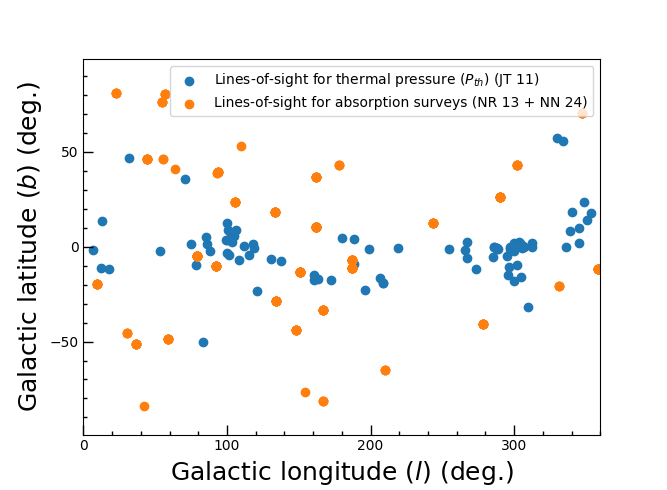}\\
 \caption{Left: The distribution of target lines of sight for detecting galactic H{\sc i} absorption used in our analysis is shown in the Mollweide projection of galactic coordinates. In the legend, NR 13 represents \citet{2013MNRAS.436.2366R} and NN 24 denotes \citet{2024MNRAS.529.4037P}. As can be seen, the sightlines are randomly distributed at Galactic latitude, thereby removing any bias in the statistical properties of the ISM. Right: Overplot of the lines of sight toward bright stars used to measure thermal pressure \citep{2011ApJ...734...65J}, shown as blue dots, together with the H{\sc i} absorption survey sight lines same as the left panel, shown as yellow dots.}
 \label{fig:fig2}
\end{figure*}

All the fitted components of the H{\sc i} absorption spectra were obtained from the works of \cite{2003MNRAS.346L..57K,2013MNRAS.436.2366R} and \cite{2024MNRAS.529.4037P}, which we have listed in Table \ref{Table:tab6} in Appendix \ref{Appendix: appendix2}. Observations were conducted with the Giant Metrewave Radio Telescope (GMRT), the Westerbork Synthesis Radio Telescope (WSRT), and the Australia Telescope Compact Array (ATCA) toward 37 radio-loud quasars. In the following, this survey is called the \texttt{GWA} survey. Of these, 30 lines of sight were observed by \citet[hereafter NR13]{2013MNRAS.436.2366R}, and 7 by \citet[hereafter NN24]{2024MNRAS.529.4037P}.  For the GMRT observations, a total bandwidth of 0.5 MHz was divided into 256 spectral channels, corresponding to a velocity coverage of approximately 105 km s$^{-1}$ and a velocity resolution of approximately 0.40 km s$^{-1}$. The WSRT observations used a bandwidth of 2.5 MHz divided into 2048 channels, providing a velocity resolution of 0.26 km s$^{-1}$ for most sources; for two sources, 1024 channels were used instead, producing a velocity resolution of 0.52 km s$^{-1}$. The ATCA observations used a bandwidth of 4.0 MHz divided into 2048 independent frequency channels, resulting in a total velocity coverage of 844 km s$^{-1}$ and a channel spacing of 0.40 km s$^{-1}$. The mean rms noise level in the optical-depth spectra ($\tau_\mathrm{rms}$) is $\sim1.07\times10^{-3}$ at the native spectral resolution and decreases to $\sim5.9\times10^{-4}$ at a spectral resolution of $\sim$ 1.0 km s$^{-1}$. Our analysis is based on the decomposed H{\sc i} absorption components from these studies. For details of the decomposition method, we refer the reader to \citet{2013MNRAS.436.2352R,2013MNRAS.436.2366R,2024MNRAS.529.4037P}. The corresponding emission spectra, used for consistency checks, were taken from the Leiden–Argentine–Bonn (LAB) survey \citep{2005A&A...440..775K,2008A&A...487..951K}, which has a slightly coarser spectral resolution of approximately 1.03 km s$^{-1}$. The total column density along each line of sight was derived using the resulting brightness temperature from the emission spectra and the optical depth profiles from the absorption data \citep{2013MNRAS.432.3074C,2013MNRAS.436.2366R,2024MNRAS.529.4037P}. The left panel of Fig. \ref{fig:fig2} shows the lines of sight of the H{\sc i} absorption survey in a Mollweide projection of Galactic coordinates. From the figure, it is evident that the lines of sight are randomly distributed, suggesting that no selection bias is present. The right panel of Fig. \ref{fig:fig2} shows the overplot between the lines of sight of the H {\sc i} absorption survey and those used in the thermal pressure study of \cite{2011ApJ...734...65J}. The close correspondence between the two sets of sight lines indicates that the observed thermal pressure is appropriately selected for the analysis.\\

In addition, we have adopted the gas components identified in the \texttt{Millennium} Survey \citep{2003ApJS..145..329H}  that are detected only in emission and have FWHM values between 20 and 30 km s$^{-1}$. The motivation for selecting these particular components is discussed in detail in a later section. We have adopted the FWHM ($\Delta V$) and column densities (N[H{\sc i}]) of these components, as listed in Table \ref{Table: tab3} in the Appendix \ref{Appendix: appendix3}. The $\tau_{\rm rms}$ in this survey is $\sim 2 \times 10^{-3}$ at 1 km s$^{-1}$, which is almost 3.4 times higher than \texttt{GWA} survey.\\

\begin{table*}
\centering
\caption{Emission and absorption derived column densities towards 37 lines of sight.}
\begin{tabular}{l c cc cc}
\hline
\multirow{2}{*}{Source} &
\multirow{2}{*}{N[H{\sc i}]$_{\text{emission}}$} &
\multicolumn{2}{c}{}N[H{\sc i}]$_{\text{absorption}}$ ( $\times$ 10$^{20}$ ) cm$^{-2}$ &
\multicolumn{2}{c}{N[H{\sc i}]$_{\text{emission-absorption}}$ ( $\times$ 10$^{20}$ ) cm$^{-2}$ } \\
\cmidrule(r{3pt}){3-4} \cmidrule(l{3pt}){5-6}
 & ( $\times$ 10$^{20}$ ) cm$^{-2}$ & $T_{\text{s}}<T_{\text{k}}$ & $T_{\text{s}}=T_{\text{k}}$ & $T_{\text{s}}<T_{\text{k}}$ & $T_{\text{s}}=T_{\text{k}}$ \\
\hline
B0023$-$263 & 1.62 & 0.33   & 0.35 & 1.29 & 1.27 \\ [0.5 ex]
B0114$-$211 & 1.42 & 0.36   & 0.38 & 1.06 & 1.04 \\ [0.5 ex]
B0117$-$155 & 1.45 & 0.21   & 0.23 & 1.24 & 1.22 \\ [0.5 ex]
B0134$+$329 & 4.33 & 2.79   & 2.91 & 1.54 & 1.42 \\ [0.5 ex]
B0202$+$149 & 4.80 & 2.89   & 3.00 & 1.91 & 1.80 \\ [0.5 ex]
B0237$-$233 & 2.13 & 2.24   & 2.46 &  --- & ---  \\ [0.5 ex] 
B0136$+$162 & 10.09 & 9.82  & 10.20  & 0.27 & --- \\ [0.5 ex]
B0316$+$413 & 13.64 & 7.16  & 7.41 & 6.48 & 6.23 \\ [0.5 ex] 
B0404$+$768 & 11.06 & 6.69  & 6.87 & 4.37 & 4.19 \\ [0.5 ex] 
B0407$-$658 & 3.41 & 3.22 & 3.46 & 0.19 & --- \\ [0.5 ex]
B0518$+$165 & 23.60 & 11.10 & 11.35 & 12.51 & 12.25 \\ [0.5 ex] 
B0531$+$194 & 28.80 & 10.50 & 10.80 & 18.30 & 18.00 \\ [0.5 ex]
B0538$+$498 & 21.31 & 12.34 & 12.71 & 8.97 & 8.60 \\ [0.5 ex] 
B0831$+$557 & 4.50 & 1.54 & 1.60 & 2.96 & 2.90 \\ [0.5 ex]
B0834$-$196 & 6.86 & 2.95 & 3.06 & 3.91 & 3.80 \\ [0.5 ex] 
B0906$+$430 & 1.28 & 1.71 & 2.12 & --- & --- \\ [0.5 ex]
B1151$-$348 & 7.06  & 2.34 & 2.42 & 4.72 & 4.64 \\ [0.5 ex]
B1245$-$197 & 3.71 & 0.64 & 0.67 & 3.07 & 3.04 \\ [0.5 ex]
B1328$+$254 & 1.11 & 0.14 & 0.15 & 0.97 & 0.96 \\ [0.5 ex]
B1328$+$307 & 1.22 & 0.75 & 0.87 & 0.47 & 0.35 \\ [0.5 ex]
B1345$+$125 & 1.92 & 1.31 & 1.35 & 0.61 & 0.57 \\ [0.5 ex]
B1611$+$343 & 1.36 & 0.21 & 0.22 & 1.15 & 1.14 \\ [0.5 ex]
B1641$+$399 & 1.08 & 0.13 & 0.14 & 0.95 & 0.94 \\ [0.5 ex]
B1814$-$637 & 6.66 & 2.92 & 3.02 & 3.74 & 3.64 \\ [0.5 ex]
B1827$-$360 & 8.13 & 4.88 & 5.04 & 3.25 & 3.09 \\ [0.5 ex]
B1921$-$293 & 7.44 & 5.70 & 6.11 & 1.74 & 1.33 \\ [0.5 ex]
B2050$+$364 & 28.53 & 11.56 & 12.10 & 16.97 & 16.43 \\ [0.5 ex]
B2200$+$420 & 18.59 & 6.64 & 6.84 & 11.95 & 11.75 \\ [0.5 ex]
B2203$-$188 & 2.38 & 2.03 & 2.42 & 0.35 & --- \\ [0.5 ex]
B2223$-$052 & 4.61 & 3.28 & 3.39 & 1.33 & 1.22 \\ [0.5 ex]
1352$+$314 & 1.20 & 0.87 & 0.94 & 0.33 & 0.26 \\ [0.5 ex]
1400$+$621 & 1.60 & 0.01 & 0.01 & 1.59 & 1.59 \\ [0.5 ex]
1609$+$266 & 3.80 & 3.26 & 3.37 & 0.54 & 0.43 \\ [0.5 ex]
1634$+$627 & 1.80 & 0.95 & 0.99 & 0.85 & 0.81 \\ [0.5 ex] 
1638$+$625 & 2.30 & 0.39 & 0.40 & 1.91 & 1.90 \\ [0.5 ex]
1927$+$739 & 7.00 & 5.86 & 5.87& 1.14 & 1.13 \\ [0.5 ex]
2137$-$207 & 3.00 & 1.62 & 1.68 & 1.38 & 1.32 \\ [0.5 ex]
\hline
\end{tabular}\\
\textbf{Notes.} The first column lists the background quasars. The second column presents the emission column density for each line of sight, estimated using the isothermal approximation \citep{2013MNRAS.432.3074C, 2013MNRAS.436.2352R, 2024MNRAS.529.4037P}. The third and\\
\hspace{2mm}fourth columns show the absorption column density for each line of sight for the two cases $T_\text{s} < T_\text{k}$ and $T_\text{s} = T_\text{k}$, respectively. The fifth and sixth columns list the differences between the emission and absorption column densities for each line of sight in these two \\
\hspace{-174mm}cases.\\
 \label{Table: tab1}
\end{table*}

\begin{figure*}
\centering 
 \includegraphics[width=3.4in,height=2.8in,angle=0]{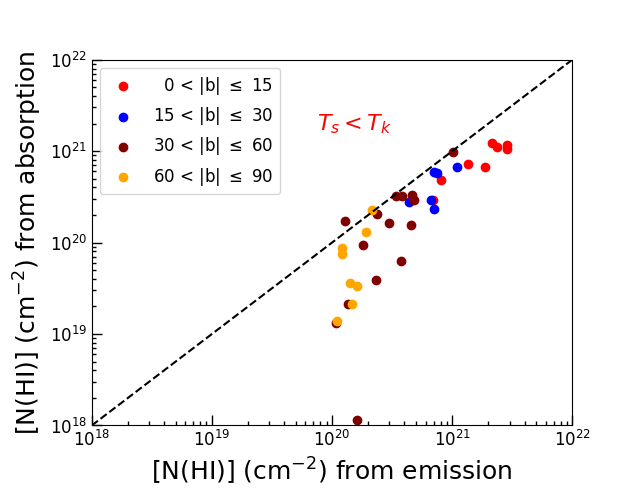}\includegraphics[width=3.4in,height=2.8in,angle=0]{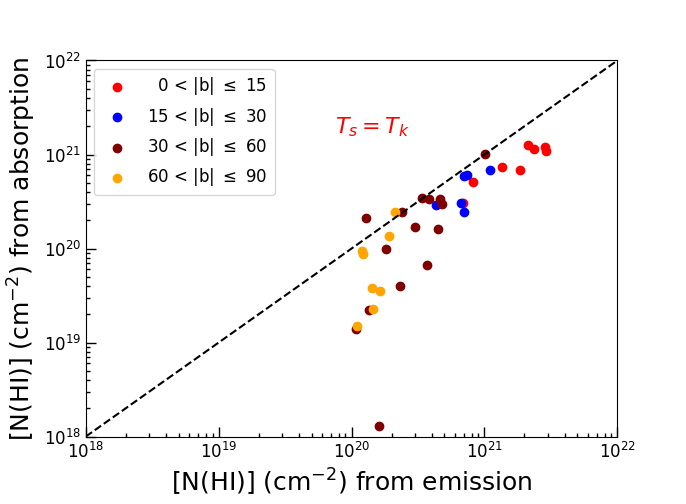}\\
 \caption{Comparisons of the emission and absorption spectra for 37 lines of sight (\texttt{GWA survey}) when $T_\text{s} < T_\text{k}$ (left) and when $T_\text{s} = T_\text{k}$ (right). The column densities from the absorption spectra are obtained using the iterative method, while the emission spectra used for comparison are taken from the \texttt{LAB survey}.}
 \label{fig:fig3}
\end{figure*}

\begin{figure*}
\centering 
 \includegraphics[width=3.0in,height=2.6in,angle=0]{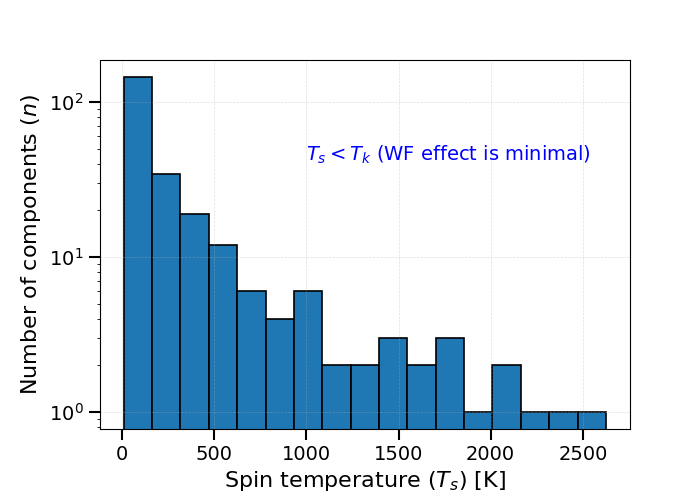}
  \includegraphics[width=3.0in,height=2.6in,angle=0]{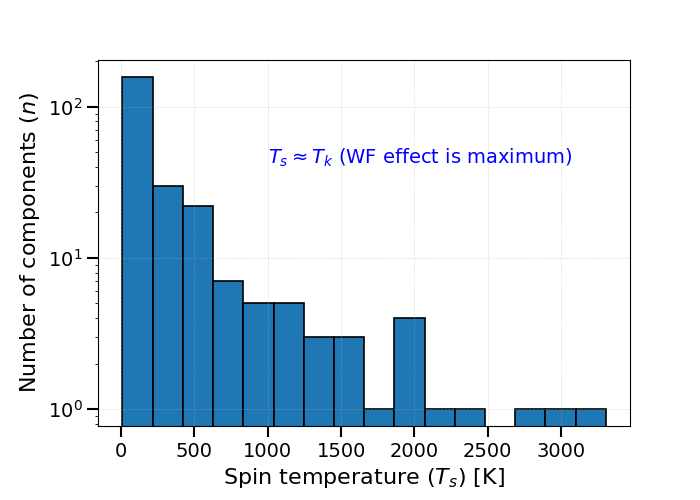}
\caption{Distribution of the spin temperature ($T_\text{s}$) of the number of components obtained from \texttt{GWA} absorption survey for two cases: (a) when $T_\text{s} < T_\text{k}$; (b) when $T_\text{s} \approx T_\text{k}$.}
 \label{fig:fig100}
 \end{figure*}

\begin{figure*}
\centering 
 \includegraphics[width=7.4in,height=4.6in]{figures/combined_figure.png}
 \caption{Emission–absorption spectra toward eight quasars. The top panel in each figure shows the emission spectrum obtained from the LAB survey, the middle panel in each figure shows the absorption spectrum from the GWA survey, and the bottom panel presents the channel-wise spin temperature derived from the comparison of the emission and absorption spectra. The emission and absorption spectra are shown at their original resolutions, and the channel-wise spin temperature is derived after spectral smoothing the absorption spectra to match the (slightly coarser) resolution of the emission spectra \citep{2013MNRAS.436.2352R, 2024MNRAS.529.4037P}. Reliable spin temperature measurements are indicated by red points with associated error bars, while lower limits are shown as blue points. The two vertical orange dashed lines mark the channel range where reliable spin temperature measurements were not possible due to the non-detection of broad wing like components in absorption. Figures adapted from \citet{2013MNRAS.436.2352R} and \citet{2024MNRAS.529.4037P}.}
 \label{fig:fig4}
\end{figure*}

\section{Results from the Iterative Method}\label{section_4}

We applied the iterative method to the ongoing GWA survey, which contains a total of 244 decomposed gas cloud components along 37 lines of sight. Using this method, we derived all the relevant physical parameters for each component, including spin temperature, kinetic temperature, column density, and characteristic length scale. After deriving the column density for each component from the absorption spectra, we sum the column densities of all the gas components along each line of sight and compared these values with the emission-derived column densities for each line of sight, specifically quantifying the differences between the emission and absorption measurements. We performed this analysis for two cases: when $T_\text{s} < T_\text{k}$ (WF effect minimal) and $T_\text{s} = T_\text{k}$ (WF effect maximum), derived from the theoretical model of \cite{2001A&A...371..698L}  A summary of these results is presented in Table \ref{Table: tab1}. We find that on average about 50\% of the total column density is not detected in absorption (in both cases when $T_\text{s} < T_\text{k}$ and $T_\text{s} = T_\text{k}$). This missing fraction is mainly attributed to the Warm Neutral Medium (WNM) or the higher-temperature components of the Unstable Neutral Medium (UNM), whose optical depths are relatively low and remain below the detection limit of the \texttt{GWA} survey \citep{2013MNRAS.436.2352R,2024MNRAS.529.4037P}. We present a comparison of the emission and absorption spectra for 37 different lines of sight in Fig. \ref{fig:fig3} for both cases. We note that, we divided the lines of sight based on the Galactic latitude ($b$) to verify that the gas content is higher toward the Galactic plane and gradually decreases with increasing latitude. In a few cases, particularly at high Galactic latitudes, the absorption spectra are significantly weaker than the emission spectra. The primary reason is that noise in the absorption spectra makes it difficult to detect all gas components compared to the emission spectra \citep{2013MNRAS.436.2352R, 2024MNRAS.529.4037P}. From this comparison, we find that most WNM components are not detected in the absorption survey. Below, we investigate the evidence for the non-detection of WNM in the \texttt{GWA} survey using two independent approaches.\\


\subsection{Non-detection of WNM gas clouds in absorption}\label{non_detection}

We first examine the spin temperature distribution of the gas components derived using the iterative method, as shown in Fig. \ref{fig:fig100}. We have shown for two cases: $T_\text{s} < T_\text{k}$ and $T_\text{s} = T_\text{k}$. In the first case, we obtained components with temperatures up to $2625~\mathrm{K}$, while in the second case, components are obtained up to $3304~\mathrm{K}$. Thus, gas components with temperatures ranging from approximately $3500$ to $10{,}000~\mathrm{K}$ are not detected in the GWA survey. These components primarily correspond to the upper-temperature portion of the UNM and the entire WNM phase. In addition to this evidence, we analyze the emission and absorption spectra from the \texttt{LAB} and \texttt{GWA} surveys. A careful inspection reveals that, in several cases, broad wing-like components clearly detected in emission are not observed in absorption. We present a few representative examples of such spectra from the \texttt{GWA} survey in Fig. \ref{fig:fig4}. We show the spectra toward the quasars 1634$+$627, 1638$+$625, 1927$+$739, B1921$-$293, B2200$+$420, B0834$-$196, B1245$-$197, and B1345$+$125. In each case, the top panel shows the emission spectra, the middle panel displays the absorption spectra, and the bottom panel illustrates the channel-wise spin temperature derived by comparing the emission and absorption spectra. In channels where the signal exceeds 3$\times$ rms noise ($\sigma_\text{rms}$) in both emission and absorption, the spin temperature is reliably measured and shown as red points with associated error bars. In contrast, in channels where only emission is detected, we derive lower limits on the spin temperature, indicated by blue points. In each figure, we mark two vertical orange dashed lines indicating the channel range where the spectra are detected only in emission and not in absorption, as supported by the blue points in the channel-wise spin temperature plots.
For the case of B1921$-$293, a few red points appear in the channel-wise spin temperature plot in between the two orange dashed lines; however, these are likely due to noise fluctuations, given the 3$\sigma_\text{rms}$ cutoff. In particular, only two consecutive channels show such red points, suggesting a high probability that they arise from noise in the absorption spectrum rather than a genuine signal. From this comparison of channel-wise emission and corresponding absorption spectra, it is evident that the broad wing-like WNM components, which have very low optical depth, are not reliably detected in the current GWA survey. As a result, the absorption-derived column densities are systematically lower than those obtained from emission along each line of sight which is portrayed in Fig. \ref{fig:fig3}.\\

\begin{figure*}
\centering 
 \includegraphics[width=3.4in,height=2.8in,angle=0]{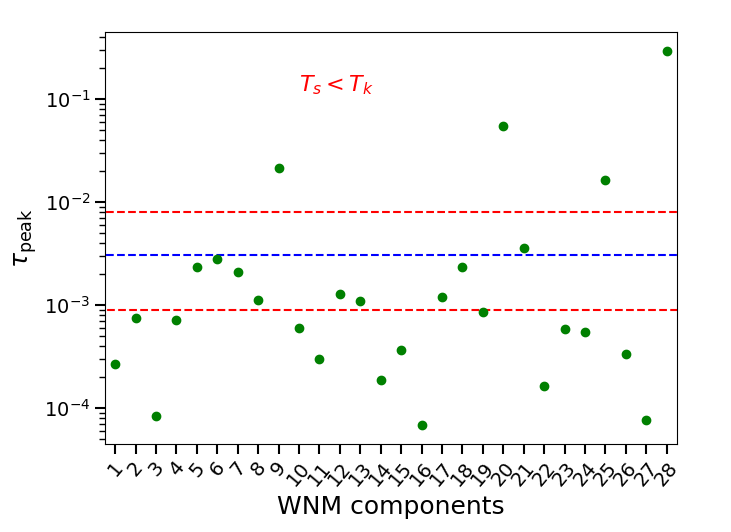}\includegraphics[width=3.4in,height=2.8in,angle=0]{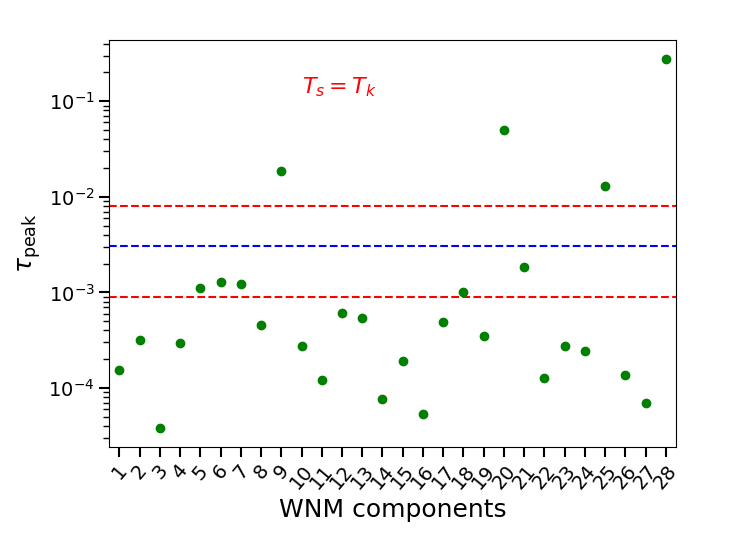}
 \caption{Left: $\tau_{\text{peak}}$ of the WNM components obtained from the \texttt{Millennium survey} where $T_{\text{k}}$ is between 5000K and 10000 K. Two red dashed lines are the  3$\tau_{\text{rms,min}}$ and the 3$\tau_{\text{rms,max}}$, whereas the blue dashed line  3$\tau_{\text{rms,mean}}$  of the GWA survey. This is the case with minimal WF effect, where $T_{\text{s}}$ $<$ $T_{\text{k}}$. Right: Same as the left panel, but for the extreme WF effect, where  $T_{\text{s}}$ = $T_{\text{k}}$.}
 \label{fig:fig5}
\end{figure*}

We adopt an additional approach to investigate the non-detection of WNM components. For this purpose, we selected specific components from the \texttt{Millennium} survey \citep{2003ApJS..145..329H}, whose full widths at half maximum (FWHM) lie between 20 and 30 km s$^{-1}$ and only detected in emission. We have excluded few extremely broad components (FWHM $\geq$ 30 km s$^{-1}$) from our analysis for two reasons which are also detected in emission. First, baseline stability can be a concern in single-dish observations, and such broad features may arise from improper baseline subtraction of the spectra. Second, as shown by \cite{1979MNRAS.186..479L}, the turbulence scaling relation at larger scales ($\geq$ 1 kpc) differs significantly from that at smaller scales: $\alpha$ becomes shallower than the Kolmogorov law of turbulence ($\sim$ 0.16), with a normalization of $A$ $\sim$ 2.5 km s$^{-1}$. At these scales, turbulence is primarily driven by large-scale processes such as galactic infall and shear. For these reasons, we removed these few components from the analysis. Likewise, we did not take into account components with FWHM $<20$ km s$^{-1}$. This is because the mean rms noise level of the optical-depth spectra ($\tau_\mathrm{rms}$) in the \texttt{Millennium} Survey is $\sim2\times10^{-3}$ at 1 km s$^{-1}$, which is approximately 3.4 times higher than that of the \texttt{GWA} survey, where $\tau_\mathrm{rms}\sim5.9\times10^{-4}$ at 1 km s$^{-1}$ \citep{2013MNRAS.436.2352R}. Consequently, several components that remain undetected in the \texttt{Millennium Survey} can be readily detected in the \texttt{GWA} survey. Therefore, to specifically examine the detectability of WNM components, we restrict our analysis to components with FWHM between 20 and 30 km s$^{-1}$. There are a total of 169 components in the \texttt{Millennium Survey} that are detected only in emission. Of those, 102 components have FWHM $<20$ km s$^{-1}$, 39 components have FWHM $>30$ km s$^{-1}$, and the remaining 28 components have FWHM between 20 and 30 km s$^{-1}$. We therefore select those 28 components for our analysis.

 To obtain both kinetic and spin temperatures of those gas components, it is necessary to consider the turbulence modeling of the WNM gas and the conversion between kinetic and spin temperatures, with and without the WF effect \citep{2001A&A...371..698L}. For the WNM medium, the turbulence scaling law is adopted from \cite{2003ApJ...587..278W}, where the parameter $A$ is 1.4 km s$^{-1}$ and $\alpha$ is 0.33, values appropriate for a subsonic medium \citep{1941DoSSR..32...16K}. After adopting the turbulence scaling relation, we solve the following nonlinear equation to obtain the length scale of the gas components:\\

\begin{equation}
    \Biggl\{L_{\text{los}} - \frac{121~ N(\mathrm{H\,{\scriptstyle I}})\bigl(\sigma^{2}_{\text{tot}}-A^{2}L_{\text{los}}^{2\alpha}\bigl) }{P_{\text{th}}}\Biggl\} = 0.
\end{equation}

Once $L_{\text{los}}$ is determined, the kinetic temperature is calculated from the following relation:

\begin{equation}
  T_{\text{k}}  = \frac{L_{\text{los}}~P_{\text{th}}}{N(\mathrm{H\,{\scriptstyle I}})}.
\end{equation}

After obtaining the kinetic temperature, we consider two scenarios. In the first case, where the WF effect is weak, $T_{\text{s}}$ is generally lower than $T_{\text{k}}$, as predicted by the empirical relation of \cite{2001A&A...371..698L}. In the second case, when the WF effect is strong, $T_{\text{s}}$ becomes nearly equal to $T_{\text{k}}$ \citep{2001A&A...371..698L}. Using the column density and the spin temperature, we calculate the peak optical depth of the line using the following formula:

\begin{equation}
    \tau_{\text{peak}} = \frac{N(\mathrm{H\,{\scriptstyle I}})}{1.823\times 10^{18}\sqrt{2\pi}~\sigma_{\text{tot}}~T_{\text{s}}}.
\end{equation}

\hspace{-5mm}The $\tau_{\text{peak}}$ values of the components for the two scenarios are listed in Table \ref{Table: tab3} in the Appendix \ref{Appendix: appendix3}. We also show the $\tau_{\text{peak}}$ values of the components for the two scenarios in Fig. \ref{fig:fig5}. Additionally, we overplot the three times the minimum, maximum, and mean of $\tau_{\text{rms}}$ (3$\tau_{\text{rms,min}}$, 3$\tau_{\text{rms,max}}$, 3$\tau_{\text{rms,mean}}$), where $\tau_{\text{rms}}$ is the noise level in the optical depth spectra of the \texttt{GWA} survey. We have listed the  back continuum flux values ($S_{1.4}$) and $\tau_{\text{rms}}$ for each line of sight in \texttt{GWA} survey in Table \ref{Table: tab4} in Appendix \ref{Appendix: appendix4}. We find that most of the components in \texttt{GWA} survey lie below the detection limits of the GMRT, ATCA, and WSRT observations. Future high-sensitivity facilities such as the Next Generation Very Large Array (ngVLA) or Square Kilometer Array (SKA1-MID) will be required to detect these components within a reasonable integration time. From a few minutes to within $\sim1.5$ hours, the WNM components can be readily detected in absorption using these telescopes. We discuss this in greater detail in Appendix \ref{Appendix: appendix4_0}.\\

\begin{figure*}
\centering 
 \includegraphics[width=6.8in,height=2.9in,angle=0]{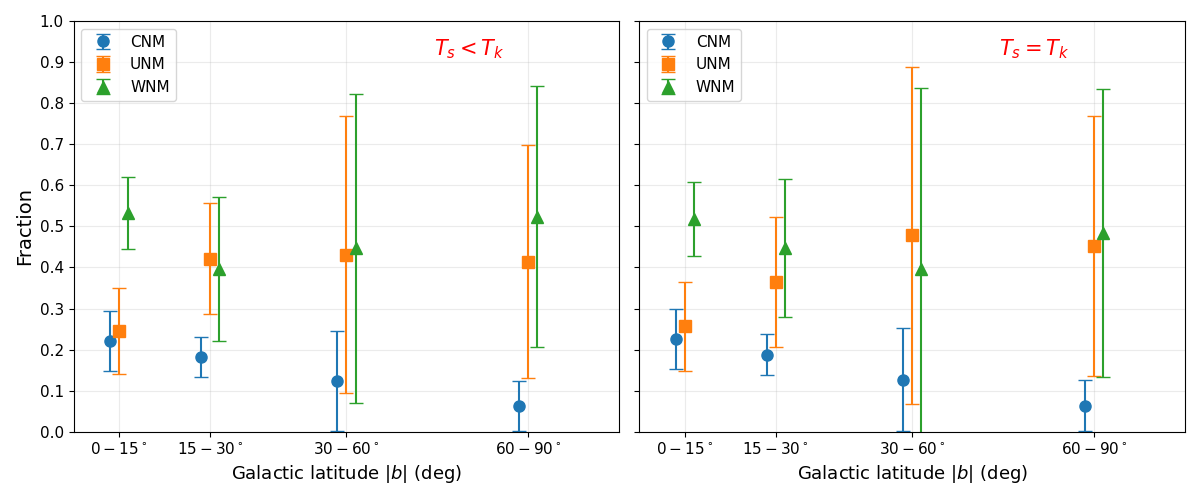}\\
\caption{Mean CNM ($f_\text{CNM}$), UNM ($f_\text{UNM}$), and WNM ($f_\text{WNM}$) fractions as a function of Galactic latitude $|b|$ for the two assumptions $T_\text{s}<T_\text{k}$ (left) and $T_\text{s}=T_\text{k}$ (right). The symbols represent the mean phase fractions within each Galactic-latitude bin, while the error bars indicate the $1\sigma$ scatter among the individual sightlines within each bin.}
 \label{fig:fig7_0}
 \end{figure*}

\begin{figure*}
\centering 
 \includegraphics[width=6.8in,height=2.9in,angle=0]{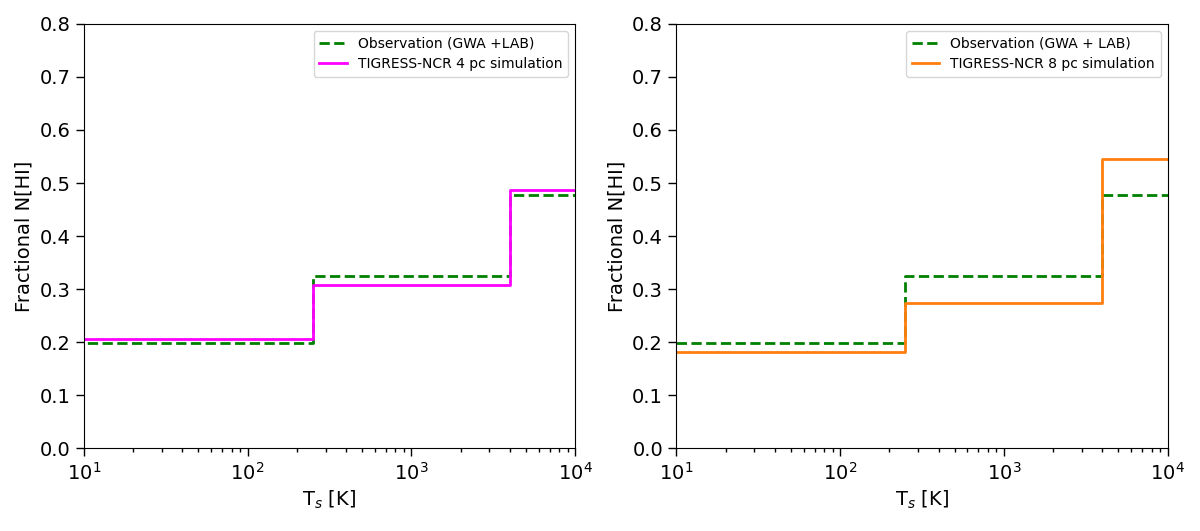}\\
  \includegraphics[width=6.8in,height=2.9in,angle=0]{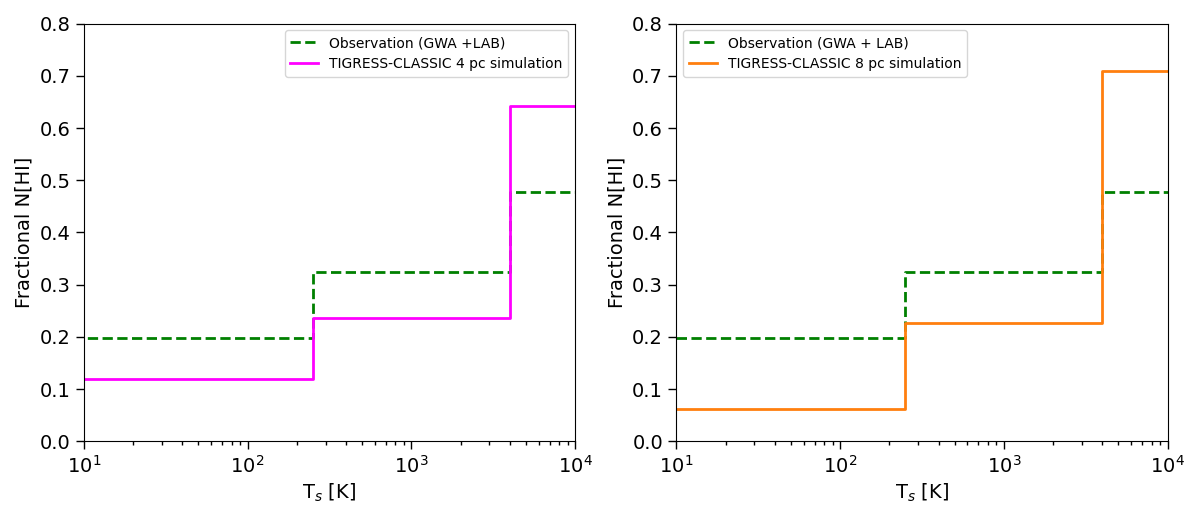}\\
\caption{Upper: Comparison of the gas phase distribution derived from our observational modeling with the R8~TIGRESS-NCR simulation \citep{2023ApJ...946....3K} at two different physical resolutions (4 pc and 8 pc). In both panels, the gas column density (N[H{\sc i}]) is shown as a function of spin temperature ($T_\text{s}$). Lower: Same as the upper panel, except the comparison is performed with the R8 TIGRESS-CLASSIC simulation \citep{KimOstriker2017}.}
 \label{fig:fig7}
 \end{figure*}


\subsection{Determination of the gas fraction}
Based on the verification of the non-detection of WNM gas clouds above, we infer that the missing column density in the absorption spectra primarily arises from WNM components. However, we note that this assumption is not strictly valid, as the UNM gas at the higher end of the temperature range may also remain undetected because of the low optical depth of the line \citep{2003ApJ...586.1067H,2018ApJS..238...14M}. In our analysis, we obtain the gas components for which the maximum values of $T_\text{s}'$s are 2625 and 3304 K when $T_\text{s} < T_\text{k}$ and $T_\text{s} = T_\text{k}$, respectively. We show the distribution of spin temperature ($T_\text{s}$) for the components in two cases: (a) $T_\text{s} < T_\text{k}$ and (b) $T_\text{s} = T_\text{k}$, in Fig.~\ref{fig:fig100}. Nevertheless, for the purposes of this analysis, we adopted the working assumption that the majority of the non-detected absorption components are associated with the WNM. Consequently, we find that $\sim$19.5\% of the gas resides in the CNM phase, $\sim$32.0\% in the UNM phase and $\sim$48.5\% in the WNM phase from the comparison of \texttt{GWA} and \texttt{LAB} surveys. This is for the case where $T_\text{s} <T_\text{k}$. For $T_\text{s} = T_\text{k}$ we find that $\sim$ 20\% of the gas is in the CNM, $\sim$33\% in the UNM, and $\sim$47\% in the WNM. The results in both cases are nearly the same. This is because most of the clouds are in the CNM and in a lower temperature range of UNM in the absorption survey. After averaging both cases, we obtain 19.75\% of the gas in the CNM, and about 32.50\% and 47.75\% in the UNM and WNM phases, respectively, in the neutral ISM. We note that the precise phase boundaries among the three phases are not well constrained by observations. Several studies suggest that, due to the effect of strong Lyman-$\alpha$ photons, the spin temperature of the UNM components can reach 4000 K, followed by the WNM component where the spin temperature can reach up to 8000 K \citep[and references therein]{1958PIRE...46..240F,2020ApJS..250....9S,2023ARA&A..61...19M}. In this study, we thus assume that the spin temperature of the UNM component can reach up to 4000 K, followed by the WNM component. \cite{2024MNRAS.527.8475B} summarized the gas phase distribution for the \texttt{SPONGE} and \texttt{Millennium} surveys using both emission and absorption spectra. According to their investigation, adopting phase boundaries of spin temperature as 250 K for CNM, 250$-$1000 K for UNM, and above 1000 K for WNM, they found that in the \texttt{SPONGE} survey approximately 28\% of the gas is in the CNM, 20\% in the UNM, and 51\% in the WNM phase. In contrast, for the \texttt{Millennium} survey, about 40\% is in the CNM, 24\% in the UNM, and 36\% in the WNM phase based on the phase boundaries of maximum kinetic temperature ($T_\text{k,max}$) of $<$ 250 K, between 250 $-$ 5000 K and $>$ 5000 K. Similarly to the \texttt{SPONGE} survey, we also find that nearly half of the gas fraction resides in the WNM phase, which remains largely undetected in current absorption studies. Furthermore, we also compared our results with the study of \cite{KalberlaHaud2018}, where they examined the gas phase distribution from the multi-Gaussian distribution of the H{\sc i} emission spectra obtained from the HI4PI survey \citep{2016A&A...594A.116H} and obtained that 24$\%$ gas in the CNM, 44$\%$  in the UNM and 32$\%$ in the WNM, respectively. These results mainly in the CNM phase are closely related to our results. However, there is a slight discrepancy between the UNM and WNM phases. We note that in our work beyond $\sim$ 2000 K fewer components are detected in absorption, and thus whatever the missing column density obtained from the comparison of emission and absorption spectra, we consider them as WNM gas clouds. However, this is not entirely correct. The higher temperature UNM clouds also belong to this. In the future, when absorption spectra will be observed with high sensitivity SKA-MID or ngVLA telescope, we hope that this UNM fraction will increase a bit and the WNM fraction will lower a bit and match well with this work. All these observational results, including our results, are summarized in Table \ref{tab:phase_fraction_comparison}.\\
 

We have also studied the distribution of the different H{\sc i} phases as a function of Galactic latitude to investigate whether variations in the Galactic environment are reflected in the relative contributions of the different gas phases. We divide the lines of sight into four different Galactic latitude bins: $0^\circ < |b| < 15^\circ$, $15^\circ < |b| < 30^\circ$, $30^\circ < |b| < 60^\circ$, and $60^\circ < |b| < 90^\circ$. The corresponding results are summarized in Table \ref{tab:phase_fraction} and Fig. \ref{fig:fig7_0} for the two assumptions, $T_\text{s}<T_\text{k}$ and $T_\text{s}=T_\text{k}$. The overall behavior is similar for both cases. In particular, the CNM fraction decreases systematically with increasing Galactic latitude, from $\sim22\%$ at $0^\circ<|b|<15^\circ$ to $\sim6\%$ at $60^\circ<|b|<90^\circ$. In contrast, the WNM contributes $\sim40$--$53\%$ of the total H{\sc i} column density across the different latitude bins, without showing a clear monotonic trend with $|b|$. The UNM fraction is relatively low near the Galactic plane, with a mean fraction of $\sim25\%$, and increases to $\sim41$--$43\%$ at higher latitudes for the $T_\text{s}<T_\text{k}$ assumption. For the $T_\text{s}=T_\text{k}$ assumption, the UNM fraction varies from $\sim26\%$ near the Galactic plane to $\sim48\%$ in the $30^\circ<|b|<60^\circ$ bin. However, the relatively large scatter, particularly at intermediate and high Galactic latitudes, indicates substantial sightline-to-sightline variation in the phase fractions. The similarity of the CNM fractions obtained under the two assumptions suggests that the observed decrease in the CNM contribution with increasing Galactic latitude is robust to the adopted spin-temperature assumption. Overall, these results indicate a systematic change in the relative contributions of the H{\sc i} phases with Galactic latitude, with the CNM becoming less dominant away from the Galactic plane and a substantial contribution from the UNM and WNM at higher latitudes.\\

Our results can be placed in the broader context of the Galactic multi-phase H{\sc i} distribution studied by \citet{KalberlaHaud2018} using the all-sky \texttt{HI4PI} survey. Their Gaussian decomposition separated the H{\sc i} emission into CNM, UNM, and WNM components and demonstrated that the relative contributions of these phases depend on the Galactic environment. In particular, their analysis showed that the CNM is more strongly associated with compact and structured H{\sc i} features, whereas the WNM is more smoothly distributed over large spatial scales. They also found that the intermediate-temperature UNM constitutes a substantial fraction of the Galactic H{\sc i}, rather than being a negligible transition component between the CNM and WNM. These results provide a useful context for our finding that the UNM and WNM remain a substantial contributor to the H{\sc i} column density over the full range of Galactic latitudes, while the CNM fraction decreases systematically away from the Galactic plane. \\

\subsection{Comparison with numerical simulation}

We also compared our observationally derived neutral ISM phase fractions with the numerical simulation results of \citet{KimOstriker2017,2023ApJ...946....3K}, who carried out high-resolution, three-dimensional magnetohydrodynamic (MHD) simulation of galactic disks within the Three-phase Interstellar Medium in Galaxies Resolving Evolution with Star Formation and Supernova Feedback (\texttt{TIGRESS}) framework to study how star formation feedback regulates the multiphase neutral interstellar medium on kpc scales. In this simulation, the evolution of the ISM is governed by the coupled effects of supernova (SN) explosions and far-ultraviolet (FUV) radiation from young stars, which together determine the dynamical and thermal state of neutral gas. Supernova feedback injects momentum and energy into the surrounding medium, driving turbulence, establishing vertical pressure support against gravity, and regulating the mid-plane pressure, while FUV photoelectric heating controls the thermal equilibrium of neutral gas through a balance with radiative cooling. As a consequence of this interplay between turbulent pressure regulation and thermal processes, the simulated ISM naturally separates into the cold neutral medium (CNM), the thermally unstable neutral medium (UNM) and the warm neutral medium (WNM), allowing a direct and physically motivated quantification of the mass fraction in each phase. Compared to \texttt{TIGRESS-CLASSIC}, \texttt{TIGRESS-NCR}; where NCR stands for Non-equilibrium Cooling and Radiation includes several additional salient features. These include explicit UV radiative transfer implemented via an adaptive ray-tracing method and direct photochemistry model. In \texttt{TIGRESS-CLASSIC}, the cooling and heating functions are spatially uniform but time-dependent. In contrast, in \texttt{TIGRESS-NCR} both cooling and heating functions depend on space and time. This leads to FUV shielding in high-density regions, resulting in a significantly enhanced cold neutral medium (CNM) at the expense of unstable neutral medium (UNM) and warm neutral medium (WNM) gas. Consequently, \texttt{TIGRESS-NCR} provides a more realistic estimate of the gas fraction in each phase. The simulation was carried out in two different density environments. One corresponds to solar-neighborhood conditions and is denoted as R8, while the other represents a higher density or pressure environment, denoted as LGR4, which is close to the molecular-gas–weighted mean conditions observed in the Physics at High Angular resolution in Nearby GalaxieS (\texttt{PHANGS}) survey. For each environment, the simulation was performed at two different spatial resolutions. In the \texttt{R8~TIGRESS-NCR} and \texttt{R8~TIGRESS-CLASSIC} simulations, the resolutions are 4 pc and 8 pc. In \texttt{R8~TIGRESS-NCR} 4 pc simulation, \cite{2023ApJ...946....3K} obtained gas of 20.5\% in CNM, 30.7\% in UNM, and 48.8\% in WNM phase, and in \texttt{R8~TIGRESS-NCR} 8 pc simulation they obtained gas of 18.0\% in CNM, 27.4\% in UNM, and 54.6\% in WNM phase, respectively within a galactic height of 300 pc. Similarly, \cite{KimOstriker2017} obtained gas of 12.0\% in CNM, 23.7\% in UNM and 64.3\% in the WNM phase in \texttt{R8~TIGRESS-CLASSIC} 4 pc simulation and in \texttt{R8~TIGRESS-CLASSIC} 8 pc simulation they obtained gas of 6.3\% in CNM, 22.7\% in UNM and 71.0\% in the WNM phase, respectively. As expected, the \texttt{R8~TIGRESS-CLASSIC} simulation contains a smaller fraction of gas in the CNM and a larger fraction in the WNM compared to the \texttt{R8~TIGRESS-NCR} simulation at both resolutions due to the absence of FUV shielding. In Fig. \ref{fig:fig7}, we overplot the gas fractions for each simulation set and the observed results obtained from \texttt{the GWA} survey. In particular, in the upper panel, we overplot \texttt{R8~TIGRESS-NCR} at 4 pc and 8 pc resolutions with observed results, and in the bottom panel, we overplot \texttt{R8~TIGRESS-CLASSIC} at 4 pc and 8 pc resolutions with observed results. From the comparison, we find that the \texttt{R8~TIGRESS-NCR} simulation provides a close match to the \texttt{GWA} results, outperforming the \texttt{R8~TIGRESS-CLASSIC} simulation at both resolutions.\\

\begin{table*}
\centering
\caption{Comparison of the CNM, UNM, and WNM fractions reported in different observational and simulation studies.}
\label{tab:phase_fraction_comparison}
\begin{tabular}{l l l c c c}
\hline
\hline
Work & Definition of phases & Method &
${f_{\rm CNM}}$ & ${f_{\rm UNM}}$ & ${f_{\rm WNM}}$ \\
\hline

\multicolumn{6}{c}{Observational works} \\

\hline
GWA (this work) &$T_{\rm s}$ (K): $<250$; $250$--$4000$; $4000$--$10000$
& Joint emission-absorption
& 19.8 & 32.5 & 47.8 \\

Millenium & $T_{\rm k,max}$ (K): $<500$; $500$--$5000$; $> 5000$
& Joint emission-absorption
& 40 & 24 & 36 \\

SPONGE & $T_{\rm s}$ (K): $<250$; $250$--$1000$; $1000$--$10000$
& Joint emission-absorption
& 28 & 20 & 51 \\

\citep{KalberlaHaud2018} & FWHM (km s$^{-1}$): 3.6$\pm$0.1; 9.6$\pm$1.3; 23.3$\pm$0.6
& Only emission
& 24 & 44 & 32 \\

\hline

\multicolumn{6}{c}{Simulation works} \\

\hline

TIGRESS-CLASSIC (4 pc) & $T_{\rm s}$ (K): $<250$; $250$--$4000$; $4000$--$10000$
&  ---
& 12.0 & 23.7 & 64.3 \\

TIGRESS-CLASSIC (8 pc) & $T_{\rm s}$ (K): $<250$; $250$--$4000$; $4000$--$10000$
&  ---
& 6.3  & 22.7 & 71.0 \\

TIGRESS-NCR(4 pc) & $T_{\rm s}$ (K): $<250$; $250$--$4000$; $4000$--$10000$
& ---
& 20.5 & 30.7 & 48.8 \\

TIGRESS-NCR(8 pc) & $T_{\rm s}$ (K): $<250$; $250$--$4000$; $4000$--$10000$
& ---
& 18.0 & 27.4 & 54.6 \\

\cite{2005AA...433....1A} & $T_{\rm k}$ (K): $<200$; $200$--$5000$; $5000$--$10000$
& ---
& 10$-$30 & 10$-$30 & 60 \\

SILCC & $T_{\rm k}$ (K): $<300$; $300$--$4000$; $4000$--$10000$
& ---
& 60 & 14 & 26 \\

\cite{2024MNRAS.527.8475B} & $T_{\rm k}$ (K): $<200$; $200$--$5000$; $5000$--$10000$
& ---
& 32 & 34 & 34 \\

\cite{2024MNRAS.527.8475B} & $T_{\rm s}$ (K): $<250$; $250$--$1000$; $1000$--$10000$
& ---
& 36 & 13 & 51 \\

\cite{2026arXiv260117255H} & $T_{\rm k}$ (K): $<200$; $200$--$5000$; $5000$--$10000$
& ---
& 21.8 & 44.3 & 34.0 \\

\hline
\hline

\end{tabular}

\textbf{Notes.} Column 1: Different observational and simulation studies, including our work. Column 2: Definitions of the gas phases adopted in each study. Column 3: Methods used to determine the gas-phase fractions. Columns 4, 5, and 6: Fractions of gas in the\\
\hspace{-84mm}CNM ($f_\text{CNM}$), UNM ($f_\text{UNM}$), and WNM ($f_\text{WNM}$) phases, respectively. \\
\end{table*}

We note that the resolutions of the \texttt{TIGRESS} simulations are 4 pc and 8 pc, thus the sub-persec to pc scale CNM structures are not well resolved. Consequently, bias may be produced in these simulations, where the associated surrounding UNM or WNM clouds affect the CNM gas cloud properties, including their spatial extent and temperature. A similar situation may also arise in the SImulating the LifeCycle of molecular Clouds  (SILCC) numerical simulations \citep{2021MNRAS.504.1039R}, which employed a spatial resolution of $\sim$ 4 pc and included the effects of supernova explosions, stellar winds, radiation, and cosmic rays (SWRC). They found that $\sim$50--60\% of the gas resides in the CNM, $\sim$15--20\% in the UNM and $\sim$25\% in the WNM. These simulation also does not properly resolve the sub-persec CNM structures. However, at the same time, it is also worth noting that these simulations have considered more realistic large scale motion produced by supernova explosion and stellar feedback, which are not considered in other high resolution numerical simulations, where a simple large scale converging flow (including turbulent motion) mimics the large scale perturbation of the WNM gas clouds. Therefore, the absence of proper modeling the large scale flow may also affect the gas phase distribution. For example, \cite{2005AA...433....1A} performed the high resolution simulation compared to the  \texttt{TIGRESS} simulations where the spatial resolution was $\sim$ 0.02 pc, thus enabling to well resolve the CNM structures. In their simulations, depending on the strength of the turbulence ($\epsilon$ = 0.5 2.0, 4.0 and 6.0) they obtained the CNM fraction from 30$\%$ to 10$\%$  and UNM from 10$\%$ to 30$\%$, where WNM is almost unchanged at 60$\%$. Similarly, \cite{2022MNRAS.514..957S} studied multiphase ISM with a spatial resolution of 0.4 pc and within a domain size of 200 pc. Here also the resolution is almost one order of magnitude higher than the \texttt{TIGRESS} simulations, however, the proper large scale effect considered in \texttt{TIGRESS} simulation is absent. Using this simulation, \cite{2024MNRAS.527.8475B} showed that almost 33$\%$ gas was obtained in each phase according to the range of $T_\text{K}$ and 36$\%$ in CNM, 13$\%$ in UNM and 51$\%$ in WNM according to the values of $T_\text{s}$. The UNM fraction obtained from $T_\text{s}$ is quite low compared to our observational results. The reason may be mainly caused by differences in the phase boundaries between them. Furthermore, \cite{2026arXiv260117255H} examined the multiphase ISM with a spatial resolution of $\sim$ 0.04 pc and obtained the gas fractions of CNM, UNM and WNM 21.79$\%$, 44.28$\%$ and 33.98$\%$, respectively. The gas phase distribution in different simulations vary and thus in the future high resolution simulation with taken more realistic large scale motions including supernova explosion and stellar feedback are necessary, so that the large scale WNM gas properties as well as small scale (sub parsec to pc) CNM structures are accurately studied. All simulation results are summarized in Table \ref{tab:phase_fraction_comparison}.

\begin{table*}
\centering
\caption{Mean CNM, UNM, and WNM fractions as a function of Galactic
latitude obtained from \texttt{GWA} survey.}
\label{tab:phase_fraction}

\begin{tabular}{c ccc | ccc}
\hline
\hline
& \multicolumn{3}{c|}{$T_\text{s}<T_\text{k}$}
& \multicolumn{3}{c}{$T_\text{s}=T_\text{k}$} \\
\cmidrule(lr){2-4}
\cmidrule(lr){5-7}
$|b|$ & $f_\text{CNM}$ & $f_\text{UNM}$ & $f_\text{WNM}$ & $f_\text{CNM}$ & $f_\text{UNM}$ & $f_\text{WNM}$ \\
\midrule

$0^\circ-15^\circ$
& $0.22\pm0.07$
& $0.25\pm0.10$
& $0.53\pm0.09$
&
$0.23\pm0.07$
& $0.26\pm0.11$
& $0.52\pm0.09$
\\

$15^\circ-30^\circ$
& $0.18\pm0.05$
& $0.42\pm0.14$
& $0.40\pm0.18$
&
$0.19\pm0.05$
& $0.36\pm0.16$
& $0.45\pm0.17$
\\

$30^\circ-60^\circ$
& $0.12\pm0.12$
& $0.43\pm0.34$
& $0.45\pm0.38$
&
$0.13\pm0.13$
& $0.48\pm0.41$
& $0.40\pm0.44$
\\

$60^\circ-90^\circ$
& $0.06\pm0.06$
& $0.41\pm0.28$
& $0.52\pm0.32$
&
$0.06\pm0.06$
& $0.45\pm0.32$
& $0.48\pm0.35$
\\

\hline
\hline
\end{tabular}\\
\textbf{Notes.} Column 1: Range of latitudes ($|b|$). Columns 2 to 7: CNM, UNM and WNM gas fractions ($f_\text{CNM}$, $f_\text{UNM}$, and $f_\text{WNM}$)\\
for the case $T_\text{s} < T_\text{k}$ and $T_\text{s} = T_\text{k}$. The uncertainties represent the $1\sigma$ scatter among the sightlines in each latitude bin.
\end{table*}

\subsection{Other cloud properties}

Although the primary aim of this work was to characterize the gas-phase distribution in the neutral ISM, we also derived several additional physical properties of the gas clouds, focusing primarily on the CNM and UNM components using the iterative method. In the following discussion, we consider the case of $T_\text{s}<T_\text{k}$. Since the results remain largely unchanged for the case of $T_\text{s}=T_\text{k}$, we present these results in Appendix \ref{other_clouds_properties}. Figure \ref{fig:fig100_0} presents the distributions of the physical properties derived using the iterative method. The probability distribution functions (PDFs) of the CNM and UNM cloud sizes show that CNM clouds span a wide range of spatial scales, from a few hundred AU to $\sim$10 pc. However, their mean size is only 0.28 pc, indicating that CNM clouds are predominantly sub-parsec structures embedded within the more extended UNM or WNM gas. In contrast, UNM components are generally larger, with spatial reaches of up to $\sim$100 pc and a mean size of $\sim$8 pc, indicating that they are mainly structures of the parsec-scale. Sub-parsec CNM clouds, as well as very small-scale CNM structures on scales of a few hundred AU, have been identified in both observations and numerical simulations \citep{2005AA...433....1A,2007A&A...465..431H,2010ApJ...722..395B,2022ApJ...930...76K,2026arXiv260314967M}. The origin of these small-scale (AU) CNM structures remains puzzling and is an active area of research \citep{2005ApJ...631..371S,2010ApJ...722..395B}.\\

We also present the PDF of the unweighted sonic Mach number, calculated from the non-thermal velocity dispersion, $\sigma_\text{nth}$, and the sound speed, $c_\text{s}$, as $M_\text{s}^\text{uw} = \sqrt{3}{\sigma_\text{nth}}/c_\text{s}$, where $c_\text{s}$ = $\sqrt{k_\text{B}T_\text{k}/\mu m_\text{H}}$. Here, $k_\text{B}=1.38\times10^{-16}$ erg K$^{-1}$ is the Boltzmann constant, $\mu=1.4$ is the mean molecular weight after accounting for the He fraction and $m_\text{H}=1.67\times10^{-24}$ g is the hydrogen mass. The resulting distributions indicate that the CNM clouds are predominantly transonic, with an unweighted mean sonic Mach number of $M_\text{s}^\text{uw}=0.91$, whereas the UNM clouds are mildly supersonic, with a mean value of $M_\text{s}^\text{uw}=2.17$. We also compare our results with the sonic Mach numbers reported by the \texttt{SPONGE}, \texttt{Millennium}, and \cite{2014ApJ...793..132S} studies. The combined sonic Mach number PDF derived from these works was presented in \cite{2023PASA...40...46K}. Since no meaningful correlation between the non-thermal velocity dispersion ($\sigma_\text{nth}$) and the component length scale ($L_\text{los}$) was found below 0.4 pc, we restricted the comparison to components with length scales larger than 0.4 pc. Therefore, we compare our $M_\text{s}^\text{uw}$ with the corresponding values of these studies. We obtain a mean $M_\text{s,mean}^\text{uw}=2.6\pm1.4$, while the corresponding value from the combined sample of previous studies is 4.2$\pm$2.7, which is approximately 1.6 times higher than our value. An important consideration is that variations in the physical environments probed along different lines of sight may contribute to the observed difference. An additional consideration is that the non-uniqueness associated with Gaussian decomposition of complex spectral profiles. Gaussian functions do not constitute an orthogonal basis and, consequently, the decomposition of a complicated spectral profile into individual Gaussian components is not unique. In particular, the number and properties of the decomposed components can depend on the adopted decomposition criteria. For example, if additional components are introduced to reproduce increasingly complex spectral structures, the number of fitted components can increase while the reduced chi-square ($\chi^{2}$) approaches unity \citep{2013MNRAS.432.3074C,2013MNRAS.436.2366R}. Similarly, decomposition based on derivative spectroscopy or on different statistical criteria can produce a different number of components for the same observed spectrum \citep{2018ApJS..238...14M}. This represents an inherent limitation of Gaussian-based spectral decomposition.\\

We have also calculated the column-density-weighted and mass-weighted sonic Mach numbers, $M_\text{s}^{N(\text{HI})}$ and $M_\text{s}^{\text{mw}}$, respectively, for both all gas and CNM components. To calculate the mass-weighted values, we assume that the gas clouds are approximately spherical. We recognize that this approximation may not be appropriate for the actual morphology of gas clouds, particularly for the CNM, where sheet-like structures have been observed \citep{2016ApJ...821..117K}. However, the three-dimensional geometry of individual CNM clouds is difficult to constrain observationally. We therefore adopt a spherical geometry as a simplifying assumption for estimating their masses. After assuming this, we obtain $M_\text{s}^{N(\mathrm{HI})}=3.4\pm1.6$ and $M_\text{s}^{\mathrm{mw}}=4.8\pm0.9$ for all gas components. For the CNM components, we obtain $M_\text{s,CNM}^{N(\mathrm{HI})}=2.4\pm1.4$ and $M_\text{s,CNM}^{\mathrm{mw}}=4.5\pm0.8$ and for the UNM gas components, we obtain $M_\text{s,UNM}^{N(\mathrm{HI})}=4.0\pm1.5$ and $M_\text{s,UNM}^{\mathrm{mw}}=4.8\pm0.9$ . Consequently, both the column-density-weighted and mass-weighted sonic Mach numbers indicate that the CNM and low temperature UNM gas clouds are supersonic in nature.\\

We further investigated the correlation between the spin temperature, $T_\text{s}$, and the line-of-sight length scale, $L_\text{los}$. We find a positive power-law correlation with an index of $0.57\pm0.03$. Although the slope is relatively well constrained, the normalization has a greater uncertainty of $\sim42\%$. The Spearman correlation coefficient is 0.74, with a corresponding $p$-value of $3.7\times10^{-4}$, indicating a statistically significant strong positive correlation between $T_\text{s}$ and $L_\text{los}$. In contrast, $T_\text{s}$ shows a negative correlation with the optical depth of the peak, $\tau_\text{peak}$. The corresponding power-law index is $-0.37\pm0.04$, while normalization is constrained to within $\sim16\%$. The Spearman coefficient of $-0.60$ and the associated $p$-value of $2.0\times10^{-20}$ further confirm the presence of a strong and statistically significant negative correlation. \cite{2024MNRAS.527.8475B} also analytically demonstrated the correlation between $T_\text{s}$ and $\tau_\text{peak}$, finding a power-law index of approximately $-0.4$, which is in good agreement with our observational result.\\

We also examined the dependence of the sonic Mach number on the line-of-sight length scale. A strong positive correlation is observed, characterized by a power-law index of $0.45\pm0.01$. In this case, normalization is well constrained, with an uncertainty of only $\sim4\%$. Using numerical simulations of the interstellar medium, \cite{2021NatAs...5..365F} showed that the sonic Mach number, $M_\text{s}$, varies with the length scale following a power-law relation, with indices of 0.39 for the subsonic regime and 0.49 for the supersonic regime. The supersonic result is particularly close to our observationally derived power-law index. Similarly, the FWHM exhibits a positive correlation with $L_\text{los}$, with a power-law index of $0.25\pm0.01$ and a normalization uncertainty of only $\sim2.6\%$. The mean FWHM of the CNM components is $2.3\pm0.1$ km s$^{-1}$, whereas that of the UNM components is $9.1\pm5.1$ km s$^{-1}$. Thus, the characteristic FWHM of the UNM components is approximately four times larger than that of the CNM components. This difference is particularly interesting because \cite{KalberlaHaud2018}, based on the FWHM constraints of the gas components, classified the CNM, UNM, and WNM phases  and estimated the corresponding gas fractions of these phases throughout the Galaxy. The FWHM also shows a positive correlation with $T_\text{s}$, with a power-law index of $0.62\pm0.01$. Normalization of this relation is limited to within $\sim7\%$. These correlations collectively indicate systematic differences in the characteristic physical scales, velocity dispersions, and thermal properties of the CNM and UNM components.\\

We additionally present the column density PDFs of the CNM, UNM, and combined CNM+UNM components, shown by the dashed blue, green, and solid red curves, respectively. The horizontal axis is defined as $\eta$= $ln$[N(H{\sc i})/<N(H{\sc i})>] while the vertical axis represents the corresponding PDF. We identified 164 components of the CNM, with a mean H{\sc i} column density of $3.04\times10^{19}$ cm$^{-2}$, and 80 components of the UNM, with a mean column density of $1.02\times10^{20}$ cm$^{-2}$. A Gaussian fit to the combined CNM+UNM PDF gives a mean of $\mu=-1.31$, a dispersion of $\sigma=1.88$, and a reduced $\chi^2=0.577$. For a perfectly log-normal distribution of $N$(H{\sc i}), the expected relation is $\mu=-\sigma^2/2$. The deviation from this relation may partly arise from the incomplete sampling of the WNM component, particularly toward the high-column-density end of the distribution. Including the missing WNM contribution could, therefore, bring the observed distribution closer to a log-normal form. The logarithmic normal column-density distribution observed in the neutral interstellar medium, together with the emergence of a power-law tail in molecular clouds, is well established and is generally interpreted as evidence of a transition from a turbulence-dominated regime to a gravity-dominated regime \citep{2015ApJ...811L..28B,2017ApJ...843...92B,2020A&A...634A.139W}.\\

Finally, we investigate the relation between the maximum optical depth, $\tau_\text{peak}$, and the maximum brightness temperature, $T_{\text{B,peak}}$, where $T_\text{B}=T_\text{s}\left(1-e^{-\tau_\text{peak}}\right)$. Using the empirical relation $T_{\rm s}=33.98\tau_{\rm peak}^{-0.36}$, we compare the optically thin approximation, $T_{\rm B}\simeq T_{\rm s}\tau_{\rm peak}=33.98\tau_{\rm peak}^{0.64}$, with the full radiative-transfer expression, $T_{\rm B}=33.98\tau_{\rm peak}^{-0.36}\left(1-e^{-\tau_{\rm peak}}\right)$. The two relations are nearly identical in the optically thin regime, but increasingly diverge as $\tau_\text{peak}$ approaches unity, where optical-depth saturation becomes significant. However, our sample contains only two components with $\tau_\text{peak}\gtrsim0.6$. Consequently, the current data set does not contain sufficient statistics to robustly distinguish between the optically thin and full radiative-transfer behaviors in the high-optical-depth regime. The channel-wise correlation between the emission-derived brightness temperature, $T_B(v)$, and the absorption-derived optical depth, $\tau(v)$, has been extensively studied by \cite{2013MNRAS.436.2352R,2024MNRAS.527.8475B}. These studies show clear evidence for the saturation of $T_B(v)$ as $\tau(v)$ approaches unity.\\

\begin{figure*}
\includegraphics[width=2.4in,height=2.1in,angle=0]{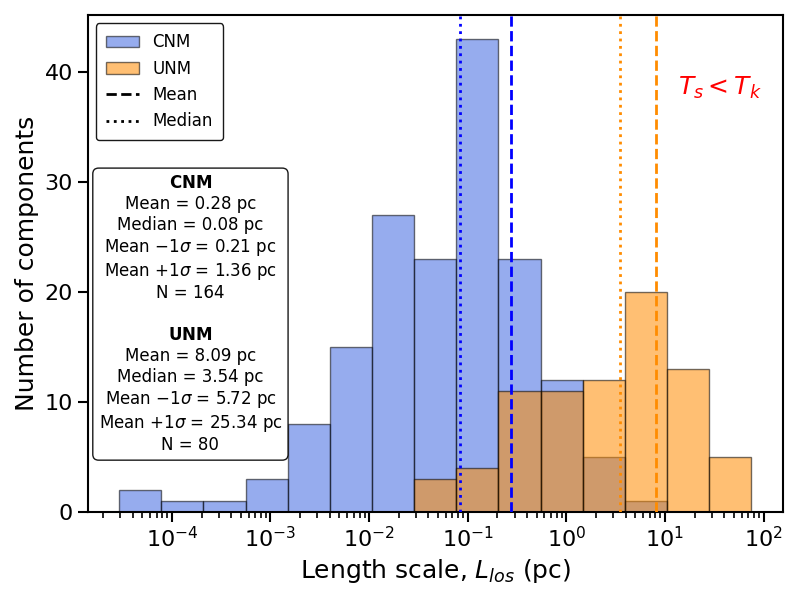}
\includegraphics[width=2.4in,height=2.1in,angle=0]{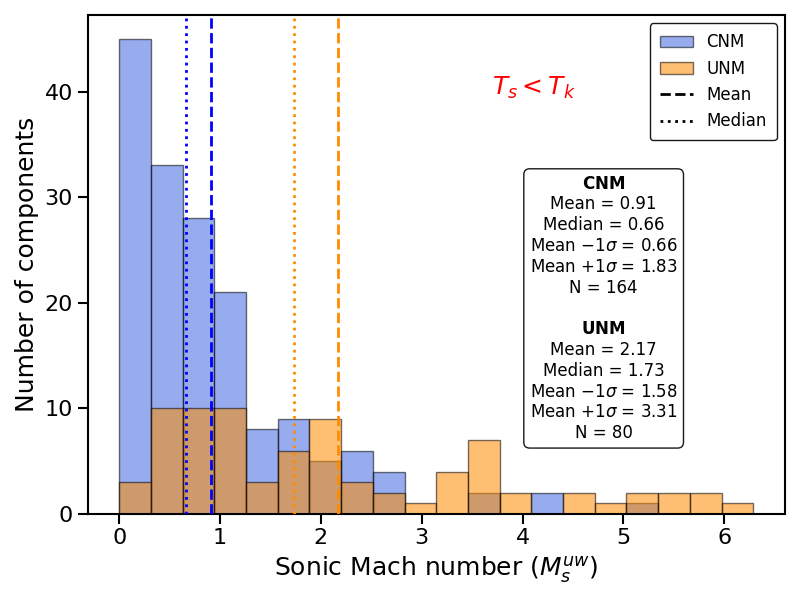}
\includegraphics[width=2.4in,height=2.1in,angle=0]{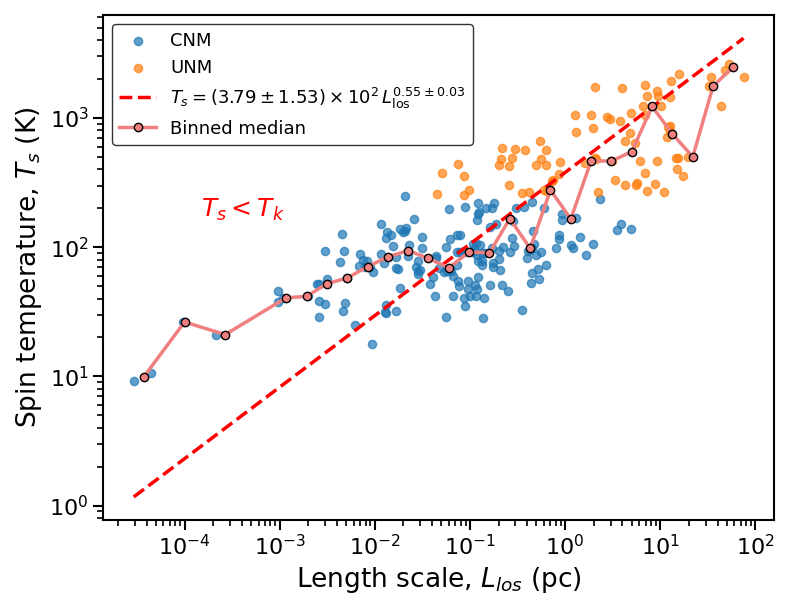}\\
\includegraphics[width=2.4in,height=2.1in,angle=0]{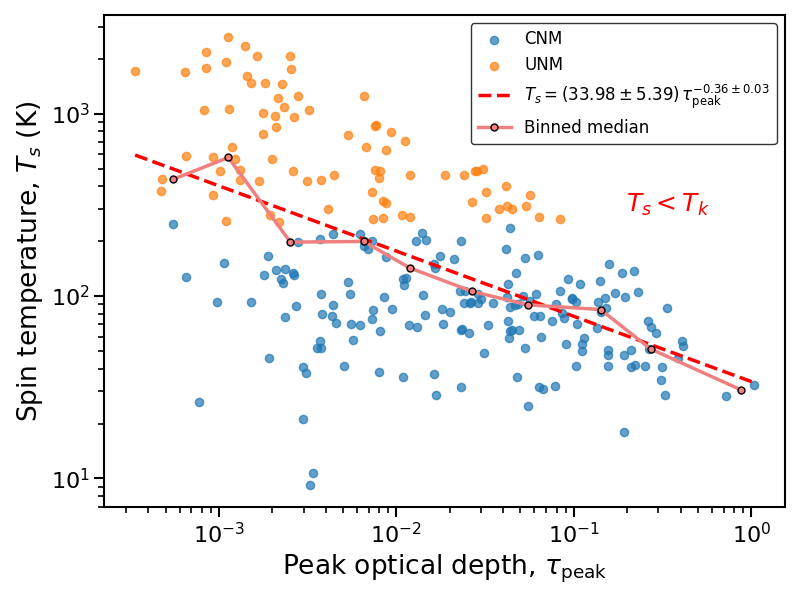}
\includegraphics[width=2.4in,height=2.1in,angle=0]{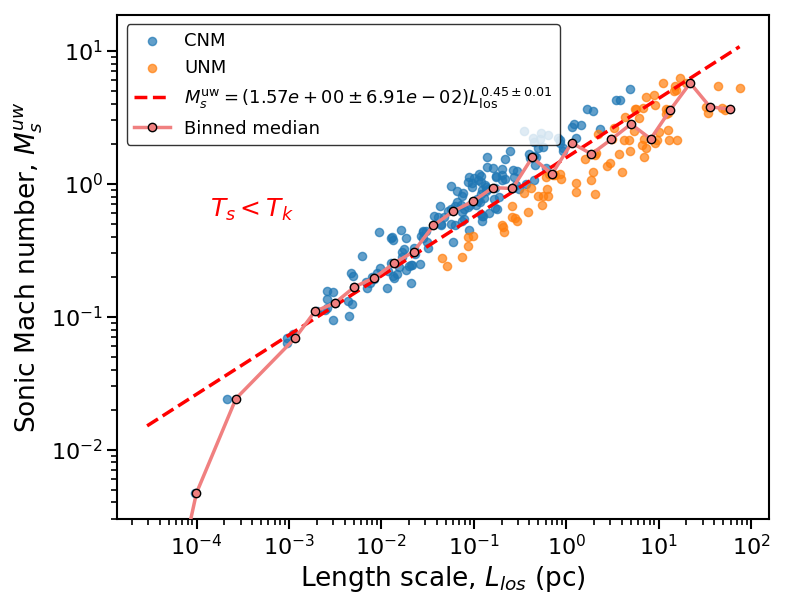}
\includegraphics[width=2.4in,height=2.1in,angle=0]{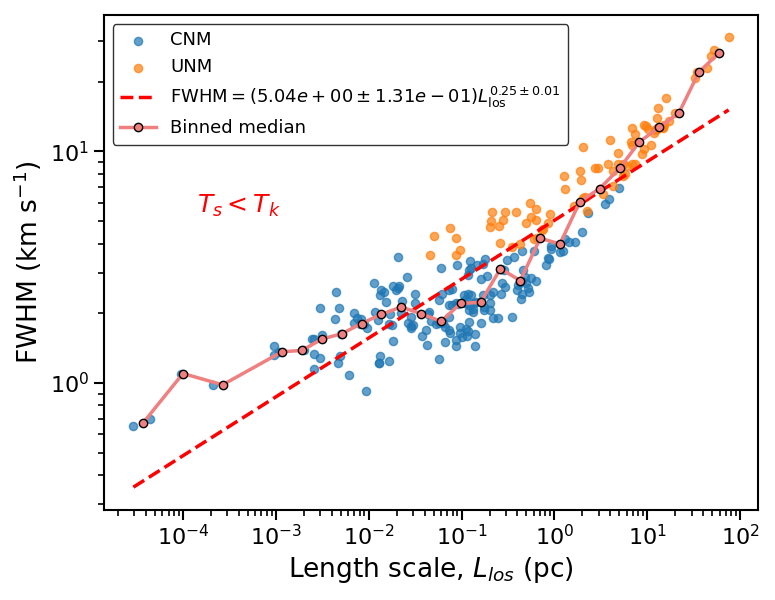}\\
\includegraphics[width=2.4in,height=2.1in,angle=0]{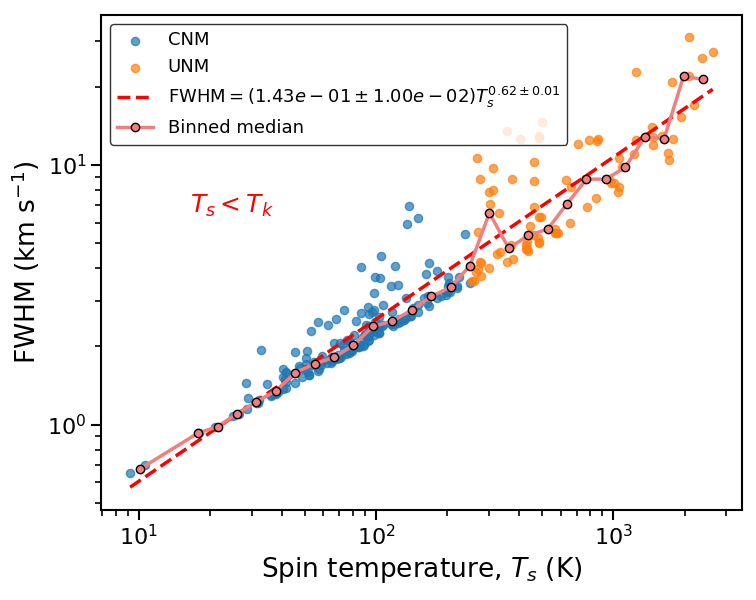}
\includegraphics[width=2.4in,height=2.1in,angle=0]{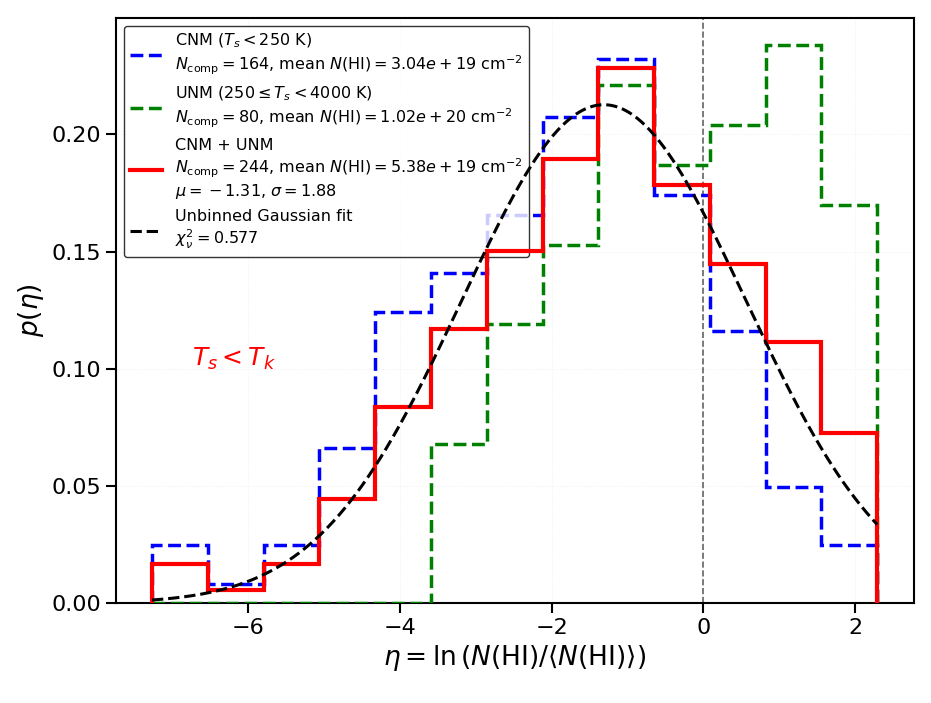}
 \includegraphics[width=2.4in,height=2.1in,angle=0]{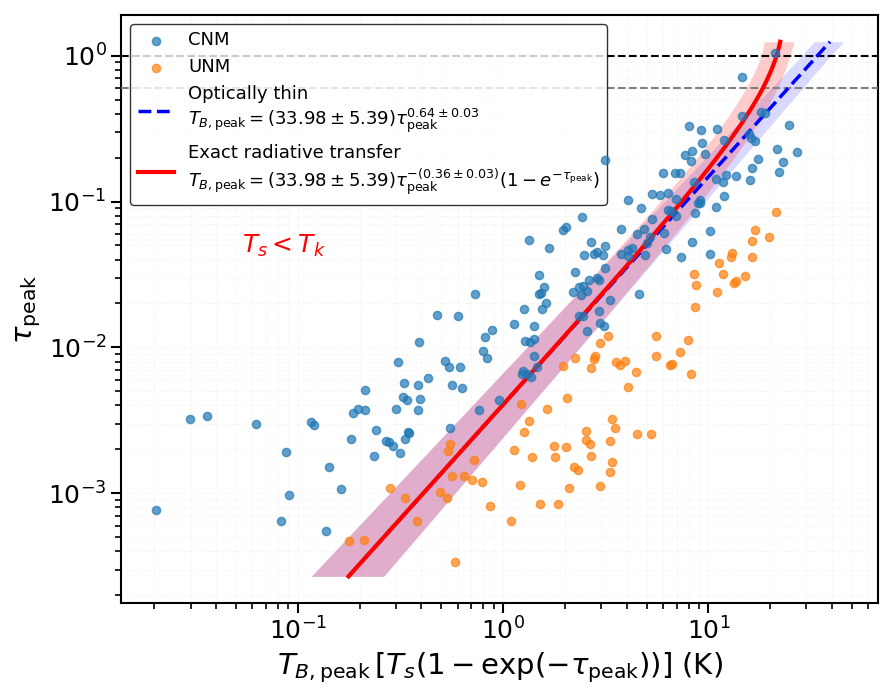}\\
\caption{Upper: Probability distribution functions (PDFs) of the length scale ($L_\text{los}$) and unweighted sonic Mach number ($M_\text{s}^\text{uw}$) of the CNM and UNM gas components, along with the correlation between spin temperature ($T_\text{s}$) and $L_\text{los}$, which follows a power-law relation with an index of 0.55. Middle: Correlations between $T_\text{s}$ and peak optical depth ($\tau_\text{peak}$), $M_\text{s}^\text{uw}$ and $L_\text{los}$, and FWHM and $L_\text{los}$, with power-law indices of $-$0.36, 0.45, and 0.25, respectively. Lower: Correlation between FWHM and $T_\text{s}$ with power law index of 0.62, the PDFs of the H{\sc i} column density, $N$(H{\sc i}), for the CNM, UNM, and combined CNM+UNM components, shown by the dashed blue, green, and solid red curves, respectively. The combined PDF is fitted with a Gaussian distribution, yielding a mean of $\mu=-1.31$ and a dispersion of $\sigma=1.88$. The final panel shows the correlation between peak brightness temperature ($T_{\text{B,peak}}$) and $\tau_\text{peak}$, comparing the optically thin approximation and the full radiative-transfer model, shown by the blue dashed and red solid curves and 1$\sigma$ uncertainty in the respective shaded areas, respectively.}
 \label{fig:fig100_0}
 \end{figure*}

\section{Discussion}\label{section_5}

The concept of a two-phase neutral interstellar medium (ISM) originated from the pioneering work of \cite{1965ApJ...142..531F}, with subsequent thermal stability analyzes showing that cold neutral medium (CNM) and warm neutral medium (WNM) can coexist in an approximate thermal pressure equilibrium \citep{2003ApJ...587..278W}. Between these stable phases lies the thermally unstable neutral medium (UNM), where small thermal perturbations drive the gas away from equilibrium toward either the CNM or the WNM. The thermal equilibrium curve is determined by the balance between radiative heating and cooling, with the net cooling rate given by $\mathcal{L}=n^{2}\Lambda(T)-cn$, where $n$ is the number density, $\Lambda(T)$ is the temperature-dependent cooling function and $c$ is the heating rate per particle. The resulting equilibrium curve gives rise to thermally stable and unstable regimes in the density--temperature and pressure--density relations. UNM gas satisfies the Field criterion \citep{2014pafd.book.....C},

\begin{equation}
\frac{\partial \mathcal{L}}{\partial T}\bigg|_{p}<0,
\end{equation}

indicating that small temperature perturbations grow, driving the gas toward the CNM or WNM. In turbulent media, when the turbulent mixing timescale becomes comparable to or shorter than the local cooling timescale, the gas is repeatedly perturbed before it can fully cool and condense, allowing a significant fraction to persist in the UNM despite classical thermal instability. In this regime, the effective adiabatic index ($\gamma'$) is negative, corresponding to a polytropic relation $P\propto\rho^{\gamma'}=\rho^{-\gamma}$, with $\gamma>0$. Here, $P$ is the pressure and $\rho$ is the mass density. Using an analytical model in which turbulence continuously perturbs the gas, \cite{2024arXiv240714199H} showed that the UNM mass fraction, $f_{\text{UNM}}$, scales with the turbulent velocity dispersion as $f_{\text{UNM}}\propto\sigma_{\text{turb}}^{2\gamma/(\gamma+1)}$. Their simulation yields $\gamma=1/3$ for the UNM, giving $f_{\text{UNM}}\propto\sigma_{\text{turb}}^{1/2}$. This indicates that the level of turbulence may have a direct impact on the gas fraction in the UNM.\\

From our observational modeling, we obtain 19.8\%, 32.5\%, and 47.8\% in the CNM, UNM and WNM phases, respectively. We compare our results with previous observational studies as well as with several numerical simulations, particularly the \texttt{TIGRESS-CLASSIC} and \texttt{TIGRESS-NCR} simulations. In the \texttt{SPONGE} survey, which is similar to our observational approach, approximately half of the H{\sc i} gas is found to reside in the WNM phase. Their CNM fraction is slightly higher than ours, compensated for by a correspondingly lower fraction of UNM gas. Compared with the results of the \texttt{HI4PI} survey presented by \citet{KalberlaHaud2018}, we find that the CNM fraction is in good agreement, while their UNM fraction is slightly higher and their WNM fraction is correspondingly lower than ours. This difference may arise from the assumption that all the H{\sc i} components that were not detected in absorption were associated with the WNM. In contrast, the results from the \texttt{Millennium} survey show somewhat different gas fractions for all three phases compared to our measurements.\\

We compare our observational results with those from different numerical simulations, particularly the \texttt{TIGRESS-NCR} and \texttt{TIGRESS-CLASSIC} simulations. Across the simulations considered, the CNM fraction ranges from $\sim6\%$ to nearly $60\%$, the UNM fraction from $\sim10\%$ to $\sim45\%$, and the WNM fraction from $\sim25\%$ to $71\%$. These differences may arise in part from the trade-off between spatial resolution and the treatment of large-scale physical processes. Some simulations achieve sufficiently high spatial resolution, reaching $\sim0.02$ to 0.4 pc \citep{2005AA...433....1A,2022MNRAS.514..957S}, to resolve the structures associated with the CNM, but do not adequately capture large-scale motions and feedback processes. Conversely, simulations that properly incorporate large-scale effects, such as supernova explosions and stellar-cluster feedback, may not have sufficient spatial resolution to resolve the sub-parsec CNM structures \citep{2021MNRAS.504.1039R, 2023ApJ...946....3K}. When comparing our results with the \texttt{TIGRESS-NCR} and \texttt{TIGRESS-CLASSIC} simulations, we find that the \texttt{TIGRESS-NCR} results are in closer agreement with our observational estimates. Although this simulation shows good agreement with our observational results, a proper comparison between numerical simulations and observations requires simulations that combine high spatial resolution with a more complete treatment of large-scale feedback and dynamical effects. This would enable for more meaningful and accurate comparisons with observations.\\

In addition to the overall gas-phase distribution, we also find that the CNM gas fraction gradually decreases from $\sim22\%$ near the Galactic plane to $\sim6\%$ at higher Galactic latitudes, consistent with the results of \citet{KalberlaHaud2018}. The decrease in the CNM fraction toward higher Galactic latitudes is also physically expected from the dependence of the thermal phases of the neutral ISM on the local Galactic environment. The thermal equilibrium calculations of \citet{2003ApJ...587..278W} show that CNM and WNM coexist under a range of physical conditions, with the relative amount of cold and warm gas depending on the local density, the radiation field and other environmental properties. The denser Galactic disk therefore provides conditions more favorable for the presence of cold H{\sc i}, consistent with the larger CNM fraction that we measure at low Galactic latitudes. Away from the Galactic plane, where the sightlines probe a more diffuse Galactic environment, the relative contribution of the CNM decreases and the warmer phases become increasingly important. The substantial fraction of UNM found in our analysis is also broadly consistent with the significant contribution of the UNM phase reported by \citet{KalberlaHaud2018}.\\

We find that the CNM gas is predominantly associated with sub-parsec-scale clouds (mean value $\sim$ 0.28 pc), whereas the UNM gas is distributed over parsec-scale structures (mean value $\sim$ 8.09 pc). This is consistent with several previous observational and numerical studies. Although sub-parsec-scale CNM clouds can form through the compression of WNM gas by converging flows, the physical origin of small-scale compact CNM structures of the AU scale remains an open question and is still a subject of active research \citep{2005A&A...436L..53B, 2005ApJ...631..371S,2010ApJ...722..395B}. \cite{2005ApJ...631..371S} identified CNM clouds with very low-column-density and discussed possible formation mechanisms involving condensation and the interaction between cold clouds and their surrounding medium, while noting that thermal evaporation imposes a lower limit on their column densities. These observations suggest that small cold structures may represent a distinct low-column-density extension of the CNM population.\\

We also find that the column-density-weighted sonic Mach number of the CNM clouds is $M_\text{s,CNM}^{N(\mathrm{HI})}=2.4\pm1.4$, while that of the UNM clouds is $M_\text{s,UNM}^{N(\mathrm{HI})}=4.0\pm1.5$. This suggests that the CNM and lower-temperature range UNM gas clouds are predominantly supersonic \citep{2003ApJ...586.1067H, 2018ApJS..238...14M,2022ApJ...930...76K,2023PASA...40...46K}. These results are consistent with those reported by other observational studies and numerical simulations. We further find a power-law relation between the spin temperature and the optical depth of the peak, $T_\text{s}\propto\tau_\text{peak}^{-0.36}$, while the sonic Mach number shows a positive correlation with the physical scale, $M_\text{s}^\text{uw}\propto L_\text{los}^{0.45}$. These scaling relations are broadly consistent with previous observational and numerical studies \citep{2021NatAs...5..365F,2024MNRAS.527.8475B}. It is also evident that the FWHM increases with physical scale following a power law indices of 0.25 and 0.62 respectively. This result is consistent with the work of \cite{KalberlaHaud2018}, who separated the gas into different phases based on the FWHM of the decomposed components and examined the gas-phase fractions in the Milky Way. The column-density probability distribution function (PDF) of the H{\sc i} clouds is well described by a lognormal distribution. A lognormal column-density PDF is commonly observed in the neutral ISM when gravity is not the dominant dynamical process, whereas a power-law tail can develop in systems where gravitational effects become important \citep{2015ApJ...811L..28B,2017ApJ...843...92B}. Finally, we investigate the $T_\text{B,peak}$--$\tau_\text{peak}$ relation, where $T_\text{B,peak}$ is estimated from the derived $T_\text{s}$ and $\tau_\text{peak}$. Although the relation shows the expected trend toward saturation with increasing optical depth, our current sample does not contain sufficient data points with high optical depth to clearly identify the saturation of $T_{\rm B,peak}$ as $\tau_{\rm peak}$ approaches unity. A larger sample with more high-optical-depth components will be required to robustly characterize this saturation effect.\\

\section{CONCLUSIONS}\label{section_6}

We investigate the gas phase distribution in the neutral interstellar medium using the publicly available GWA survey, which is mainly based on H{\sc i} absorption observations obtained with the GMRT, WSRT, and ATCA telescopes, and compare these results with publicly available H{\sc i} emission observations obtained from the LAB survey. The main findings of our analysis are summarized as follows.

\renewcommand{\labelenumi}{(\roman{enumi})}
\begin{enumerate}

    \item We find that $19.8\%$, $32.5\%$, and $47.8\%$ of the H{\sc i} gas is in the CNM, UNM, and WNM phases, respectively.\\

    \item The CNM gas fraction is $\sim22\%$ towards the Galactic plane and gradually decreases with increasing Galactic latitude and reaches $\sim6\%$. In contrast, we do not find any clear trend for the UNM and WNM fractions, which appear to be more widely distributed throughout the Galaxy. \\

    \item We obtain mean physical scales of $\sim0.28$ pc for CNM clouds and $\sim8.09$ pc for UNM clouds, indicating that the CNM is predominantly associated with sub-parsec-scale structures, whereas the UNM is distributed over larger, parsec-scale structures.\\
    
    \item The column-density-weighted sonic Mach numbers of the CNM and UNM clouds are $M_\text{s,CNM}^{N(\mathrm{HI})}=2.4\pm1.4$ and $M_\text{s,UNM}^{N(\mathrm{HI})}= 4.0\pm1.5$, respectively.\\

    \item We find power-law correlations between $T_{\rm s}$ and $\tau_{\rm peak}$, $M_\text{s}$ and $L_\text{los}$, and FWHM and $L_\text{los}$, with power-law indices of $-0.36$, $0.45$, and $0.25$, respectively.\\

    \item The column-density PDF of the H{\sc i} clouds is well described by a nearly lognormal distribution. \\

    \item Although we compare the gas-phase fraction distributions with several numerical simulations, including \texttt{TIGRESS-NCR} and \texttt{TIGRESS-CLASSIC}, and find that the \texttt{TIGRESS-NCR} results are in good agreement with our observations, we expect that future high-resolution simulations capable of resolving the small-scale CNM structures while simultaneously incorporating the relevant large-scale processes, including supernova explosions and stellar-cluster feedback, will provide a more appropriate framework for comparison with high-sensitivity H{\sc i} emission--absorption observations.\\

\end{enumerate}

\begin{acknowledgements}
The author expresses his sincere gratitude to Harvey Liszt (National Radio Astronomy Observatory, USA) and C. G. Kim (Princeton University, USA) for generously sharing their analytical and simulation results. Author also sincerely thank Nirupam Roy (IISc Bangalore, India \& New Mexico Institute of Mining and Technology, USA), Narendra Nath Patra (IIT Indore, India) and Soumyadeep Bhattacharjee (Caltech, USA). AK acknowledges support from the Fondecyt postdoctoral project (project id: 3250070, 2025).
\end{acknowledgements}




\bibliographystyle{aa} 
\bibliography{a.bib} 

@ARTICLE{1965ApJ...142..531F,
       author = {{Field}, George B.},
        title = "{Thermal Instability.}",
      journal = {\apj},
         year = 1965,
        month = aug,
       volume = {142},
        pages = {531},
          doi = {10.1086/148317},
       adsurl = {https://ui.adsabs.harvard.edu/abs/1965ApJ...142..531F}
}

@ARTICLE{1969ApJ...155L.149F,
       author = {{Field}, G.~B. and {Goldsmith}, D.~W. and {Habing}, H.~J.},
        title = "{Cosmic-Ray Heating of the Interstellar Gas}",
      journal = {\apjl},
         year = 1969,
        month = mar,
       volume = {155},
        pages = {L149},
          doi = {10.1086/180324},
       adsurl = {https://ui.adsabs.harvard.edu/abs/1969ApJ...155L.149F}
}

@ARTICLE{2022MNRAS.514..957S,
       author = {{Seta}, Amit and {Federrath}, Christoph},
        title = "{Turbulent dynamo in the two-phase interstellar medium}",
      journal = {\mnras},
         year = 2022,
        month = jul,
       volume = {514},
       number = {1},
        pages = {957-976},
          doi = {10.1093/mnras/stac1400},
archivePrefix = {arXiv},
       eprint = {2202.08324},
 primaryClass = {astro-ph.GA},
       adsurl = {https://ui.adsabs.harvard.edu/abs/2022MNRAS.514..957S}
}

@ARTICLE{2026arXiv260117255H,
       author = {{Hu}, Yue and {Truong}, Bao and {Hoang}, Thiem and {Tram}, Le Ngoc},
        title = "{Galactic Dust Polarization in Turbulent Multiphase ISM: On the Origin of the $EE/BB$ Asymmetry}",
      journal = {arXiv e-prints},
         year = 2026,
        month = jan,
          eid = {arXiv:2601.17255},
        pages = {arXiv:2601.17255},
          doi = {10.48550/arXiv.2601.17255},
archivePrefix = {arXiv},
       eprint = {2601.17255},
 primaryClass = {astro-ph.GA},
       adsurl = {https://ui.adsabs.harvard.edu/abs/2026arXiv260117255H}
}

@article{KalberlaHaud2018,
  author  = {{Kalberla}, P.~M. and {Haud}, U.},
  title   = {Properties of cold and warm H I gas phases derived from a
             Gaussian decomposition of HI4PI data},
  journal = {Astronomy \& Astrophysics},
  volume  = {619},
  pages   = {A58},
  year    = {2018}
}

@ARTICLE{2016A&A...594A.116H,
       author = {{HI4PI Collaboration} and {Ben Bekhti}, N. and {Fl{\"o}er}, L. and {Keller}, R. and {Kerp}, J. and {Lenz}, D. and {Winkel}, B. and {Bailin}, J. and {Calabretta}, M.~R. and {Dedes}, L. and {Ford}, H.~A. and {Gibson}, B.~K. and {Haud}, U. and {Janowiecki}, S. and {Kalberla}, P.~M.~W. and {Lockman}, F.~J. and {McClure-Griffiths}, N.~M. and {Murphy}, T. and {Nakanishi}, H. and {Pisano}, D.~J. and {Staveley-Smith}, L.},
        title = "{HI4PI: A full-sky H I survey based on EBHIS and GASS}",
      journal = {\aap},
         year = 2016,
        month = oct,
       volume = {594},
          eid = {A116},
        pages = {A116},
          doi = {10.1051/0004-6361/201629178},
archivePrefix = {arXiv},
       eprint = {1610.06175},
 primaryClass = {astro-ph.GA},
       adsurl = {https://ui.adsabs.harvard.edu/abs/2016A&A...594A.116H}
}

@ARTICLE{2017A&A...606A..15C,
       author = {{Cao}, Shuo and {Zheng}, Xiaogang and {Biesiada}, Marek and {Qi}, Jingzhao and {Chen}, Yun and {Zhu}, Zong-Hong},
        title = "{Ultra-compact structure in intermediate-luminosity radio quasars: building a sample of standard cosmological rulers and improving the dark energy constraints up to z   3}",
      journal = {\aap},
         year = 2017,
        month = sep,
       volume = {606},
          eid = {A15},
        pages = {A15},
          doi = {10.1051/0004-6361/201730551},
archivePrefix = {arXiv},
       eprint = {1708.08635},
 primaryClass = {astro-ph.CO},
       adsurl = {https://ui.adsabs.harvard.edu/abs/2017A&A...606A..15C}
}

@ARTICLE{1999A&A...342..378G,
       author = {{Gurvits}, L.~I. and {Kellermann}, K.~I. and {Frey}, S.},
        title = "{The ``angular size - redshift'' relation for compact radio structures in quasars and radio galaxies}",
      journal = {\aap},
         year = 1999,
        month = feb,
       volume = {342},
        pages = {378-388},
          doi = {10.48550/arXiv.astro-ph/9812018},
archivePrefix = {arXiv},
       eprint = {astro-ph/9812018},
 primaryClass = {astro-ph},
       adsurl = {https://ui.adsabs.harvard.edu/abs/1999A&A...342..378G}
}

@ARTICLE{2021MNRAS.504.1039R,
       author = {{Rathjen}, Tim-Eric and {Naab}, Thorsten and {Girichidis}, Philipp and {Walch}, Stefanie and {W{\"u}nsch}, Richard and {Dinnbier}, Frantis̆ek and {Seifried}, Daniel and {Klessen}, Ralf S. and {Glover}, Simon C.~O.},
        title = "{SILCC VI - Multiphase ISM structure, stellar clustering, and outflows with supernovae, stellar winds, ionizing radiation, and cosmic rays}",
      journal = {\mnras},
         year = 2021,
        month = jun,
       volume = {504},
       number = {1},
        pages = {1039-1061},
          doi = {10.1093/mnras/stab900},
archivePrefix = {arXiv},
       eprint = {2103.14128},
 primaryClass = {astro-ph.GA},
       adsurl = {https://ui.adsabs.harvard.edu/abs/2021MNRAS.504.1039R}
}

@ARTICLE{2007A&A...465..431H,
       author = {{Hennebelle}, P. and {Audit}, E.},
        title = "{On the structure of the turbulent interstellar atomic hydrogen. I. Physical characteristics. Influence and nature of turbulence in a thermally bistable flow}",
      journal = {\aap},
         year = 2007,
        month = apr,
       volume = {465},
       number = {2},
        pages = {431-443},
          doi = {10.1051/0004-6361:20066139},
archivePrefix = {arXiv},
       eprint = {astro-ph/0612778},
 primaryClass = {astro-ph},
       adsurl = {https://ui.adsabs.harvard.edu/abs/2007A&A...465..431H}
}

@ARTICLE{2026arXiv260314967M,
       author = {{Matsuzuki}, Yamato and {Yamamoto}, Hiroaki and {Tachihara}, Kengo},
        title = "{Mapping of the Cold Neutral Medium via HI Phase Separation in an Atomic Cloud Undergoing Molecular Cloud Formation}",
      journal = {arXiv e-prints},
         year = 2026,
        month = mar,
          eid = {arXiv:2603.14967},
        pages = {arXiv:2603.14967},
          doi = {10.48550/arXiv.2603.14967},
archivePrefix = {arXiv},
       eprint = {2603.14967},
 primaryClass = {astro-ph.GA},
       adsurl = {https://ui.adsabs.harvard.edu/abs/2026arXiv260314967M}
}

@ARTICLE{2020A&A...634A.139W,
       author = {{Wang}, Y. and {Bihr}, S. and {Beuther}, H. and {Rugel}, M.~R. and {Soler}, J.~D. and {Ott}, J. and {Kainulainen}, J. and {Schneider}, N. and {Klessen}, R.~S. and {Glover}, S.~C.~O. and {McClure-Griffiths}, N.~M. and {Goldsmith}, P.~F. and {Johnston}, K.~G. and {Menten}, K.~M. and {Ragan}, S. and {Anderson}, L.~D. and {Urquhart}, J.~S. and {Linz}, H. and {Roy}, N. and {Smith}, R.~J. and {Bigiel}, F. and {Henning}, T. and {Longmore}, S.~N.},
        title = "{Cloud formation in the atomic and molecular phase: H I self absorption (HISA) towards a giant molecular filament}",
      journal = {\aap},
         year = 2020,
        month = feb,
       volume = {634},
          eid = {A139},
        pages = {A139},
          doi = {10.1051/0004-6361/201935866},
archivePrefix = {arXiv},
       eprint = {2001.00953},
 primaryClass = {astro-ph.SR},
       adsurl = {https://ui.adsabs.harvard.edu/abs/2020A&A...634A.139W}
}

@ARTICLE{2017ApJ...843...92B,
       author = {{Bialy}, Shmuel and {Burkhart}, Blakesley and {Sternberg}, Amiel},
        title = "{The H I-to-H$_{2}$ Transition in a Turbulent Medium}",
      journal = {\apj},
         year = 2017,
        month = jul,
       volume = {843},
       number = {2},
          eid = {92},
        pages = {92},
          doi = {10.3847/1538-4357/aa7854},
archivePrefix = {arXiv},
       eprint = {1703.08549},
 primaryClass = {astro-ph.GA},
       adsurl = {https://ui.adsabs.harvard.edu/abs/2017ApJ...843...92B}
}

@ARTICLE{2025AJ....169..284C,
       author = {{Chen}, Hongxing and {Stanimirovi{\'c}}, Sne{\v{z}}ana and {Pingel}, Nickolas M. and {Dempsey}, James and {Buckland-Willis}, Frances and {Clark}, Susan E. and {D{\'e}nes}, Helga and {Dickey}, John M. and {Gibson}, Steven and {Jameson}, Katherine and {Kemp}, Ian and {Leahy}, Denis and {Lee}, Min-Young and {Lynn}, Callum and {Ma}, Yik Ki and {McClure-Griffiths}, N.~M. and {Murray}, Claire E. and {Nguyen}, Hiep and {Uscanga}, Lucero and {van Loon}, Jacco Th. and {V{\'a}zquez-Semadeni}, Enrique},
        title = "{A Neutral Hydrogen Absorption Study of Cold Gas in the Outskirts of the Magellanic Clouds Using the GASKAP-H I Survey}",
      journal = {\aj},
         year = 2025,
        month = may,
       volume = {169},
       number = {5},
          eid = {284},
        pages = {284},
          doi = {10.3847/1538-3881/adc67c},
archivePrefix = {arXiv},
       eprint = {2504.00968},
 primaryClass = {astro-ph.GA},
       adsurl = {https://ui.adsabs.harvard.edu/abs/2025AJ....169..284C}
}

@ARTICLE{2026arXiv260821115C,
       author = {{Chen}, Hongxing and {Stanimirovi\{{\'c}\}}, Sne\{\textbackslashvz\}ana and {Dempsey}, James and {van Loon}, Jacco Th. and {Ma}, Yik Ki and {Nguyen}, Hiep and {Pingel}, Nickolas M. and {Dickey}, John M. and {Lee}, Min-Young and {McClure-Griffiths}, N.~M. and {Murray}, Claire E. and {Staveley-Smith}, Lister and {Leahy}, Denis and {Lynn}, Callum and {Marchal}, Antoine and {Rybarczyk}, Daniel R. and {D\{{\'e}\}nes}, Helga and {Gibson}, Steven and {Jameson}, Katherine and {Kemp}, Ian and {Stelea}, Ioana A.},
        title = "{The GASKAP-HI Survey towards the Magellanic Clouds: Cold Atomic Gas Survival and Evolution in the Large Magellanic Cloud}",
      journal = {arXiv e-prints},
         year = 2026,
        month = aug,
          eid = {arXiv:2608.21115},
        pages = {arXiv:2608.21115},
archivePrefix = {arXiv},
       eprint = {2608.21115},
 primaryClass = {astro-ph.GA},
       adsurl = {https://ui.adsabs.harvard.edu/abs/2026arXiv260821115C}
}

@ARTICLE{2021NatAs...5..365F,
       author = {{Federrath}, Christoph and {Klessen}, Ralf S. and {Iapichino}, Luigi and {Beattie}, James R.},
        title = "{The sonic scale of interstellar turbulence}",
      journal = {Nature Astronomy},
         year = 2021,
        month = jan,
       volume = {5},
        pages = {365-371},
          doi = {10.1038/s41550-020-01282-z},
archivePrefix = {arXiv},
       eprint = {2011.06238},
 primaryClass = {astro-ph.GA},
       adsurl = {https://ui.adsabs.harvard.edu/abs/2021NatAs...5..365F}
}

@ARTICLE{2015ApJ...811L..28B,
       author = {{Burkhart}, Blakesley and {Lee}, Min-Young and {Murray}, Claire E. and {Stanimirovi{\'c}}, Snezana},
        title = "{The Lognormal Probability Distribution Function of the Perseus Molecular Cloud: A Comparison of HI and Dust}",
      journal = {\apjl},
         year = 2015,
        month = oct,
       volume = {811},
       number = {2},
          eid = {L28},
        pages = {L28},
          doi = {10.1088/2041-8205/811/2/L28},
archivePrefix = {arXiv},
       eprint = {1509.02889},
 primaryClass = {astro-ph.GA},
       adsurl = {https://ui.adsabs.harvard.edu/abs/2015ApJ...811L..28B}
}

@ARTICLE{2016ApJ...821..117K,
       author = {{Kalberla}, P.~M.~W. and {Kerp}, J. and {Haud}, U. and {Winkel}, B. and {Ben Bekhti}, N. and {Fl{\"o}er}, L. and {Lenz}, D.},
        title = "{Cold Milky Way HI Gas in Filaments}",
      journal = {\apj},
         year = 2016,
        month = apr,
       volume = {821},
       number = {2},
          eid = {117},
        pages = {117},
          doi = {10.3847/0004-637X/821/2/117},
archivePrefix = {arXiv},
       eprint = {1602.07604},
 primaryClass = {astro-ph.GA},
       adsurl = {https://ui.adsabs.harvard.edu/abs/2016ApJ...821..117K}
}

@ARTICLE{2005ApJ...631..371S,
       author = {{Stanimirovi{\'c}}, Sne{\v{z}}ana and {Heiles}, Carl},
        title = "{The Thinnest Cold H I Clouds in the Diffuse Interstellar Medium?}",
      journal = {\apj},
         year = 2005,
        month = sep,
       volume = {631},
       number = {1},
        pages = {371-375},
          doi = {10.1086/432533},
archivePrefix = {arXiv},
       eprint = {astro-ph/0505106},
 primaryClass = {astro-ph},
       adsurl = {https://ui.adsabs.harvard.edu/abs/2005ApJ...631..371S}
}

@ARTICLE{2005AA...433....1A,
       author = {{Audit}, E. and {Hennebelle}, P.},
        title = "{Thermal condensation in a turbulent atomic hydrogen flow}",
      journal = {\aap},
         year = 2005,
        month = apr,
       volume = {433},
       number = {1},
        pages = {1-13},
          doi = {10.1051/0004-6361:20041474},
archivePrefix = {arXiv},
       eprint = {astro-ph/0410062},
 primaryClass = {astro-ph},
       adsurl = {https://ui.adsabs.harvard.edu/abs/2005A&A...433....1A}
}

@ARTICLE{2003ApJS..145..329H,
       author = {{Heiles}, Carl and {Troland}, T.~H.},
        title = "{The Millennium Arecibo 21 Centimeter Absorption-Line Survey. I. Techniques and Gaussian Fits}",
      journal = {\apjs},
         year = 2003,
        month = apr,
       volume = {145},
       number = {2},
        pages = {329-354},
          doi = {10.1086/367785},
archivePrefix = {arXiv},
       eprint = {astro-ph/0207104},
 primaryClass = {astro-ph},
       adsurl = {https://ui.adsabs.harvard.edu/abs/2003ApJS..145..329H}
}

@ARTICLE{2003ApJ...586.1067H,
       author = {{Heiles}, Carl and {Troland}, T.~H.},
        title = "{The Millennium Arecibo 21 Centimeter Absorption-Line Survey. II. Properties of the Warm and Cold Neutral Media}",
      journal = {\apj},
         year = 2003,
        month = apr,
       volume = {586},
       number = {2},
        pages = {1067-1093},
          doi = {10.1086/367828},
archivePrefix = {arXiv},
       eprint = {astro-ph/0207105},
 primaryClass = {astro-ph},
       adsurl = {https://ui.adsabs.harvard.edu/abs/2003ApJ...586.1067H}
}

@ARTICLE{2018ApJS..238...14M,
       author = {{Murray}, Claire E. and {Stanimirovi{\'c}}, Sne{\v{z}}ana and {Goss}, W.~M. and {Heiles}, Carl and {Dickey}, John M. and {Babler}, Brian and {Kim}, Chang-Goo},
        title = "{The 21-SPONGE H I Absorption Line Survey. I. The Temperature of Galactic H I}",
      journal = {\apjs},
         year = 2018,
        month = oct,
       volume = {238},
       number = {2},
          eid = {14},
        pages = {14},
          doi = {10.3847/1538-4365/aad81a},
archivePrefix = {arXiv},
       eprint = {1806.06065},
 primaryClass = {astro-ph.GA},
       adsurl = {https://ui.adsabs.harvard.edu/abs/2018ApJS..238...14M}
}

@ARTICLE{2013MNRAS.436.2352R,
       author = {{Roy}, Nirupam and {Kanekar}, Nissim and {Braun}, Robert and {Chengalur}, Jayaram N.},
        title = "{The temperature of the diffuse H I in the Milky Way - I. High resolution H I-21 cm absorption studies}",
      journal = {\mnras},
         year = 2013,
        month = dec,
       volume = {436},
       number = {3},
        pages = {2352-2365},
          doi = {10.1093/mnras/stt1743},
archivePrefix = {arXiv},
       eprint = {1309.4098},
 primaryClass = {astro-ph.GA},
       adsurl = {https://ui.adsabs.harvard.edu/abs/2013MNRAS.436.2352R}
}

@ARTICLE{2013MNRAS.436.2366R,
       author = {{Roy}, Nirupam and {Kanekar}, Nissim and {Chengalur}, Jayaram N.},
        title = "{The temperature of the diffuse H I in the Milky Way - II. Gaussian decomposition of the H I-21 cm absorption spectra}",
      journal = {\mnras},
         year = 2013,
        month = dec,
       volume = {436},
       number = {3},
        pages = {2366-2385},
          doi = {10.1093/mnras/stt1746},
archivePrefix = {arXiv},
       eprint = {1309.4099},
 primaryClass = {astro-ph.GA},
       adsurl = {https://ui.adsabs.harvard.edu/abs/2013MNRAS.436.2366R}
}

@ARTICLE{2019MNRAS.483..593K,
       author = {{Koley}, Atanu and {Roy}, Nirupam},
        title = "{Estimating the kinetic temperature from H I 21-cm absorption studies: correction for turbulence broadening}",
      journal = {\mnras},
         year = 2019,
        month = feb,
       volume = {483},
       number = {1},
        pages = {593-598},
          doi = {10.1093/mnras/sty3152},
archivePrefix = {arXiv},
       eprint = {1811.07352},
 primaryClass = {astro-ph.GA},
       adsurl = {https://ui.adsabs.harvard.edu/abs/2019MNRAS.483..593K}
}

@article{KimOstriker2017,
  author = {Kim, Chang-Goo and Ostriker, Eve C.},
  title = {Three-phase Interstellar Medium in Galaxies Resolving Evolution with Star Formation and Supernova Feedback (TIGRESS): Algorithms, Fiducial Model, and Convergence},
  journal = {The Astrophysical Journal},
  volume = {846},
  number = {2},
  pages = {133},
  year = {2017},
  doi = {10.3847/1538-4357/aa8599}
}

@ARTICLE{2011ApJ...734...65J,
       author = {{Jenkins}, Edward B. and {Tripp}, Todd M.},
        title = "{The Distribution of Thermal Pressures in the Diffuse, Cold Neutral Medium of Our Galaxy. II. An Expanded Survey of Interstellar C I Fine-structure Excitations}",
      journal = {\apj},
         year = 2011,
        month = jun,
       volume = {734},
       number = {1},
          eid = {65},
        pages = {65},
          doi = {10.1088/0004-637X/734/1/65},
archivePrefix = {arXiv},
       eprint = {1104.2323},
 primaryClass = {astro-ph.GA},
       adsurl = {https://ui.adsabs.harvard.edu/abs/2011ApJ...734...65J}
}

@ARTICLE{1941DoSSR..32...16K,
       author = {{Kolmogorov}, Andrey Nikolaevich},
        title = "{Dissipation of Energy in Locally Isotropic Turbulence}",
      journal = {Akademiia Nauk SSSR Doklady},
         year = 1941,
        month = apr,
       volume = {32},
        pages = {16},
       adsurl = {https://ui.adsabs.harvard.edu/abs/1941DoSSR..32...16K}
}

@BOOK{1995tlan.book.....F,
       author = {{Frisch}, Uriel},
        title = "{Turbulence. The legacy of A.N. Kolmogorov}",
         year = 1995,
       adsurl = {https://ui.adsabs.harvard.edu/abs/1995tlan.book.....F}
}

@ARTICLE{2007ApJ...666L..69K,
       author = {{Kowal}, Grzegorz and {Lazarian}, A.},
        title = "{Scaling Relations of Compressible MHD Turbulence}",
      journal = {\apjl},
         year = 2007,
        month = sep,
       volume = {666},
       number = {2},
        pages = {L69-L72},
          doi = {10.1086/521788},
archivePrefix = {arXiv},
       eprint = {0705.2464},
 primaryClass = {astro-ph},
       adsurl = {https://ui.adsabs.harvard.edu/abs/2007ApJ...666L..69K}
}

@ARTICLE{2022ApJ...930...76K,
       author = {{Kobayashi}, Masato I.~N. and {Inoue}, Tsuyoshi and {Tomida}, Kengo and {Iwasaki}, Kazunari and {Nakatsugawa}, Hiroki},
        title = "{Nature of Supersonic Turbulence and Density Distribution Function in the Multiphase Interstellar Medium}",
      journal = {\apj},
         year = 2022,
        month = may,
       volume = {930},
       number = {1},
          eid = {76},
        pages = {76},
          doi = {10.3847/1538-4357/ac5a54},
archivePrefix = {arXiv},
       eprint = {2203.00699},
 primaryClass = {astro-ph.GA},
       adsurl = {https://ui.adsabs.harvard.edu/abs/2022ApJ...930...76K}
}

@ARTICLE{2023PASA...40...46K,
       author = {{Koley}, Atanu},
        title = "{Turbulence measurements in the neutral ISM from HI-21 cm emission-absorption spectra}",
      journal = {\pasa},
         year = 2023,
        month = sep,
       volume = {40},
          eid = {e046},
        pages = {e046},
          doi = {10.1017/pasa.2023.43},
archivePrefix = {arXiv},
       eprint = {2308.01808},
 primaryClass = {astro-ph.GA},
       adsurl = {https://ui.adsabs.harvard.edu/abs/2023PASA...40...46K}
}

@ARTICLE{2007ApJ...665..416K,
       author = {{Kritsuk}, Alexei G. and {Norman}, Michael L. and {Padoan}, Paolo and {Wagner}, Rick},
        title = "{The Statistics of Supersonic Isothermal Turbulence}",
      journal = {\apj},
         year = 2007,
        month = aug,
       volume = {665},
       number = {1},
        pages = {416-431},
          doi = {10.1086/519443},
archivePrefix = {arXiv},
       eprint = {0704.3851},
 primaryClass = {astro-ph},
       adsurl = {https://ui.adsabs.harvard.edu/abs/2007ApJ...665..416K}
}

@ARTICLE{2010A&A...512A..81F,
       author = {{Federrath}, C. and {Roman-Duval}, J. and {Klessen}, R.~S. and {Schmidt}, W. and {Mac Low}, M.-M.},
        title = "{Comparing the statistics of interstellar turbulence in simulations and observations. Solenoidal versus compressive turbulence forcing}",
      journal = {\aap},
         year = 2010,
        month = mar,
       volume = {512},
          eid = {A81},
        pages = {A81},
          doi = {10.1051/0004-6361/200912437},
archivePrefix = {arXiv},
       eprint = {0905.1060},
 primaryClass = {astro-ph.SR},
       adsurl = {https://ui.adsabs.harvard.edu/abs/2010A&A...512A..81F}
}

@article{gupta2017ugmrt,
  author       = {Gupta, Yashwant and Ajithkumar, B. and Kale, H. S. and Nayak, S. and Sabhapathy, S. and Sureshkumar, S. and Swami, R. V. and Chengalur, J. N. and Ghosh, S. K. and Ishwara-Chandra, C. H. and Joshi, B. C. and Kanekar, N. and Lal, D. V. and Roy, S.},
  title        = {The upgraded GMRT: Opening new windows on the radio Universe},
  journal      = {Current Science},
  volume       = {113},
  number       = {4},
  pages        = {707--714},
  year         = {2017},
  doi          = {10.18520/cs/v113/i04/707-714}
}

@ARTICLE{1979MNRAS.186..479L,
       author = {{Larson}, R.~B.},
        title = "{Stellar kinematics and interstellar turbulence.}",
      journal = {\mnras},
         year = 1979,
        month = feb,
       volume = {186},
        pages = {479-490},
          doi = {10.1093/mnras/186.3.479},
       adsurl = {https://ui.adsabs.harvard.edu/abs/1979MNRAS.186..479L}
}

@ARTICLE{2023ARA&A..61...19M,
       author = {{McClure-Griffiths}, Naomi M. and {Stanimirovi{\'c}}, Sne{\v{z}}ana and {Rybarczyk}, Daniel R.},
        title = "{Atomic Hydrogen in the Milky Way: A Stepping Stone in the Evolution of Galaxies}",
      journal = {\araa},
         year = 2023,
        month = aug,
       volume = {61},
        pages = {19-63},
          doi = {10.1146/annurev-astro-052920-104851},
archivePrefix = {arXiv},
       eprint = {2307.08464},
 primaryClass = {astro-ph.GA},
       adsurl = {https://ui.adsabs.harvard.edu/abs/2023ARA&A..61...19M}
}

@ARTICLE{2001A&A...371..698L,
       author = {{Liszt}, H.},
        title = "{The spin temperature of warm interstellar H I}",
      journal = {\aap},
         year = 2001,
        month = may,
       volume = {371},
        pages = {698-707},
          doi = {10.1051/0004-6361:20010395},
archivePrefix = {arXiv},
       eprint = {astro-ph/0103246},
 primaryClass = {astro-ph},
       adsurl = {https://ui.adsabs.harvard.edu/abs/2001A&A...371..698L}
}

@ARTICLE{1958PIRE...46..240F,
       author = {{Field}, George B.},
        title = "{Excitation of the Hydrogen 21-CM Line}",
      journal = {Proceedings of the IRE},
         year = 1958,
        month = jan,
       volume = {46},
        pages = {240-250},
          doi = {10.1109/JRPROC.1958.286741},
       adsurl = {https://ui.adsabs.harvard.edu/abs/1958PIRE...46..240F}
}

@ARTICLE{2020ApJS..250....9S,
       author = {{Seon}, Kwang-il and {Kim}, Chang-Goo},
        title = "{Ly{\ensuremath{\alpha}} Radiative Transfer: Monte Carlo Simulation of the Wouthuysen-Field Effect}",
      journal = {\apjs},
         year = 2020,
        month = sep,
       volume = {250},
       number = {1},
          eid = {9},
        pages = {9},
          doi = {10.3847/1538-4365/aba2d6},
archivePrefix = {arXiv},
       eprint = {2005.00238},
 primaryClass = {astro-ph.GA},
       adsurl = {https://ui.adsabs.harvard.edu/abs/2020ApJS..250....9S}
}

@ARTICLE{2004ApJ...615L..45H,
       author = {{Heyer}, Mark H. and {Brunt}, Christopher M.},
        title = "{The Universality of Turbulence in Galactic Molecular Clouds}",
      journal = {\apjl},
         year = 2004,
        month = nov,
       volume = {615},
       number = {1},
        pages = {L45-L48},
          doi = {10.1086/425978},
archivePrefix = {arXiv},
       eprint = {astro-ph/0409420},
 primaryClass = {astro-ph},
       adsurl = {https://ui.adsabs.harvard.edu/abs/2004ApJ...615L..45H}
}

@ARTICLE{2025A&A...702A.133K,
       author = {{Koley}, A. and {Stutz}, A.~M. and {Louvet}, F. and {Motte}, F. and {Ginsburg}, A. and {Galv{\'a}n-Madrid}, R. and {{\'A}lvarez-Guti{\'e}rrez}, R.~H. and {Sanhueza}, P. and {Baug}, T. and {Sandoval-Garrido}, N. and {Salinas}, J. and {Busquet}, G. and {Braine}, J. and {Liu}, H.-L. and {Csengeri}, T. and {Gusdorf}, A. and {Fern{\'a}ndez-L{\'o}pez}, M. and {Cunningham}, N. and {Bronfman}, L. and {Bonfand}, M.},
        title = "{ALMA-IMF: XIX. C$^{18}$O (J = 2─1): Measurements of turbulence in 15 massive protoclusters}",
      journal = {\aap},
         year = 2025,
        month = oct,
       volume = {702},
          eid = {A133},
        pages = {A133},
          doi = {10.1051/0004-6361/202553830},
archivePrefix = {arXiv},
       eprint = {2507.14502},
 primaryClass = {astro-ph.GA},
       adsurl = {https://ui.adsabs.harvard.edu/abs/2025A&A...702A.133K}
}

@ARTICLE{1987ApJ...319..730S,
       author = {{Solomon}, P.~M. and {Rivolo}, A.~R. and {Barrett}, J. and {Yahil}, A.},
        title = "{Mass, Luminosity, and Line Width Relations of Galactic Molecular Clouds}",
      journal = {\apj},
         year = 1987,
        month = aug,
       volume = {319},
        pages = {730},
          doi = {10.1086/165493},
       adsurl = {https://ui.adsabs.harvard.edu/abs/1987ApJ...319..730S}
}

@ARTICLE{2024MNRAS.529.4037P,
       author = {{Patra}, Narendra Nath and {Roy}, Nirupam},
        title = "{The temperature of the neutral interstellar medium in the Galaxy}",
      journal = {\mnras},
         year = 2024,
        month = apr,
       volume = {529},
       number = {4},
        pages = {4037-4049},
          doi = {10.1093/mnras/stae771},
archivePrefix = {arXiv},
       eprint = {2403.11653},
 primaryClass = {astro-ph.GA},
       adsurl = {https://ui.adsabs.harvard.edu/abs/2024MNRAS.529.4037P}
}

@ARTICLE{2003ApJ...587..278W,
       author = {{Wolfire}, Mark G. and {McKee}, Christopher F. and {Hollenbach}, David and {Tielens}, A.~G.~G.~M.},
        title = "{Neutral Atomic Phases of the Interstellar Medium in the Galaxy}",
      journal = {\apj},
         year = 2003,
        month = apr,
       volume = {587},
       number = {1},
        pages = {278-311},
          doi = {10.1086/368016},
archivePrefix = {arXiv},
       eprint = {astro-ph/0207098},
 primaryClass = {astro-ph},
       adsurl = {https://ui.adsabs.harvard.edu/abs/2003ApJ...587..278W}
}

@ARTICLE{2024arXiv240714199H,
       author = {{Ho}, Ka Wai and {Yuen}, Ka Ho and {Lazarian}, Alex},
        title = "{The stable ``Unstable Natural Media'' due to the presence of turbulence}",
      journal = {arXiv e-prints},
         year = 2024,
        month = jul,
          eid = {arXiv:2407.14199},
        pages = {arXiv:2407.14199},
          doi = {10.48550/arXiv.2407.14199},
archivePrefix = {arXiv},
       eprint = {2407.14199},
 primaryClass = {astro-ph.GA},
       adsurl = {https://ui.adsabs.harvard.edu/abs/2024arXiv240714199H}
}

@BOOK{2014pafd.book.....C,
       author = {{Clarke}, Cathie and {Carswell}, Bob},
        title = "{Principles of Astrophysical Fluid Dynamics}",
         year = 2014,
       adsurl = {https://ui.adsabs.harvard.edu/abs/2014pafd.book.....C}
}

@ARTICLE{2014ApJ...793..132S,
       author = {{Stanimirovi{\'c}}, Sne{\v{z}}ana and {Murray}, Claire E. and {Lee}, Min-Young and {Heiles}, Carl and {Miller}, Jesse},
        title = "{Cold and Warm Atomic Gas around the Perseus Molecular Cloud. I. Basic Properties}",
      journal = {\apj},
         year = 2014,
        month = oct,
       volume = {793},
       number = {2},
          eid = {132},
        pages = {132},
          doi = {10.1088/0004-637X/793/2/132},
archivePrefix = {arXiv},
       eprint = {1407.7778},
 primaryClass = {astro-ph.GA},
       adsurl = {https://ui.adsabs.harvard.edu/abs/2014ApJ...793..132S}
}

@ARTICLE{2023ApJ...946....3K,
       author = {{Kim}, Chang-Goo and {Kim}, Jeong-Gyu and {Gong}, Munan and {Ostriker}, Eve C.},
        title = "{Introducing TIGRESS-NCR. I. Coregulation of the Multiphase Interstellar Medium and Star Formation Rates}",
      journal = {\apj},
         year = 2023,
        month = mar,
       volume = {946},
       number = {1},
          eid = {3},
        pages = {3},
          doi = {10.3847/1538-4357/acbd3a},
archivePrefix = {arXiv},
       eprint = {2211.13293},
 primaryClass = {astro-ph.GA},
       adsurl = {https://ui.adsabs.harvard.edu/abs/2023ApJ...946....3K}
}

@INPROCEEDINGS{2010ASPC..438..126B,
       author = {{Begum}, A. and {Stanimirovi{\'c}}, S. and {Peek}, J.~E.~G. and {Ballering}, N. and {Heiles}, C. and {Douglas}, K.~A. and {Putman}, M. and {Gibson}, S.~J. and {Grcevich}, J. and {Korpela}, E.~J. and {Lee}, M. and {Saul}, D. and {Gallagher}, J.},
        title = "{Galactic Small Scale Structure Revealed by the GALFA-H I Survey}",
    booktitle = {The Dynamic Interstellar Medium: A Celebration of the Canadian Galactic Plane Survey},
         year = 2010,
       editor = {{Kothes}, R. and {Landecker}, T.~L. and {Willis}, A.~G.},
       series = {Astronomical Society of the Pacific Conference Series},
       volume = {438},
        month = dec,
        pages = {126},
archivePrefix = {arXiv},
       eprint = {1008.3185},
 primaryClass = {astro-ph.GA},
       adsurl = {https://ui.adsabs.harvard.edu/abs/2010ASPC..438..126B}
}

@ARTICLE{2010ApJ...722..395B,
       author = {{Begum}, Ayesha and {Stanimirovi{\'c}}, Sne{\v{z}}ana and {Peek}, Joshua E. and {Ballering}, Nicholas P. and {Heiles}, Carl and {Douglas}, Kevin A. and {Putman}, Mary and {Gibson}, Steven J. and {Grcevich}, Jana and {Korpela}, Eric J. and {Lee}, Min-Young and {Saul}, Destry and {Gallagher}, John S., III},
        title = "{Compact H I Clouds from the GALFA-H I Survey}",
      journal = {\apj},
         year = 2010,
        month = oct,
       volume = {722},
       number = {1},
        pages = {395-411},
          doi = {10.1088/0004-637X/722/1/395},
archivePrefix = {arXiv},
       eprint = {1008.1364},
 primaryClass = {astro-ph.GA},
       adsurl = {https://ui.adsabs.harvard.edu/abs/2010ApJ...722..395B}
}

@ARTICLE{2005A&A...436L..53B,
       author = {{Braun}, R. and {Kanekar}, N.},
        title = "{Tiny H I clouds in the local ISM}",
      journal = {\aap},
         year = 2005,
        month = jun,
       volume = {436},
       number = {3},
        pages = {L53-L56},
          doi = {10.1051/0004-6361:200500122},
archivePrefix = {arXiv},
       eprint = {astro-ph/0505055},
 primaryClass = {astro-ph},
       adsurl = {https://ui.adsabs.harvard.edu/abs/2005A&A...436L..53B}
}

@ARTICLE{2019ApJ...880..141N,
       author = {{Nguyen}, Hiep and {Dawson}, J.~R. and {Lee}, Min-Young and {Murray}, Claire E. and {Stanimirovi{\'c}}, Sne{\v{z}}ana and {Heiles}, Carl and {Miville-Desch{\^e}nes}, M.-A. and {Petzler}, Anita},
        title = "{Exploring the Properties of Warm and Cold Atomic Hydrogen in the Taurus and Gemini Regions}",
      journal = {\apj},
         year = 2019,
        month = aug,
       volume = {880},
       number = {2},
          eid = {141},
        pages = {141},
          doi = {10.3847/1538-4357/ab2b9f},
archivePrefix = {arXiv},
       eprint = {1906.06846},
 primaryClass = {astro-ph.GA},
       adsurl = {https://ui.adsabs.harvard.edu/abs/2019ApJ...880..141N}
}

@ARTICLE{2024MNRAS.527.8475B,
       author = {{Bhattacharjee}, Soumyadeep and {Roy}, Nirupam and {Sharma}, Prateek and {Seta}, Amit and {Federrath}, Christoph},
        title = "{Multiphase neutral interstellar medium: analysing simulation with H I 21cm observational data analysis techniques}",
      journal = {\mnras},
         year = 2024,
        month = jan,
       volume = {527},
       number = {3},
        pages = {8475-8496},
          doi = {10.1093/mnras/stad3682},
archivePrefix = {arXiv},
       eprint = {2309.05000},
 primaryClass = {astro-ph.GA},
       adsurl = {https://ui.adsabs.harvard.edu/abs/2024MNRAS.527.8475B}
}

@ARTICLE{2008A&A...487..951K,
       author = {{Kalberla}, P.~M.~W. and {Dedes}, L.},
        title = "{Global properties of the H I distribution in the outer Milky Way. Planar and extra-planar gas}",
      journal = {\aap},
         year = 2008,
        month = sep,
       volume = {487},
       number = {3},
        pages = {951-963},
          doi = {10.1051/0004-6361:20079240},
archivePrefix = {arXiv},
       eprint = {0804.4831},
 primaryClass = {astro-ph},
       adsurl = {https://ui.adsabs.harvard.edu/abs/2008A&A...487..951K}
}

@incollection{BURGERS1948171,
booktitle = {A Mathematical Model Illustrating the Theory of Turbulence},
editor = {Richard {Von Mises} and Theodore {Von Kármán}},
series = {Advances in Applied Mechanics},
publisher = {Elsevier},
volume = {1},
pages = {171-199},
year = {1948},
issn = {0065-2156},
doi = {https://doi.org/10.1016/S0065-2156(08)70100-5},
url = {https://www.sciencedirect.com/science/article/pii/S0065215608701005},
author = {J.M. Burgers}
}

@ARTICLE{2005A&A...440..775K,
       author = {{Kalberla}, P.~M.~W. and {Burton}, W.~B. and {Hartmann}, Dap and {Arnal}, E.~M. and {Bajaja}, E. and {Morras}, R. and {P{\"o}ppel}, W.~G.~L.},
        title = "{The Leiden/Argentine/Bonn (LAB) Survey of Galactic HI. Final data release of the combined LDS and IAR surveys with improved stray-radiation corrections}",
      journal = {\aap},
         year = 2005,
        month = sep,
       volume = {440},
       number = {2},
        pages = {775-782},
          doi = {10.1051/0004-6361:20041864},
archivePrefix = {arXiv},
       eprint = {astro-ph/0504140},
 primaryClass = {astro-ph},
       adsurl = {https://ui.adsabs.harvard.edu/abs/2005A&A...440..775K}
}

@ARTICLE{2013MNRAS.432.3074C,
       author = {{Chengalur}, Jayaram N. and {Kanekar}, Nissim and {Roy}, Nirupam},
        title = "{Accurate measurement of the H I column density from H I 21 cm absorption-emission spectroscopy}",
      journal = {\mnras},
         year = 2013,
        month = jul,
       volume = {432},
       number = {4},
        pages = {3074-3079},
          doi = {10.1093/mnras/stt658},
archivePrefix = {arXiv},
       eprint = {1305.0951},
 primaryClass = {astro-ph.GA},
       adsurl = {https://ui.adsabs.harvard.edu/abs/2013MNRAS.432.3074C}
}

@techreport{ngVLA_memo17_2018,
  author       = {Selina, R. J. and Murphy, E. J. and others},
  title        = {ngVLA Reference Design Development and Performance Estimates},
  institution  = {National Radio Astronomy Observatory},
  number       = {ngVLA Memo No. 17},
  year         = {2018},
  url          = {https://ngvla.nrao.edu/page/memos}
}

@article{Braun2019SKA1,
  author  = {Braun, R. and Bonaldi, A. and Bourke, T. and Keane, E. and Wagg, J.},
  title   = {Anticipated Performance of the Square Kilometre Array -- Phase 1 (SKA1)},
  journal = {arXiv e-prints},
  year    = {2019},
  eprint  = {1912.12699},
  archivePrefix = {arXiv},
  primaryClass  = {astro-ph.IM}
}

@ARTICLE{2003MNRAS.346L..57K,
       author = {{Kanekar}, Nissim and {Subrahmanyan}, Ravi and {Chengalur}, Jayaram N. and {Safouris}, Vicky},
        title = "{The temperature of the warm neutral medium in the Milky Way}",
      journal = {\mnras},
         year = 2003,
        month = dec,
       volume = {346},
       number = {4},
        pages = {L57-L61},
          doi = {10.1111/j.1365-2966.2003.07333.x},
archivePrefix = {arXiv},
       eprint = {astro-ph/0310761},
 primaryClass = {astro-ph},
       adsurl = {https://ui.adsabs.harvard.edu/abs/2003MNRAS.346L..57K}
}

@ARTICLE{2025A&A...694L..11K,
       author = {{Kalberla}, P.~M.~W.},
        title = "{The cold neutral medium in filaments at high Galactic latitudes}",
      journal = {\aap},
         year = 2025,
        month = feb,
       volume = {694},
          eid = {L11},
        pages = {L11},
          doi = {10.1051/0004-6361/202452771},
archivePrefix = {arXiv},
       eprint = {2502.01223},
 primaryClass = {astro-ph.GA},
       adsurl = {https://ui.adsabs.harvard.edu/abs/2025A&A...694L..11K}
}


             
              
            
\vspace{-20mm}

\appendix

\section{{\color{black}Turbulence from Koley A. 2023}}\label{Appendix: appendix1}
{\color{black}We have adopted the turbulence scaling relation from \cite{2023PASA...40...46K}, where they studied turbulence using H{\sc i} data from the \texttt{Millenium} \citep{2003ApJ...586.1067H} and \texttt{SPONGE} \citep{2018ApJS..238...14M} surveys, as well as from the work of \cite{2014ApJ...793..132S}. From their analysis, they found that below $\sim$ 0.4 pc, no significant scaling relation is observed. There are multiple reasons that can lead to the breakdown of the power law. In the following, we outline the plausible reasons for this.\\

\hspace{-5mm}a) These observations are made mainly using the Arecibo single-dish telescope, which has a beam size of $\sim$ 3 arcminutes. This corresponds to a physical scale of approximately 0.44 pc at a distance of 0.50 kpc or 0.87 pc at a distance of 1.0 kpc. Consequently, tiny clouds that reside at a distance of $\sim$ 0.50 kpc with sizes below 0.44 pc or $\sim$ 1.0 kpc with sizes below 0.87 pc are not resolved. Consequently, the beam fill factor $f_\text{beam}$ for these small components of the gas cloud is less than one, which reduces the brightness temperature, while leaving the line width unchanged:

\begin{eqnarray}
 T_\text{B,obs} = f_\text{beam}~T_\text{B,true}.   
\end{eqnarray}

\hspace{-5mm}For example, if $f_\text{beam}$ = 0.2 and $T_\text{B,true}$ = 23 K, then $T_\text{B,obs}$ = 4.6 K. From the work of \cite{2023PASA...40...46K}, the median value of $T_\text{B,obs}$ is $\sim$ 23 K for the CNM clouds for which $L_\text{los} > $ 0.4 pc and  $T_\text{B,obs}$ $\sim$ 4 K for the CNM clouds for which $L_\text{los} < $ 0.4 pc. Thus, the beam filling factor can be a plausible reason for the decrease in brightness temperature.\\

For absorption measurements, we observe the absorption spectra against compact background quasars, typically on the order of a few milli-arcseconds \citep{1999A&A...342..378G,2017A&A...606A..15C}, allowing us to probe structures on scales comparable to the angular size of the background source. In contrast, the corresponding H{\sc i} emission is observed with a substantially larger telescope sizes of several arcminutes \citep{2003ApJ...586.1067H,2018ApJS..238...14M}, and therefore the emission from compact structures can be significantly affected by beam dilution. Consequently, a direct comparison between the absorption and emission spectra can sometimes be ambiguous. Although the optical depth, $\tau$, can be measured relatively accurately along the pencil-beam line of sight toward the background quasar, the brightness temperature, $T_{\rm B}$, may be reduced because of the beam-filling factor of the emitting structure. If the measured $T_{\rm B}$ is primarily affected by beam dilution, the spin temperature derived from the following equation

\begin{eqnarray}
    T_\text{s} = \frac{T_\text{B,obs}}{\biggl[1-e^{-\tau_\text{obs}}\biggl]},
\end{eqnarray}

will also be underestimated. Since the total velocity width of the line is largely unaffected by the beam-filling factor, an underestimated $T_\text{s}$ can lead to an overestimate of the non-thermal velocity dispersion, $\sigma_\text{nth}$, when it is derived by separating the thermal contribution from the observed total velocity dispersion.  The beam mismatch issue has also been discussed in recent studies \citep{2025AJ....169..284C,2026arXiv260821115C}.\\

\hspace{-5mm}This reduction in $T_\text{s}$ influences the kinetic temperature $T_\text{k}$. In our case, the median values of $T_\text{k}$ are $\sim$ 30 K and $\sim$ 75 K for line-of-sight sizes $L_\text{los}$ $<$ 0.4 pc and $>$ 0.4 pc, respectively. Due to the decrease in $T_\text{s}$, the column density N(H{\sc i}) will also decrease, which, in turn, reduces the inferred $L_\text{los}$ of the cloud. The median N(H{\sc i}) of the gas clouds whose $L_\text{los} < $ 0.4 pc is $\sim$ 2.0 $\times$ 10$^{19}$ cm$^{-2}$ and $\sim$ 2.0 $\times$ 10$^{20}$ cm$^{-2}$ for $L_\text{los} > $ 0.4 pc.\\

\hspace{-5mm}b) Another possibility is that these small cloud components are real and their turbulent or may be thermal properties differ from those of larger structures. \cite{2010ApJ...722..395B, 2010ASPC..438..126B} reported the presence of small CNM cloud components in the GALFA$-$H{\sc i} survey. \cite{2023PASA...40...46K} discussed this in great detail about the small components of CNM gas clouds.\\

\hspace{-5mm}However, regardless of the underlying reasons, more detailed studies are required to better understand the origin and nature of cold CNM gas cloud components. This problem can be effectively addressed only through high spectral and spatial resolution observations of both emission and absorption spectra. High-sensitive, high-resolution absorption, and emission observations with next-generation facilities such as the ngVLA or SKA1-MID (may be feathering with single-dish data) will be essential to resolve this issue. In this study, we do not attempt to complicate the analysis; instead, we simply adopt the turbulence scaling law derived from fitting the data points.\\



\section{Acquisition of observational data}\label{Appendix: appendix2}

We adopted all the physical parameters of the decomposed H{\sc i} absorption components from the \texttt{GWA} survey \citep{2003MNRAS.346L..57K,2013MNRAS.436.2366R,2024MNRAS.529.4037P}. In Table \ref{Table:tab6}, we list the fitted parameters for each component, including the Galactic longitude ($l$) and latitude ($b$) of the gas clouds. The fitted parameters are peak optical depth ($\tau_\text{peak}$), center velocity ($v_\text{c}$), and the broadening parameter ($b_\text{broad}$).\\

\begin{table*}[ht!]\label{Table:tab6}
    \centering
    \caption{Physical parameters of the H {\sc i} spectra obtained from GWA survey \citep{2003MNRAS.346L..57K,2013MNRAS.436.2352R,2024MNRAS.529.4037P}.}
    \begin{tabular}{ c c c c c c c }
    
    \hline  
    Source   & \hspace{-1.3mm} Longitude & \hspace{-1.3mm} Latitude & \hspace{-1.3mm} Peak optical depth & \hspace{-1.3mm} Center velocity & \hspace{-1.3mm} Broadening parameter & \hspace{-1.3mm} References  \\ [0.5 ex]
     name  & \hspace{-1.3mm} ($l$) (deg.)  & \hspace{-1.3mm} ($b$) (deg.) &  \hspace{-1.3mm} ($\tau_{\text{peak}}$) & \hspace{-1.3mm}($v_{\text{c}})$ (km s$^{-1}$)  & \hspace{-1.3mm} ($b_\text{broad}$) (km s$^{-1}$)&  \hspace{-1.3mm} NR13/NN24 \\ [0.5 ex]
    \hline 
    
    B0023-263  & \hspace{-1.3mm}42.3 & \hspace{-1.3mm}-84.2 & \hspace{-1.3mm} 0.00207 $\pm$ 0.00027 & \hspace{-1.3mm}-5.09 $\pm$ 0.54 & \hspace{-1.3mm}5.10 $\pm$ 0.77& \hspace{-1.3mm}NR13 \\
    \hline
    B0114-211  & \hspace{-1.3mm}167.1 & \hspace{-1.3mm}-81.5 & \hspace{-1.3mm} 0.0327 $\pm$ 0.0037 & \hspace{-1.3mm}-7.727 $\pm$ 0.038 & \hspace{-1.3mm}1.109 $\pm$ 0.086& \hspace{-1.3mm}NR13 \\
    B0114-211 & \hspace{-1.3mm}167.1 & \hspace{-1.3mm}-81.5 & \hspace{-1.3mm} 0.00620 $\pm$ 0.00089  & \hspace{-1.3mm}-2.07 $\pm$ 0.14 & \hspace{-1.3mm}1.08 $\pm$ 0.20& \hspace{-1.3mm}NR13 \\
    B0114-211  & \hspace{-1.3mm}167.1 & \hspace{-1.3mm}-81.5 & \hspace{-1.3mm} 0.0119 $\pm$ 0.0038 & \hspace{-1.3mm}-7.10 $\pm$ 0.25 & \hspace{-1.3mm}2.53 $\pm$ 0.35& \hspace{-1.3mm}NR13 \\
    \hline
    B0117-155 & \hspace{-1.3mm}154.2 & \hspace{-1.3mm}-76.4 & \hspace{-1.3mm} 0.00448 $\pm$ 0.00035  & \hspace{-1.3mm}-6.33 $\pm$ 0.21 & \hspace{-1.3mm}3.22 $\pm$ 0.29& \hspace{-1.3mm}NR13 \\
    \hline
    B0134+329  & \hspace{-1.3mm}134.0 & \hspace{-1.3mm}-28.7 & \hspace{-1.3mm} 0.00065 $\pm$ 0.00024 & \hspace{-1.3mm}19.99 $\pm$ 0.45 & \hspace{-1.3mm}1.49 $\pm$ 0.64& \hspace{-1.3mm}NR13 \\
    B0134+329 & \hspace{-1.3mm}134.0 & \hspace{-1.3mm}-28.7 & \hspace{-1.3mm} 0.0518 $\pm$ 0.0011  & \hspace{-1.3mm}2.258 $\pm$ 0.013 & \hspace{-1.3mm}1.489 $\pm$ 0.025& \hspace{-1.3mm}NR13 \\
    B0134+329  & \hspace{-1.3mm}134.0 & \hspace{-1.3mm}-28.7 & \hspace{-1.3mm} 0.0131 $\pm$ 0.0014  & \hspace{-1.3mm}-0.663 $\pm$ 0.050 & \hspace{-1.3mm}1.072 $\pm$ 0.094& \hspace{-1.3mm}NR13 \\
    B0134+329 & \hspace{-1.3mm}134.0 & \hspace{-1.3mm}-28.7 & \hspace{-1.3mm} 0.0109 $\pm$ 0.0013  & \hspace{-1.3mm}-2.413 $\pm$ 0.051 & \hspace{-1.3mm}0.774 $\pm$ 0.083& \hspace{-1.3mm}NR13 \\
    B0134+329 & \hspace{-1.3mm}134.0 & \hspace{-1.3mm}-28.7 & \hspace{-1.3mm} 0.0077 $\pm$ 0.0019  & \hspace{-1.3mm}-2.83 $\pm$ 0.67 & \hspace{-1.3mm}7.60 $\pm$ 0.94& \hspace{-1.3mm}NR13 \\
    B0134+329  & \hspace{-1.3mm}134.0 & \hspace{-1.3mm}-28.7 & \hspace{-1.3mm} 0.0140 $\pm$ 0.0017 & \hspace{-1.3mm}-4.56 $\pm$ 0.11 & \hspace{-1.3mm}2.23 $\pm$ 0.23& \hspace{-1.3mm}NR13 \\
    B0134+329 & \hspace{-1.3mm}134.0 & \hspace{-1.3mm}-28.7  & \hspace{-1.3mm} 0.01651 $\pm$ 0.00074  & \hspace{-1.3mm}-11.104 $\pm$ 0.038 & \hspace{-1.3mm}1.685 $\pm$ 0.070& \hspace{-1.3mm}NR13 \\
    B0134+329  & \hspace{-1.3mm}134.0 & \hspace{-1.3mm}-28.7 & \hspace{-1.3mm} 0.00756 $\pm$ 0.00040 & \hspace{-1.3mm}-15.33 $\pm$ 0.20 & \hspace{-1.3mm}3.79 $\pm$ 0.23& \hspace{-1.3mm}NR13 \\
    B0134+329 & \hspace{-1.3mm}134.0 & \hspace{-1.3mm}-28.7 & \hspace{-1.3mm} 0.00093 $\pm$ 0.00017  & \hspace{-1.3mm}-29.27 $\pm$ 0.48 & \hspace{-1.3mm}3.29 $\pm$ 0.68& \hspace{-1.3mm}NR13 \\
    \hline
    B0202+149 & \hspace{-1.3mm}147.9 & \hspace{-1.3mm}-44.0  & \hspace{-1.3mm} 0.0570 $\pm$ 0.0045  & \hspace{-1.3mm}-10.761 $\pm$ 0.042 & \hspace{-1.3mm}1.434 $\pm$ 0.052& \hspace{-1.3mm}NR13 \\
    B0202+149  & \hspace{-1.3mm}147.9 & \hspace{-1.3mm}-44.0 & \hspace{-1.3mm} 0.0428 $\pm$ 0.0059  & \hspace{-1.3mm}-5.749 $\pm$ 0.039 & \hspace{-1.3mm}1.63 $\pm$ 0.10& \hspace{-1.3mm}NR13 \\
    B0202+149  & \hspace{-1.3mm}147.9 & \hspace{-1.3mm}-44.0 & \hspace{-1.3mm} 0.0164 $\pm$ 0.0019 & \hspace{-1.3mm}-0.120 $\pm$ 0.071 & \hspace{-1.3mm}1.74 $\pm$ 0.14& \hspace{-1.3mm}NR13 \\
    B0202+149  & \hspace{-1.3mm}147.9 & \hspace{-1.3mm}-44.0 & \hspace{-1.3mm} 0.0129 $\pm$ 0.0060 & \hspace{-1.3mm}-13.53 $\pm$ 0.37 & \hspace{-1.3mm}2.05 $\pm$ 0.47& \hspace{-1.3mm}NR13 \\
    B0202+149  & \hspace{-1.3mm}147.9 & \hspace{-1.3mm}-44.0 & \hspace{-1.3mm} 0.00189 $\pm$ 0.00042  & \hspace{-1.3mm}-52.76 $\pm$ 0.31 & \hspace{-1.3mm}1.72 $\pm$ 0.44& \hspace{-1.3mm}NR13 \\
    B0202+149  & \hspace{-1.3mm}147.9 & \hspace{-1.3mm}-44.0 & \hspace{-1.3mm} 0.0446 $\pm$ 0.0050 & \hspace{-1.3mm}-7.09 $\pm$ 0.54 & \hspace{-1.3mm}4.71 $\pm$ 0.86& \hspace{-1.3mm}NR13 \\
    B0202+149  & \hspace{-1.3mm}147.9 & \hspace{-1.3mm}-44.0 & \hspace{-1.3mm} 0.00230 $\pm$ 0.00054 & \hspace{-1.3mm}3.79 $\pm$ 0.32 & \hspace{-1.3mm}1.44 $\pm$ 0.45& \hspace{-1.3mm}NR13 \\
    \hline
    B0237-233  & \hspace{-1.3mm}209.8 & \hspace{-1.3mm}-65.1 & \hspace{-1.3mm} 0.11190 $\pm$ 0.00072 & \hspace{-1.3mm}-0.0689 $\pm$ 0.0051 & \hspace{-1.3mm}0.9871 $\pm$ 0.0078& \hspace{-1.3mm}NR13 \\
    B0237-233  & \hspace{-1.3mm}209.8 & \hspace{-1.3mm}-65.1 & \hspace{-1.3mm} 0.00836$\pm$ 0.00051 & \hspace{-1.3mm}-10.01 $\pm$ 0.13 & \hspace{-1.3mm}2.78 $\pm$ 0.21& \hspace{-1.3mm}NR13 \\
    B0237-233  & \hspace{-1.3mm}209.8 & \hspace{-1.3mm}-65.1 & \hspace{-1.3mm} 0.00255$\pm$ 0.00034 & \hspace{-1.3mm}-4.5 $\pm$ 1.1 & \hspace{-1.3mm}12.5 $\pm$ 1.2& \hspace{-1.3mm}NR13 \\
     \hline
    B0136+162  & \hspace{-1.3mm}166.6 & \hspace{-1.3mm}-33.6 & \hspace{-1.3mm} 0.2614 $\pm$ 0.0061  & \hspace{-1.3mm}-2.666 $\pm$ 0.050 & \hspace{-1.3mm}1.660 $\pm$ 0.034& \hspace{-1.3mm}NR13 \\
    B0136+162  & \hspace{-1.3mm}166.6 & \hspace{-1.3mm}-33.6 & \hspace{-1.3mm} 0.0627 $\pm$ 0.0028 & \hspace{-1.3mm}6.007 $\pm$ 0.065 & \hspace{-1.3mm}2.518 $\pm$ 0.085& \hspace{-1.3mm}NR13 \\
    B0136+162  & \hspace{-1.3mm}166.6 & \hspace{-1.3mm}-33.6  & \hspace{-1.3mm} 0.0313 $\pm$ 0.0018  & \hspace{-1.3mm}7.025 $\pm$ 0.023 & \hspace{-1.3mm}0.913 $\pm$ 0.044& \hspace{-1.3mm}NR13 \\
    B0136+162 & \hspace{-1.3mm}166.6 & \hspace{-1.3mm}-33.6 & \hspace{-1.3mm} 0.01829 $\pm$ 0.00056  & \hspace{-1.3mm}-10.188 $\pm$ 0.026 & \hspace{-1.3mm}1.096 $\pm$ 0.042& \hspace{-1.3mm}NR13 \\
    B0136+162 & \hspace{-1.3mm}166.6 & \hspace{-1.3mm}-33.6 & \hspace{-1.3mm} 0.0066 $\pm$ 0.0015  & \hspace{-1.3mm}-1.13 $\pm$ 0.55 & \hspace{-1.3mm}13.7 $\pm$ 1.0& \hspace{-1.3mm}NR13 \\
    B0136+162 & \hspace{-1.3mm}166.6 & \hspace{-1.3mm}-33.6 & \hspace{-1.3mm} 0.0540 $\pm$ 0.0028 & \hspace{-1.3mm}0.74 $\pm$ 0.23 & \hspace{-1.3mm}5.83 $\pm$ 0.25& \hspace{-1.3mm}NR13 \\
    B0136+162 & \hspace{-1.3mm}166.6 & \hspace{-1.3mm}-33.6 & \hspace{-1.3mm} 0.0780 $\pm$ 0.0095  & \hspace{-1.3mm}-1.731 $\pm$ 0.019 & \hspace{-1.3mm}0.747 $\pm$ 0.043& \hspace{-1.3mm}NR13 \\
    B0136+162 & \hspace{-1.3mm}166.6 & \hspace{-1.3mm}-33.6 & \hspace{-1.3mm} 0.4122 $\pm$ 0.0050 & \hspace{-1.3mm}-0.125 $\pm$ 0.013 & \hspace{-1.3mm}1.383 $\pm$ 0.013& \hspace{-1.3mm}NR13 \\
    \hline
     B0316+413 & \hspace{-1.3mm}150.6 & \hspace{-1.3mm}-13.3 & \hspace{-1.3mm} 0.00109 $\pm$ 0.00029  & \hspace{-1.3mm}-34.51 $\pm$ 0.44 & \hspace{-1.3mm}2.15 $\pm$ 0.76& \hspace{-1.3mm}NR13 \\
    B0316+413 & \hspace{-1.3mm}150.6 & \hspace{-1.3mm}-13.3  & \hspace{-1.3mm} 0.00047 $\pm$ 0.00016 & \hspace{-1.3mm}-55.78 $\pm$ 0.72 & \hspace{-1.3mm}2.6 $\pm$ 1.0& \hspace{-1.3mm}NR13 \\
    B0316+413 & \hspace{-1.3mm}150.6 & \hspace{-1.3mm}-13.3 & \hspace{-1.3mm} 0.00055 $\pm$ 0.00019 & \hspace{-1.3mm}-39.78 $\pm$ 0.67 & \hspace{-1.3mm}2.1 $\pm$ 1.0& \hspace{-1.3mm}NR13 \\
    \hline

     
    \end{tabular}
    \vspace{3mm}
    
    \end{table*}


\begin{table*}[hbt!]
    \centering
   
    \begin{tabular}{ c c c c c c c }
    \textbf{Continued.}\\
    \hline  
    Source   & \hspace{-1.3mm} Longitude & \hspace{-1.3mm} Latitude & \hspace{-1.3mm} Peak optical depth & \hspace{-1.3mm} Center velocity & \hspace{-1.3mm} Broadening parameter & \hspace{-1.3mm} References  \\ [0.5 ex]
     name & \hspace{-1.3mm} ($l$) (deg.)  & \hspace{-1.3mm} ($b$) (deg.)  &  \hspace{-1.3mm} ($\tau_{\text{peak}}$) & \hspace{-1.3mm}($v_{\text{c}})$ (km s$^{-1}$)  & \hspace{-1.3mm} ($b_\text{broad}$) (km s$^{-1}$)&  \hspace{-1.3mm} NR13/NN24 \\ [0.5 ex]

    \hline
    B0316+413 & \hspace{-1.3mm}150.6 & \hspace{-1.3mm}-13.3 & \hspace{-1.3mm} 0.0235 $\pm$ 0.0091 & \hspace{-1.3mm}3.378 $\pm$ 0.094 & \hspace{-1.3mm}1.07 $\pm$ 0.15& \hspace{-1.3mm}NR13 \\
   
    B0316+413 & \hspace{-1.3mm}150.6 & \hspace{-1.3mm}-13.3 & \hspace{-1.3mm} 0.0492 $\pm$ 0.0037 & \hspace{-1.3mm}-2.155 $\pm$ 0.065 & \hspace{-1.3mm}1.094 $\pm$ 0.092& \hspace{-1.3mm}NR13 \\
    B0316+413 & \hspace{-1.3mm}150.6 & \hspace{-1.3mm}-13.3 & \hspace{-1.3mm} 0.0283 $\pm$ 0.0032 & \hspace{-1.3mm}0.08 $\pm$ 0.16 & \hspace{-1.3mm}7.79 $\pm$ 0.21& \hspace{-1.3mm}NR13 \\
    B0316+413 & \hspace{-1.3mm}150.6 & \hspace{-1.3mm}-13.3 & \hspace{-1.3mm} 0.01769 $\pm$ 0.00067  & \hspace{-1.3mm}8.978 $\pm$ 0.046 & \hspace{-1.3mm}1.871 $\pm$ 0.068& \hspace{-1.3mm}NR13 \\
    B0316+413 & \hspace{-1.3mm}150.6 & \hspace{-1.3mm}-13.3 & \hspace{-1.3mm} 0.1952 $\pm$ 0.0068 & \hspace{-1.3mm}2.20 $\pm$ 0.10 & \hspace{-1.3mm}2.222 $\pm$ 0.084& \hspace{-1.3mm}NR13 \\
    B0316+413 & \hspace{-1.3mm}150.6 & \hspace{-1.3mm}-13.3 & \hspace{-1.3mm} 0.1139 $\pm$ 0.0098 & \hspace{-1.3mm}-0.380 $\pm$ 0.024 & \hspace{-1.3mm}1.107 $\pm$ 0.060& \hspace{-1.3mm}NR13 \\
    B0316+413 & \hspace{-1.3mm}150.6 & \hspace{-1.3mm}-13.3 & \hspace{-1.3mm} 0.0168 $\pm$ 0.0018 & \hspace{-1.3mm}-4.811 $\pm$ 0.027 & \hspace{-1.3mm}0.689 $\pm$ 0.054& \hspace{-1.3mm}NR13 \\
    B0316+413 & \hspace{-1.3mm}150.6 & \hspace{-1.3mm}-13.3 & \hspace{-1.3mm} 0.0146 $\pm$ 0.0014 & \hspace{-1.3mm}-21.794 $\pm$ 0.032 & \hspace{-1.3mm}2.10 $\pm$ 0.12& \hspace{-1.3mm}NR13 \\
    B0316+413 & \hspace{-1.3mm}150.6 & \hspace{-1.3mm}-13.3 & \hspace{-1.3mm} 0.00535 $\pm$ 0.00026  & \hspace{-1.3mm}-26.11 $\pm$ 0.82 & \hspace{-1.3mm}5.3 $\pm$ 1.1& \hspace{-1.3mm}NR13 \\
    B0316+413 & \hspace{-1.3mm}150.6 & \hspace{-1.3mm}-13.3 & \hspace{-1.3mm} 0.0528 $\pm$ 0.0020 & \hspace{-1.3mm}-5.302 $\pm$ 0.072 & \hspace{-1.3mm}2.276 $\pm$ 0.059& \hspace{-1.3mm}NR13 \\
    B0316+413 & \hspace{-1.3mm}150.6 & \hspace{-1.3mm}-13.3 & \hspace{-1.3mm} 0.00313 $\pm$ 0.00027 & \hspace{-1.3mm}-15.30 $\pm$ 0.28 & \hspace{-1.3mm}2.96 $\pm$ 0.38& \hspace{-1.3mm}NR13 \\
    \hline
    B0404+768 & \hspace{-1.3mm}133.4 & \hspace{-1.3mm}+18.3 & \hspace{-1.3mm} 0.00261 $\pm$ 0.00053  & \hspace{-1.3mm}11.57 $\pm$ 0.26 & \hspace{-1.3mm}1.54 $\pm$ 0.38& \hspace{-1.3mm}NR13 \\
    B0404+768 & \hspace{-1.3mm}133.4 & \hspace{-1.3mm}+18.3 & \hspace{-1.3mm} 0.06115 $\pm$ 0.00092  & \hspace{-1.3mm}3.183 $\pm$ 0.029 & \hspace{-1.3mm}1.566 $\pm$ 0.040& \hspace{-1.3mm}NR13 \\
    B0404+768 & \hspace{-1.3mm}133.4 & \hspace{-1.3mm}+18.3 & \hspace{-1.3mm} 0.2533 $\pm$ 0.0081 & \hspace{-1.3mm}-0.4963 $\pm$ 0.0082 & \hspace{-1.3mm}0.959 $\pm$ 0.017& \hspace{-1.3mm}NR13 \\

    B0404+768 & \hspace{-1.3mm}133.4 & \hspace{-1.3mm}+18.3 & \hspace{-1.3mm} 0.08553 $\pm$ 0.00097 & \hspace{-1.3mm}-10.897 $\pm$ 0.017 & \hspace{-1.3mm}1.328 $\pm$ 0.027& \hspace{-1.3mm}NR13 \\
    B0404+768 & \hspace{-1.3mm}133.4 & \hspace{-1.3mm}+18.3 & \hspace{-1.3mm} 0.04637 $\pm$ 0.00082 & \hspace{-1.3mm}-13.606 $\pm$ 0.031 & \hspace{-1.3mm}1.326 $\pm$ 0.040& \hspace{-1.3mm}NR13 \\
    
    B0404+768 & \hspace{-1.3mm}133.4 & \hspace{-1.3mm}+18.3 & \hspace{-1.3mm} 0.1497 $\pm$ 0.0076 & \hspace{-1.3mm}-0.901 $\pm$ 0.035 & \hspace{-1.3mm}1.934 $\pm$ 0.045& \hspace{-1.3mm}NR13 \\
    B0404+768 & \hspace{-1.3mm}133.4 & \hspace{-1.3mm}+18.3 & \hspace{-1.3mm} 0.156 $\pm$ 0.0013 & \hspace{-1.3mm}-6.490 $\pm$ 0.046 & \hspace{-1.3mm}1.03 $\pm$ 0.11& \hspace{-1.3mm}NR13 \\
    B0404+768 & \hspace{-1.3mm}133.4 & \hspace{-1.3mm}+18.3 & \hspace{-1.3mm} 0.0307 $\pm$ 0.0014 & \hspace{-1.3mm}-4.49 $\pm$ 0.17 & \hspace{-1.3mm}8.82 $\pm$ 0.17& \hspace{-1.3mm}NR13 \\
    \hline
    B0407-658 & \hspace{-1.3mm}278.6 & \hspace{-1.3mm}-40.9 & \hspace{-1.3mm} 0.1117 $\pm$ 0.0027 & \hspace{-1.3mm}16.3923 $\pm$ 0.0089 & \hspace{-1.3mm}1.047 $\pm$ 0.018& \hspace{-1.3mm}NR13 \\
    B0407-658 & \hspace{-1.3mm}278.6 & \hspace{-1.3mm}-40.9 & \hspace{-1.3mm} 0.00228 $\pm$ 0.00057 & \hspace{-1.3mm}19.8 $\pm$ 1.4 & \hspace{-1.3mm}8.4 $\pm$ 1.6& \hspace{-1.3mm}NR13 \\
    B0407-658 & \hspace{-1.3mm}278.6 & \hspace{-1.3mm}-40.9  & \hspace{-1.3mm} 0.02588 $\pm$ 0.00077  & \hspace{-1.3mm}0.497 $\pm$ 0.029 & \hspace{-1.3mm}1.304 $\pm$ 0.049& \hspace{-1.3mm}NR13 \\
    B0407-658 & \hspace{-1.3mm}278.6 & \hspace{-1.3mm}-40.9 & \hspace{-1.3mm} 0.00144 $\pm$ 0.00025 & \hspace{-1.3mm}49.1 $\pm$ 1.1 & \hspace{-1.3mm}7.8 $\pm$ 1.6& \hspace{-1.3mm}NR13 \\
    B0407-658 & \hspace{-1.3mm}278.6 & \hspace{-1.3mm}-40.9 & \hspace{-1.3mm} 0.00280 $\pm$ 0.00046 & \hspace{-1.3mm}-0.52 $\pm$ 0.75 & \hspace{-1.3mm}7.5 $\pm$ 1.2& \hspace{-1.3mm}NR13 \\
    B0407-658 & \hspace{-1.3mm}278.6 & \hspace{-1.3mm}-40.9 & \hspace{-1.3mm} 0.0417 $\pm$ 0.0026 & \hspace{-1.3mm}17.053 $\pm$ 0.058 & \hspace{-1.3mm}2.348 $\pm$ 0.076& \hspace{-1.3mm}NR13 \\
    B0407-658 & \hspace{-1.3mm}278.6 & \hspace{-1.3mm}-40.9 & \hspace{-1.3mm} 0.00265 $\pm$ 0.00057 & \hspace{-1.3mm}-18.26 $\pm$ 0.27 & \hspace{-1.3mm}1.52 $\pm$ 0.38& \hspace{-1.3mm}NR13 \\
    \hline
    B0518+165 & \hspace{-1.3mm}187.4 & \hspace{-1.3mm}-11.3 & \hspace{-1.3mm} 0.02574 $\pm$ 0.00076  & \hspace{-1.3mm}-22.138 $\pm$ 0.021 & \hspace{-1.3mm}1.039 $\pm$ 0.038& \hspace{-1.3mm}NR13 \\
    B0518+165 & \hspace{-1.3mm}187.4 & \hspace{-1.3mm}-11.3 & \hspace{-1.3mm} 0.01885 $\pm$ 0.00054  & \hspace{-1.3mm}-21.585 $\pm$ 0.066 & \hspace{-1.3mm}5.21 $\pm$ 0.12& \hspace{-1.3mm}NR13 \\
    B0518+165 & \hspace{-1.3mm}187.4 & \hspace{-1.3mm}-11.3 & \hspace{-1.3mm} 0.00379 $\pm$ 0.00059 & \hspace{-1.3mm}-12.89 $\pm$ 0.15 & \hspace{-1.3mm}1.15 $\pm$ 0.23& \hspace{-1.3mm}NR13 \\
    B0518+165 & \hspace{-1.3mm}187.4 & \hspace{-1.3mm}-11.3 & \hspace{-1.3mm} 0.221 $\pm$ 0.021  & \hspace{-1.3mm}-0.675 $\pm$ 0.030 & \hspace{-1.3mm}0.945 $\pm$ 0.048& \hspace{-1.3mm}NR13 \\
    B0518+165 & \hspace{-1.3mm}187.4 & \hspace{-1.3mm}-11.3 & \hspace{-1.3mm} 0.0266 $\pm$ 0.0020 & \hspace{-1.3mm}-6.12 $\pm$ 0.24 & \hspace{-1.3mm}3.93 $\pm$ 0.20& \hspace{-1.3mm}NR13 \\
    B0518+165 & \hspace{-1.3mm}187.4 & \hspace{-1.3mm}-11.3 & \hspace{-1.3mm} 0.136 $\pm$ 0.013 & \hspace{-1.3mm}-2.01 $\pm$ 0.12 & \hspace{-1.3mm}1.242 $\pm$ 0.080& \hspace{-1.3mm}NR13 \\
    B0518+165 & \hspace{-1.3mm}187.4 & \hspace{-1.3mm}-11.3 & \hspace{-1.3mm} 0.2653 $\pm$ 0.0057 & \hspace{-1.3mm}1.051 $\pm$ 0.027 & \hspace{-1.3mm}1.151 $\pm$ 0.029& \hspace{-1.3mm}NR13 \\
    B0518+165 & \hspace{-1.3mm}187.4 & \hspace{-1.3mm}-11.3 & \hspace{-1.3mm} 0.0841 $\pm$ 0.0027 & \hspace{-1.3mm}4.85 $\pm$ 0.34 & \hspace{-1.3mm}6.41 $\pm$ 0.29& \hspace{-1.3mm}NR13 \\
    B0518+165 & \hspace{-1.3mm}187.4 & \hspace{-1.3mm}-11.3  & \hspace{-1.3mm} 1.0378 $\pm$ 0.0073 & \hspace{-1.3mm}5.968 $\pm$ 0.012 & \hspace{-1.3mm}1.162 $\pm$ 0.011& \hspace{-1.3mm}NR13 \\
    B0518+165 & \hspace{-1.3mm}187.4 & \hspace{-1.3mm}-11.3  & \hspace{-1.3mm} 0.327 $\pm$ 0.017 & \hspace{-1.3mm}7.354 $\pm$ 0.018 & \hspace{-1.3mm}0.762 $\pm$ 0.024& \hspace{-1.3mm}NR13 \\
    B0518+165 & \hspace{-1.3mm}187.4 & \hspace{-1.3mm}-11.3 & \hspace{-1.3mm} 0.4060 $\pm$ 0.0042 & \hspace{-1.3mm}8.965 $\pm$ 0.018 & \hspace{-1.3mm}1.488 $\pm$ 0.019& \hspace{-1.3mm}NR13 \\
    B0518+165 & \hspace{-1.3mm}187.4 & \hspace{-1.3mm}-11.3 & \hspace{-1.3mm} 0.00102 $\pm$ 0.00033  & \hspace{-1.3mm}21.91 $\pm$ 0.83 & \hspace{-1.3mm}3.0 $\pm$ 1.3& \hspace{-1.3mm}NR13 \\
    B0518+165 & \hspace{-1.3mm}187.4 & \hspace{-1.3mm}-11.3 & \hspace{-1.3mm} 0.00107 $\pm$ 0.00042  & \hspace{-1.3mm}28.13 $\pm$ 0.56 & \hspace{-1.3mm}1.63 $\pm$ 0.81& \hspace{-1.3mm}NR13 \\
    \hline
    B0531+194  & \hspace{-1.3mm}186.8 & \hspace{-1.3mm}-7.1 & \hspace{-1.3mm} 0.0021 $\pm$ 0.0011 & \hspace{-1.3mm}-10.3 $\pm$ 1.9 & \hspace{-1.3mm}4.5 $\pm$ 1.7& \hspace{-1.3mm}NR13 \\
    B0531+194 & \hspace{-1.3mm}186.8 & \hspace{-1.3mm}-7.1 & \hspace{-1.3mm} 0.0111 $\pm$ 0.0030  & \hspace{-1.3mm}-2.43 $\pm$ 0.17 & \hspace{-1.3mm}1.46 $\pm$ 0.27& \hspace{-1.3mm}NR13 \\
    B0531+194 & \hspace{-1.3mm}186.8 & \hspace{-1.3mm}-7.1 & \hspace{-1.3mm} 0.00199 $\pm$ 0.00051  & \hspace{-1.3mm}-27.6 $\pm$ 1.8 & \hspace{-1.3mm}3.4 $\pm$ 1.8& \hspace{-1.3mm}NR13 \\
    B0531+194 & \hspace{-1.3mm}186.8 & \hspace{-1.3mm}-7.1 & \hspace{-1.3mm} 0.0084 $\pm$ 0.0012 & \hspace{-1.3mm}-23.13 $\pm$ 0.27 & \hspace{-1.3mm}2.39 $\pm$ 0.25& \hspace{-1.3mm}NR13 \\
    B0531+194 & \hspace{-1.3mm}186.8 & \hspace{-1.3mm}-7.1 & \hspace{-1.3mm} 0.387 $\pm$ 0.014 & \hspace{-1.3mm}1.4574 $\pm$ 0.0085 & \hspace{-1.3mm}1.142 $\pm$ 0.018& \hspace{-1.3mm}NR13 \\
    B0531+194 & \hspace{-1.3mm}186.8 & \hspace{-1.3mm}-7.1 & \hspace{-1.3mm} 0.229 $\pm$ 0.020 & \hspace{-1.3mm}2.55 $\pm$ 0.16 & \hspace{-1.3mm}2.69 $\pm$ 0.17& \hspace{-1.3mm}NR13 \\
    B0531+194 & \hspace{-1.3mm}186.8 & \hspace{-1.3mm}-7.1 & \hspace{-1.3mm} 0.057 $\pm$ 0.017 & \hspace{-1.3mm}4.29 $\pm$ 0.34 & \hspace{-1.3mm}8.10 $\pm$ 0.81& \hspace{-1.3mm}NR13 \\
    B0531+194 & \hspace{-1.3mm}186.8 & \hspace{-1.3mm}-7.1 & \hspace{-1.3mm} 0.151 $\pm$ 0.017  & \hspace{-1.3mm}6.202 $\pm$ 0.054& \hspace{-1.3mm}1.62 $\pm$ 0.11& \hspace{-1.3mm}NR13 \\
    B0531+194 & \hspace{-1.3mm}186.8 & \hspace{-1.3mm}-7.1 & \hspace{-1.3mm} 0.2714 $\pm$ 0.0092 & \hspace{-1.3mm}9.213 $\pm$ 0.016& \hspace{-1.3mm}1.537 $\pm$ 0.031& \hspace{-1.3mm}NR13 \\
    B0531+194 & \hspace{-1.3mm}186.8 & \hspace{-1.3mm}-7.1 & \hspace{-1.3mm} 0.0301 $\pm$ 0.0040 & \hspace{-1.3mm}12.050 $\pm$ 0.099& \hspace{-1.3mm}1.38 $\pm$ 0.13& \hspace{-1.3mm}NR13 \\
    B0531+194 & \hspace{-1.3mm}186.8 & \hspace{-1.3mm}-7.1 & \hspace{-1.3mm} 0.00624 $\pm$ 0.00048 & \hspace{-1.3mm}21.05 $\pm$ 0.13& \hspace{-1.3mm}2.06 $\pm$ 0.21& \hspace{-1.3mm}NR13 \\
    \hline
    B0538+498  & \hspace{-1.3mm}161.7 & \hspace{-1.3mm}+10.3 & \hspace{-1.3mm} 0.0645 $\pm$ 0.0024 & \hspace{-1.3mm}4.720 $\pm$ 0.022& \hspace{-1.3mm}1.251 $\pm$ 0.023& \hspace{-1.3mm}NR13 \\
    B0538+498 & \hspace{-1.3mm}161.7 & \hspace{-1.3mm}+10.3 & \hspace{-1.3mm} 0.336 $\pm$ 0.010  & \hspace{-1.3mm}0.79 $\pm$ 0.12& \hspace{-1.3mm}2.44 $\pm$ 0.10& \hspace{-1.3mm}NR13 \\
    B0538+498 & \hspace{-1.3mm}161.7 & \hspace{-1.3mm}+10.3 & \hspace{-1.3mm} 0.211 $\pm$ 0.016 & \hspace{-1.3mm}-0.4550 $\pm$ 0.0094& \hspace{-1.3mm}1.087 $\pm$ 0.024& \hspace{-1.3mm}NR13 \\
    B0538+498 & \hspace{-1.3mm}161.7 & \hspace{-1.3mm}+10.3 & \hspace{-1.3mm} 0.0278 $\pm$ 0.0024 & \hspace{-1.3mm}-1.94 $\pm$ 0.54& \hspace{-1.3mm}7.66 $\pm$ 0.37& \hspace{-1.3mm}NR13 \\
    B0538+498 & \hspace{-1.3mm}161.7 & \hspace{-1.3mm}+10.3 & \hspace{-1.3mm} 0.084 $\pm$ 0.018 & \hspace{-1.3mm}-2.32 $\pm$ 0.13& \hspace{-1.3mm}1.730 $\pm$ 0.083& \hspace{-1.3mm}NR13 \\
    B0538+498 & \hspace{-1.3mm}161.7 & \hspace{-1.3mm}+10.3 & \hspace{-1.3mm} 0.718 $\pm$ 0.080 & \hspace{-1.3mm}-8.091 $\pm$ 0.042& \hspace{-1.3mm}0.869 $\pm$ 0.015& \hspace{-1.3mm}NR13 \\
    B0538+498 & \hspace{-1.3mm}161.7 & \hspace{-1.3mm}+10.3 & \hspace{-1.3mm} 0.210 $\pm$ 0.065  & \hspace{-1.3mm}-8.92 $\pm$ 0.19& \hspace{-1.3mm}0.917 $\pm$ 0.086& \hspace{-1.3mm}NR13 \\
    B0538+498 & \hspace{-1.3mm}161.7 & \hspace{-1.3mm}+10.3 & \hspace{-1.3mm} 0.1561 $\pm$ 0.0059 & \hspace{-1.3mm}-11.022 $\pm$ 0.010& \hspace{-1.3mm}0.989 $\pm$ 0.026& \hspace{-1.3mm}NR13 \\
    
    \hline

     \hline
    \end{tabular}
    \vspace{3mm}
    
    \end{table*}

\begin{table*}[hbt!]
    \centering
   
    \begin{tabular}{ c c c c c c c }
    \textbf{Continued.}\\
    \hline  
    Source  & \hspace{-1.3mm} Longitude & \hspace{-1.3mm} Latitude & \hspace{-1.3mm} Peak optical depth  & \hspace{-1.3mm} Center velocity & \hspace{-1.3mm} Broadening parameter & \hspace{-1.3mm} References  \\ [0.5 ex]
     name  & \hspace{-1.3mm} ($l$) (deg.)  & \hspace{-1.3mm} ($b$) (deg.) &  \hspace{-1.3mm} ($\tau_{\text{peak}}$) & \hspace{-1.3mm}($v_{\text{c}})$ (km s$^{-1}$)  & \hspace{-1.3mm} ($b_\text{broad}$) (km s$^{-1}$)&  \hspace{-1.3mm} NR13/NN24 \\ [0.5 ex]
    \hline
    B0538+498 & \hspace{-1.3mm}161.7 & \hspace{-1.3mm}+10.3 & \hspace{-1.3mm} 0.1584 $\pm$ 0.0066 & \hspace{-1.3mm}-11.267 $\pm$ 0.080& \hspace{-1.3mm}3.743 $\pm$ 0.061& \hspace{-1.3mm}NR13 \\
    B0538+498 & \hspace{-1.3mm}161.7 & \hspace{-1.3mm}+10.3 & \hspace{-1.3mm} 0.0449 $\pm$ 0.0028 & \hspace{-1.3mm}-13.855 $\pm$ 0.030& \hspace{-1.3mm}1.089 $\pm$ 0.043& \hspace{-1.3mm}NR13 \\
    
    B0538+498 & \hspace{-1.3mm}161.7 & \hspace{-1.3mm}+10.3 & \hspace{-1.3mm} 0.01407 $\pm$ 0.00025 & \hspace{-1.3mm}-19.664 $\pm$ 0.019& \hspace{-1.3mm}1.367 $\pm$ 0.031& \hspace{-1.3mm}NR13 \\
    B0538+498 & \hspace{-1.3mm}161.7 & \hspace{-1.3mm}+10.3 & \hspace{-1.3mm} 0.00232 $\pm$ 0.00013  & \hspace{-1.3mm}-27.42 $\pm$ 0.43& \hspace{-1.3mm}5.89 $\pm$ 0.78& \hspace{-1.3mm}NR13 \\
    B0538+498 & \hspace{-1.3mm}161.7 & \hspace{-1.3mm}+10.3 & \hspace{-1.3mm} 0.00236 $\pm$ 0.00030  & \hspace{-1.3mm}-34.51 $\pm$ 0.13& \hspace{-1.3mm}1.58 $\pm$ 0.23& \hspace{-1.3mm}NR13 \\
    B0538+498 & \hspace{-1.3mm}161.7 & \hspace{-1.3mm}+10.3 & \hspace{-1.3mm} 0.00216 $\pm$ 0.00018  & \hspace{-1.3mm}-41.16 $\pm$ 0.71& \hspace{-1.3mm}6.6 $\pm$ 1.4& \hspace{-1.3mm}NR13 \\
    B0538+498 & \hspace{-1.3mm}161.7 & \hspace{-1.3mm}+10.3 & \hspace{-1.3mm} 0.00273 $\pm$ 0.00028  & \hspace{-1.3mm}-43.591 $\pm$ 0.087& \hspace{-1.3mm}1.21 $\pm$ 0.15& \hspace{-1.3mm}NR13 \\
    B0538+498 & \hspace{-1.3mm}161.7 & \hspace{-1.3mm}+10.3 & \hspace{-1.3mm} 0.00077 $\pm$ 0.00030 & \hspace{-1.3mm}-49.44 $\pm$ 0.21& \hspace{-1.3mm}0.66 $\pm$ 0.32& \hspace{-1.3mm}NR13 \\
    B0538+498 & \hspace{-1.3mm}161.7 & \hspace{-1.3mm}+10.3 & \hspace{-1.3mm} 0.000644 $\pm$ 0.000095 & \hspace{-1.3mm}-55.8 $\pm$ 1.5& \hspace{-1.3mm}6.7 $\pm$ 2.4& \hspace{-1.3mm}NR13 \\
    B0538+498 & \hspace{-1.3mm}161.7 & \hspace{-1.3mm}+10.3 & \hspace{-1.3mm} 0.000337 $\pm$ 0.000090 & \hspace{-1.3mm}-73.0 $\pm$ 1.5& \hspace{-1.3mm}6.3 $\pm$ 2.3& \hspace{-1.3mm}NR13 \\
    
    B0538+498 & \hspace{-1.3mm}161.7 & \hspace{-1.3mm}+10.3 & \hspace{-1.3mm} 0.00065 $\pm$ 0.00012  & \hspace{-1.3mm}-115.83 $\pm$ 0.51& \hspace{-1.3mm}3.29 $\pm$ 0.73& \hspace{-1.3mm}NR13 \\
    B0538+498 & \hspace{-1.3mm}161.7 & \hspace{-1.3mm}+10.3 & \hspace{-1.3mm} 0.00373 $\pm$ 0.00016 & \hspace{-1.3mm}-124.236 $\pm$ 0.068& \hspace{-1.3mm}1.941 $\pm$ 0.097& \hspace{-1.3mm}NR13 \\

    \hline
    B0831+557 & \hspace{-1.3mm}162.2 & \hspace{-1.3mm}+36.6 & \hspace{-1.3mm} 0.00194 $\pm$ 0.00047 & \hspace{-1.3mm}-5.59 $\pm$ 0.63& \hspace{-1.3mm}2.25 $\pm$ 0.84& \hspace{-1.3mm}NR13 \\
    B0831+557 & \hspace{-1.3mm}162.2 & \hspace{-1.3mm}+36.6 & \hspace{-1.3mm} 0.0596 $\pm$ 0.0065  & \hspace{-1.3mm}3.355 $\pm$ 0.035& \hspace{-1.3mm}1.238 $\pm$ 0.063& \hspace{-1.3mm}NR13 \\
    B0831+557 & \hspace{-1.3mm}162.2 & \hspace{-1.3mm}+36.6 & \hspace{-1.3mm} 0.0289 $\pm$ 0.0072 & \hspace{-1.3mm}0.959 $\pm$ 0.074& \hspace{-1.3mm}1.32 $\pm$ 0.13& \hspace{-1.3mm}NR13 \\
    B0831+557 & \hspace{-1.3mm}162.2 & \hspace{-1.3mm}+36.6 & \hspace{-1.3mm} 0.0321 $\pm$ 0.0094 & \hspace{-1.3mm}1.94 $\pm$ 0.11& \hspace{-1.3mm}3.31 $\pm$ 0.31& \hspace{-1.3mm}NR13 \\
    B0831+557 & \hspace{-1.3mm}162.2 & \hspace{-1.3mm}+36.6 & \hspace{-1.3mm} 0.02651 $\pm$ 0.00056 & \hspace{-1.3mm}-37.950 $\pm$ 0.023& \hspace{-1.3mm}1.329 $\pm$ 0.032& \hspace{-1.3mm}NR13 \\
    B0831+557 & \hspace{-1.3mm}162.2 & \hspace{-1.3mm}+36.6 & \hspace{-1.3mm} 0.00377 $\pm$ 0.00038 & \hspace{-1.3mm}-56.20 $\pm$ 0.25& \hspace{-1.3mm}3.03 $\pm$ 0.35& \hspace{-1.3mm}NR13 \\
    B0831+557 & \hspace{-1.3mm}162.2 & \hspace{-1.3mm}+36.6 & \hspace{-1.3mm} 0.00375 $\pm$ 0.00055 & \hspace{-1.3mm}-65.76 $\pm$ 0.16& \hspace{-1.3mm}1.35 $\pm$ 0.23& \hspace{-1.3mm}NR13 \\
    \hline
    B0834-196 & \hspace{-1.3mm}243.3 & \hspace{-1.3mm}+12.6 & \hspace{-1.3mm} 0.00124 $\pm$ 0.00044 & \hspace{-1.3mm}50.73 $\pm$ 0.94& \hspace{-1.3mm}3.3 $\pm$ 1.3& \hspace{-1.3mm}NR13 \\
    B0834-196 & \hspace{-1.3mm}243.3 & \hspace{-1.3mm}+12.6 & \hspace{-1.3mm} 0.0120 $\pm$ 0.0014 & \hspace{-1.3mm}11.07 $\pm$ 0.16& \hspace{-1.3mm}4.15 $\pm$ 0.38& \hspace{-1.3mm}NR13 \\
    B0834-196 & \hspace{-1.3mm}243.3 & \hspace{-1.3mm}+12.6 & \hspace{-1.3mm} 0.0200 $\pm$ 0.0015 & \hspace{-1.3mm}11.177 $\pm$ 0.042& \hspace{-1.3mm}1.197 $\pm$ 0.092& \hspace{-1.3mm}NR13 \\
    B0834-196 & \hspace{-1.3mm}243.3 & \hspace{-1.3mm}+12.6 & \hspace{-1.3mm} 0.0477 $\pm$ 0.0034 & \hspace{-1.3mm}2.739 $\pm$ 0.022& \hspace{-1.3mm}0.784 $\pm$ 0.046& \hspace{-1.3mm}NR13 \\
    B0834-196 & \hspace{-1.3mm}243.3 & \hspace{-1.3mm}+12.6 & \hspace{-1.3mm} 0.0420 $\pm$ 0.0058 & \hspace{-1.3mm}-0.721 $\pm$ 0.065& \hspace{-1.3mm}1.44 $\pm$ 0.11& \hspace{-1.3mm}NR13 \\
    
    B0834-196 & \hspace{-1.3mm}243.3 & \hspace{-1.3mm}+12.6 & \hspace{-1.3mm} 0.1414 $\pm$ 0.0045 & \hspace{-1.3mm}2.308 $\pm$ 0.055& \hspace{-1.3mm}2.442 $\pm$ 0.071& \hspace{-1.3mm}NR13 \\
    B0834-196  & \hspace{-1.3mm}243.3 & \hspace{-1.3mm}+12.6 & \hspace{-1.3mm} 0.0081 $\pm$ 0.0023 & \hspace{-1.3mm}-2.8 $\pm$ 1.8& \hspace{-1.3mm}3.8 $\pm$ 1.3& \hspace{-1.3mm}NR13 \\
    \hline
     
    B0906+430  & \hspace{-1.3mm}178.3 & \hspace{-1.3mm}+42.8 & \hspace{-1.3mm} 0.00262 $\pm$ 0.00030 & \hspace{-1.3mm}12.47 $\pm$ 0.27& \hspace{-1.3mm}3.14 $\pm$ 0.45& \hspace{-1.3mm}NR13 \\
    B0906+430 & \hspace{-1.3mm}178.3 & \hspace{-1.3mm}+42.8 & \hspace{-1.3mm} 0.00355 $\pm$ 0.00050  & \hspace{-1.3mm}1.97 $\pm$ 0.10& \hspace{-1.3mm}0.93 $\pm$ 0.16& \hspace{-1.3mm}NR13 \\
    B0906+430  & \hspace{-1.3mm}178.3 & \hspace{-1.3mm}+42.8 & \hspace{-1.3mm} 0.00113 $\pm$ 0.00017 & \hspace{-1.3mm}3.1 $\pm$ 1.9& \hspace{-1.3mm}16.4 $\pm$ 2.0& \hspace{-1.3mm}NR13 \\
     \hline   
    B1151-348  & \hspace{-1.3mm} 289.9 & \hspace{-1.3mm} +26.3 & \hspace{-1.3mm} 0.00217 $\pm$ 0.00043 & \hspace{-1.3mm}-35.66 $\pm$ 0.50 & \hspace{-1.3mm}2.15 $\pm$ 0.72& \hspace{-1.3mm}NR13 \\
    B1151-348 & \hspace{-1.3mm} 289.9 & \hspace{-1.3mm} +26.3 & \hspace{-1.3mm} 0.00438 $\pm$ 0.00044 & \hspace{-1.3mm}-31.05 $\pm$ 0.24 & \hspace{-1.3mm}2.02 $\pm$ 0.34& \hspace{-1.3mm}NR13 \\
    B1151-348 & \hspace{-1.3mm} 289.9 & \hspace{-1.3mm} +26.3 & \hspace{-1.3mm} 0.00442 $\pm$ 0.00059 & \hspace{-1.3mm}-18.62 $\pm$ 0.17 & \hspace{-1.3mm}1.22 $\pm$ 0.24& \hspace{-1.3mm}NR13 \\
    B1151-348 & \hspace{-1.3mm} 289.9 & \hspace{-1.3mm} +26.3 & \hspace{-1.3mm} 0.00658 $\pm$ 0.00052 & \hspace{-1.3mm}-14.87 $\pm$ 0.26 & \hspace{-1.3mm}1.88 $\pm$ 0.44& \hspace{-1.3mm}NR13 \\
    B1151-348 & \hspace{-1.3mm} 289.9 & \hspace{-1.3mm} +26.3 & \hspace{-1.3mm} 0.00872 $\pm$ 0.00073 & \hspace{-1.3mm}-11.28 $\pm$ 0.19 & \hspace{-1.3mm}1.76 $\pm$ 0.31& \hspace{-1.3mm}NR13 \\
    B1151-348 & \hspace{-1.3mm} 289.9 & \hspace{-1.3mm} +26.3 & \hspace{-1.3mm} 0.0880 $\pm$ 0.0027 & \hspace{-1.3mm}-5.583 $\pm$ 0.015 & \hspace{-1.3mm}1.274 $\pm$ 0.029& \hspace{-1.3mm}NR13 \\
    B1151-348 & \hspace{-1.3mm} 289.9   & \hspace{-1.3mm} +26.3 & \hspace{-1.3mm} 0.0381 $\pm$ 0.0034 & \hspace{-1.3mm}-4.198 $\pm$ 0.079 & \hspace{-1.3mm}4.27 $\pm$ 0.17& \hspace{-1.3mm}NR13 \\
    B1151-348 & \hspace{-1.3mm} 289.9   & \hspace{-1.3mm} +26.3 & \hspace{-1.3mm} 0.0430 $\pm$ 0.0025  & \hspace{-1.3mm}-2.425 $\pm$ 0.019 & \hspace{-1.3mm}1.021 $\pm$ 0.047& \hspace{-1.3mm}NR13 \\
    B1151-348 & \hspace{-1.3mm}289.9  & \hspace{-1.3mm} +26.3 & \hspace{-1.3mm} 0.02396 $\pm$ 0.00057 & \hspace{-1.3mm}6.605 $\pm$ 0.029 & \hspace{-1.3mm}1.303 $\pm$ 0.043& \hspace{-1.3mm}NR13 \\
    B1151-348 & \hspace{-1.3mm}289.9  & \hspace{-1.3mm} +26.3 & \hspace{-1.3mm} 0.00531 $\pm$ 0.00053  & \hspace{-1.3mm}10.08 $\pm$ 0.13 & \hspace{-1.3mm}1.46 $\pm$ 0.20& \hspace{-1.3mm}NR13 \\
     \hline  
    B1245-197 & \hspace{-1.3mm} 302.0 & \hspace{-1.3mm} +42.9 & \hspace{-1.3mm} 0.00309 $\pm$ 0.00094 & \hspace{-1.3mm}-17.11 $\pm$ 0.20 & \hspace{-1.3mm}0.79 $\pm$ 0.28& \hspace{-1.3mm}NR13 \\
    B1245-197 & \hspace{-1.3mm} 302.0 & \hspace{-1.3mm} +42.9 & \hspace{-1.3mm} 0.00296 $\pm$ 0.00095 & \hspace{-1.3mm}-12.19 $\pm$ 0.22 & \hspace{-1.3mm}0.82 $\pm$ 0.32& \hspace{-1.3mm}NR13 \\
    B1245-197 & \hspace{-1.3mm} 302.0 & \hspace{-1.3mm} +42.9 & \hspace{-1.3mm} 0.00728 $\pm$ 0.00067 & \hspace{-1.3mm}-6.80 $\pm$ 0.57 & \hspace{-1.3mm}2.95 $\pm$ 0.69& \hspace{-1.3mm}NR13 \\
    B1245-197 & \hspace{-1.3mm} 302.0 & \hspace{-1.3mm} +42.9 & \hspace{-1.3mm} 0.0243 $\pm$ 0.0081  & \hspace{-1.3mm}-3.39 $\pm$ 0.11 & \hspace{-1.3mm}1.45 $\pm$ 0.20& \hspace{-1.3mm}NR13 \\
    B1245-197 & \hspace{-1.3mm} 302.0 & \hspace{-1.3mm} +42.9 & \hspace{-1.3mm} 0.0108 $\pm$ 0.0034 & \hspace{-1.3mm}-1.2 $\pm$ 1.2 & \hspace{-1.3mm}2.52 $\pm$ 0.84& \hspace{-1.3mm}NR13 \\
    B1245-197 & \hspace{-1.3mm} 302.0 & \hspace{-1.3mm} +42.9  & \hspace{-1.3mm} 0.00373 $\pm$ 0.00084 & \hspace{-1.3mm}9.15 $\pm$ 0.18 & \hspace{-1.3mm}0.97 $\pm$ 0.25& \hspace{-1.3mm}NR13 \\
    \hline
    B1328+254 & \hspace{-1.3mm}22.5 & \hspace{-1.3mm}+81.0 & \hspace{-1.3mm} 0.00131 $\pm$ 0.00016 & \hspace{-1.3mm}-23.47 $\pm$ 0.29 & \hspace{-1.3mm}2.84 $\pm$ 0.44& \hspace{-1.3mm}NR13 \\
    B1328+254 & \hspace{-1.3mm}22.5 & \hspace{-1.3mm}+81.0 & \hspace{-1.3mm} 0.00132 $\pm$ 0.00015  & \hspace{-1.3mm}-3.29 $\pm$ 0.28 & \hspace{-1.3mm}3.04 $\pm$ 0.40& \hspace{-1.3mm}NR13 \\
    B1328+254 & \hspace{-1.3mm}22.5 & \hspace{-1.3mm}+81.0 & \hspace{-1.3mm} 0.00048 $\pm$ 0.00015  & \hspace{-1.3mm}11.23 $\pm$ 0.74 & \hspace{-1.3mm}2.8 $\pm$ 1.0& \hspace{-1.3mm}NR13 \\
    B1328+254 & \hspace{-1.3mm}22.5 & \hspace{-1.3mm}+81.0 & \hspace{-1.3mm} 0.00098 $\pm$ 0.00023 & \hspace{-1.3mm}-17.95 $\pm$ 0.26 & \hspace{-1.3mm}1.26 $\pm$ 0.37& \hspace{-1.3mm}NR13 \\
    \hline
    B1328+307 & \hspace{-1.3mm}56.5 & \hspace{-1.3mm}+80.7 & \hspace{-1.3mm} 0.00654 $\pm$ 0.00022 & \hspace{-1.3mm}-7.371 $\pm$ 0.045 & \hspace{-1.3mm}1.949 $\pm$ 0.085& \hspace{-1.3mm}NR13 \\
    B1328+307 & \hspace{-1.3mm}56.5 & \hspace{-1.3mm}+80.7  & \hspace{-1.3mm} 0.00730 $\pm$ 0.00024 & \hspace{-1.3mm}-14.281 $\pm$ 0.029 & \hspace{-1.3mm}1.122 $\pm$ 0.047& \hspace{-1.3mm}NR13 \\
    B1328+307 & \hspace{-1.3mm}56.5 & \hspace{-1.3mm}+80.7  & \hspace{-1.3mm} 0.00549 $\pm$ 0.00021 & \hspace{-1.3mm}-28.785 $\pm$ 0.041 & \hspace{-1.3mm}1.357 $\pm$ 0.059& \hspace{-1.3mm}NR13 \\
    B1328+307 & \hspace{-1.3mm}56.5 & \hspace{-1.3mm}+80.7  & \hspace{-1.3mm} 0.00109 $\pm$ 0.00019 & \hspace{-1.3mm}-8.41 $\pm$ 0.68 & \hspace{-1.3mm}9.2 $\pm$ 1.0& \hspace{-1.3mm}NR13 \\
   
    \hline

     \hline
    \end{tabular}
    \vspace{3mm}
    
    \end{table*}

\begin{table*}[hbt!]
    \centering
   
    \begin{tabular}{ c c c c c c c }
    \textbf{Continued.}\\
    \hline  
    Source & \hspace{-1.3mm} Longitude & \hspace{-1.3mm} Latitude  & \hspace{-1.3mm} Peak optical depth & \hspace{-1.3mm} Center velocity & \hspace{-1.3mm} Broadening parameter & \hspace{-1.3mm} References  \\ [0.5 ex]
     name  & \hspace{-1.3mm} ($l$) (deg.)  & \hspace{-1.3mm} ($b$) (deg.) &  \hspace{-1.3mm} ($\tau_{\text{peak}}$) & \hspace{-1.3mm}($v_{\text{c}})$ (km s$^{-1}$)  & \hspace{-1.3mm} ($b_\text{broad}$) (km s$^{-1}$)&  \hspace{-1.3mm} NR13/NN24 \\ [0.5 ex]
     \hline
    B1345+125 & \hspace{-1.3mm}347.2 & \hspace{-1.3mm}+70.2 & \hspace{-1.3mm} 0.00298 $\pm$ 0.00085 & \hspace{-1.3mm}-48.06 $\pm$ 0.14 & \hspace{-1.3mm}0.59 $\pm$ 0.19& \hspace{-1.3mm}NR13 \\
    B1345+125 & \hspace{-1.3mm} 347.2 & \hspace{-1.3mm} +70.2  & \hspace{-1.3mm} 0.0080 $\pm$ 0.0010 & \hspace{-1.3mm}-3.930 $\pm$ 0.074 & \hspace{-1.3mm}0.80 $\pm$ 0.12& \hspace{-1.3mm}NR13 \\
    B1345+125 & \hspace{-1.3mm} 347.2 & \hspace{-1.3mm} +70.2 & \hspace{-1.3mm} 0.0799 $\pm$ 0.0011 & \hspace{-1.3mm}-1.173 $\pm$ 0.011 & \hspace{-1.3mm}1.454 $\pm$ 0.020& \hspace{-1.3mm}NR13 \\
    B1345+125 & \hspace{-1.3mm} 347.2 &\hspace{0.0mm} +70.2 & \hspace{-1.3mm} 0.00876 $\pm$ 0.00099 & \hspace{-1.3mm}-3.08 $\pm$ 0.22 & \hspace{-1.3mm}5.28 $\pm$ 0.32& \hspace{-1.3mm}NR13 \\
    B1345+125 & \hspace{-1.3mm}347.2 & \hspace{-1.3mm} +70.2 & \hspace{-1.3mm} 0.00180 $\pm$ 0.00053 & \hspace{-1.3mm}19.52 $\pm$ 0.37 & \hspace{-1.3mm}1.52 $\pm$ 0.52& \hspace{-1.3mm}NR13 \\
    \hline
      B1611+343 & \hspace{-1.3mm} 55.2 & \hspace{-1.3mm} +46.4 & \hspace{-1.3mm} 0.00177 $\pm$ 0.00021 & \hspace{-1.3mm}-4.10 $\pm$ 0.40 & \hspace{-1.3mm}4.14 $\pm$ 0.56& \hspace{-1.3mm}NR13 \\
    B1611+343 & \hspace{-1.3mm} 55.2 & \hspace{-1.3mm} +46.4 & \hspace{-1.3mm} 0.00093 $\pm$ 0.00025 & \hspace{-1.3mm}-30.32 $\pm$ 0.56 & \hspace{-1.3mm}2.54 $\pm$ 0.80& \hspace{-1.3mm}NR13 \\
     \hline
    B1641+399 & \hspace{-1.3mm} 63.5 & \hspace{-1.3mm} +40.9 & \hspace{-1.3mm} 0.00082 $\pm$ 0.00013  & \hspace{-1.3mm}-1.25 $\pm$ 0.61 & \hspace{-1.3mm}4.72 $\pm$ 0.86& \hspace{-1.3mm}NR13 \\
    \hline
    B1814-637  & \hspace{-1.3mm} 330.9 & \hspace{-1.3mm} -20.8 & \hspace{-1.3mm} 0.3083 $\pm$ 0.0051 & \hspace{-1.3mm}-0.9024 $\pm$ 0.0035 & \hspace{-1.3mm}0.8638 $\pm$ 0.0095& \hspace{-1.3mm}NR13 \\
    B1814-637  & \hspace{-1.3mm} 330.9 & \hspace{-1.3mm} -20.8 & \hspace{-1.3mm} 0.1088 $\pm$ 0.0049 & \hspace{-1.3mm}-1.055 $\pm$ 0.017 & \hspace{-1.3mm}2.052 $\pm$ 0.051& \hspace{-1.3mm}NR13 \\
    B1814-637  & \hspace{-1.3mm} 330.9 & \hspace{-1.3mm} -20.8 & \hspace{-1.3mm} 0.00927 $\pm$ 0.00092 & \hspace{-1.3mm}-2.37 $\pm$ 0.25 & \hspace{-1.3mm}7.53 $\pm$ 0.42& \hspace{-1.3mm}NR13 \\
    \hline
    B1827-360  & \hspace{-1.3mm} 358.3 & \hspace{-1.3mm} -11.8 & \hspace{-1.3mm} 0.00152 $\pm$ 0.00048 & \hspace{-1.3mm}-15.29 $\pm$ 0.32 & \hspace{-1.3mm}1.27 $\pm$ 0.49& \hspace{-1.3mm}NR13 \\
    B1827-360  & \hspace{-1.3mm} 358.3 & \hspace{-1.3mm} -11.8 & \hspace{-1.3mm} 0.00266 $\pm$ 0.00080  & \hspace{-1.3mm}-6.1 $\pm$ 2.9 & \hspace{-1.3mm}5.3 $\pm$ 2.6& \hspace{-1.3mm}NR13 \\
    B1827-360  & \hspace{-1.3mm} 358.3 & \hspace{-1.3mm}  -11.8 & \hspace{-1.3mm} 0.0057 $\pm$ 0.0019 & \hspace{-1.3mm}-3.86 $\pm$ 0.27 & \hspace{-1.3mm}0.98 $\pm$ 0.28& \hspace{-1.3mm}NR13 \\
   
    B1827-360  & \hspace{-1.3mm} 358.3 & \hspace{-1.3mm}  -11.8 & \hspace{-1.3mm} 0.0349 $\pm$ 0.0017 & \hspace{-1.3mm}-1.944 $\pm$ 0.057 & \hspace{-1.3mm}1.329 $\pm$ 0.088& \hspace{-1.3mm}NR13 \\
    B1827-360  & \hspace{-1.3mm} 358.3 & \hspace{-1.3mm}  -11.8 & \hspace{-1.3mm} 0.1371 $\pm$ 0.0054 & \hspace{-1.3mm}4.395 $\pm$ 0.098 & \hspace{-1.3mm}1.71 $\pm$ 0.11& \hspace{-1.3mm}NR13 \\
    B1827-360  & \hspace{-1.3mm} 358.3 & \hspace{-1.3mm}  -11.8 & \hspace{-1.3mm} 0.0640 $\pm$ 0.0061 & \hspace{-1.3mm}4.44 $\pm$ 0.27 & \hspace{-1.3mm}5.29 $\pm$ 0.24& \hspace{-1.3mm}NR13 \\
    B1827-360  & \hspace{-1.3mm} 358.3 & \hspace{-1.3mm}  -11.8 & \hspace{-1.3mm} 0.0231 $\pm$ 0.0076 & \hspace{-1.3mm}4.017 $\pm$ 0.042 & \hspace{-1.3mm}0.73 $\pm$ 0.11& \hspace{-1.3mm}NR13 \\
    B1827-360  & \hspace{-1.3mm} 358.3 & \hspace{-1.3mm}  -11.8 & \hspace{-1.3mm} 0.1023 $\pm$ 0.0086 & \hspace{-1.3mm}6.865 $\pm$ 0.090 & \hspace{-1.3mm}1.596 $\pm$ 0.078& \hspace{-1.3mm}NR13 \\
    B1827-360 & \hspace{-1.3mm} 358.3 & \hspace{-1.3mm}  -11.8  & \hspace{-1.3mm} 0.0439 $\pm$ 0.0013 & \hspace{-1.3mm}10.961 $\pm$ 0.019 & \hspace{-1.3mm}1.076 $\pm$ 0.032& \hspace{-1.3mm}NR13 \\
    \hline
   
    B1921-293 & \hspace{-1.3mm} 9.3 & \hspace{-1.3mm} -19.6 & \hspace{-1.3mm} 0.191 $\pm$ 0.015  & \hspace{-1.3mm}4.736 $\pm$ 0.016 & \hspace{-1.3mm}1.012 $\pm$ 0.030& \hspace{-1.3mm}NR13 \\
    B1921-293  & \hspace{-1.3mm} 9.3  & \hspace{-1.3mm} -19.6 & \hspace{-1.3mm} 0.0640 $\pm$ 0.0083 & \hspace{-1.3mm}3.163 $\pm$ 0.030 & \hspace{-1.3mm}0.736 $\pm$ 0.048& \hspace{-1.3mm}NR13 \\
    B1921-293  & \hspace{-1.3mm} 9.3 & \hspace{-1.3mm} -19.6 & \hspace{-1.3mm} 0.0240 $\pm$ 0.0028 & \hspace{-1.3mm}4.43 $\pm$ 0.26 & \hspace{-1.3mm}6.16 $\pm$ 0.39& \hspace{-1.3mm}NR13 \\
    B1921-293  & \hspace{-1.3mm} 9.3 & \hspace{-1.3mm}  -19.6 & \hspace{-1.3mm} 0.170 $\pm$ 0.015 & \hspace{-1.3mm}4.171 $\pm$ 0.025 & \hspace{-1.3mm}2.202 $\pm$ 0.086& \hspace{-1.3mm}NR13 \\
    B1921-293  & \hspace{-1.3mm} 9.3  & \hspace{-1.3mm} -19.6 & \hspace{-1.3mm} 0.00800 $\pm$ 0.00098 & \hspace{-1.3mm}13.31 $\pm$ 0.36 & \hspace{-1.3mm}3.49 $\pm$ 0.40& \hspace{-1.3mm}NR13 \\
    B1921-293  & \hspace{-1.3mm} 9.3 & \hspace{-1.3mm} -19.6 & \hspace{-1.3mm} 0.00279 $\pm$ 0.00053 & \hspace{-1.3mm}-13.96 $\pm$ 0.30 & \hspace{-1.3mm}1.89 $\pm$ 0.42& \hspace{-1.3mm}NR13 \\

    B1921-293 & \hspace{-1.3mm} 9.3 & \hspace{-1.3mm} -19.6  & \hspace{-1.3mm} 0.00164 $\pm$ 0.00020 & \hspace{-1.3mm}42.4 $\pm$ 1.3 & \hspace{-1.3mm}13.2 $\pm$ 1.9& \hspace{-1.3mm}NR13 \\
    \hline
    B2050+364 & \hspace{-1.3mm} 78.9 & \hspace{-1.3mm} -5.1 & \hspace{-1.3mm} 0.00085 $\pm$ 0.00025 & \hspace{-1.3mm}-76.6 $\pm$ 2.8 & \hspace{-1.3mm}10.2 $\pm$ 4.3& \hspace{-1.3mm}NR13 \\
    B2050+364 & \hspace{-1.3mm} 78.9  & \hspace{-1.3mm} -5.1 & \hspace{-1.3mm} 0.00085 $\pm$ 0.00029  & \hspace{-1.3mm}-39.7 $\pm$ 2.2 & \hspace{-1.3mm}7.6 $\pm$ 3.5& \hspace{-1.3mm}NR13 \\
    B2050+364 & \hspace{-1.3mm} 78.9 & \hspace{-1.3mm} -5.1 & \hspace{-1.3mm} 0.00168 $\pm$ 0.00047 & \hspace{-1.3mm}-52.79 $\pm$ 0.68 & \hspace{-1.3mm}2.85 $\pm$ 0.99& \hspace{-1.3mm}NR13 \\
    B2050+364 & \hspace{-1.3mm} 78.9 & \hspace{-1.3mm} -5.1 & \hspace{-1.3mm} 0.00224 $\pm$ 0.00064 & \hspace{-1.3mm}30.31 $\pm$ 0.35 & \hspace{-1.3mm}1.48 $\pm$ 0.52& \hspace{-1.3mm}NR13 \\
    B2050+364 & \hspace{-1.3mm} 78.9 & \hspace{-1.3mm} -5.1 & \hspace{-1.3mm} 0.00689 $\pm$ 0.00062 & \hspace{-1.3mm}-64.59 $\pm$ 0.13 & \hspace{-1.3mm}1.85 $\pm$ 0.21& \hspace{-1.3mm}NR13 \\
    B2050+364 & \hspace{-1.3mm} 78.9 & \hspace{-1.3mm} -5.1 & \hspace{-1.3mm} 0.00742 $\pm$ 0.00054 & \hspace{-1.3mm}-21.73 $\pm$ 0.14 & \hspace{-1.3mm}2.33 $\pm$ 0.20& \hspace{-1.3mm}NR13 \\
    B2050+364 & \hspace{-1.3mm} 78.9 & \hspace{-1.3mm} -5.1 & \hspace{-1.3mm} 0.01135 $\pm$ 0.00089 & \hspace{-1.3mm}-12.214 $\pm$ 0.088 & \hspace{-1.3mm}1.53 $\pm$ 0.16& \hspace{-1.3mm}NR13 \\
    B2050+364 & \hspace{-1.3mm} 78.9 & \hspace{-1.3mm} -5.1 & \hspace{-1.3mm} 0.0164 $\pm$ 0.0014  & \hspace{-1.3mm}-6.931 $\pm$ 0.051 & \hspace{-1.3mm}0.788 $\pm$ 0.084& \hspace{-1.3mm}NR13 \\
    B2050+364 & \hspace{-1.3mm}78.9 & \hspace{-1.3mm} -5.1 & \hspace{-1.3mm} 0.0973 $\pm$ 0.0026 & \hspace{-1.3mm}-1.600 $\pm$ 0.023 & \hspace{-1.3mm}1.637 $\pm$ 0.044& \hspace{-1.3mm}NR13 \\
    B2050+364 & \hspace{-1.3mm} 78.9 & \hspace{-1.3mm} -5.1 & \hspace{-1.3mm} 0.0438 $\pm$ 0.0036 & \hspace{-1.3mm}1.625 $\pm$ 0.048 & \hspace{-1.3mm}1.299 $\pm$ 0.097& \hspace{-1.3mm}NR13 \\
    B2050+364 & \hspace{-1.3mm} 78.9 & \hspace{-1.3mm} -5.1 & \hspace{-1.3mm} 0.1559 $\pm$ 0.0043  & \hspace{-1.3mm}8.202 $\pm$ 0.012 & \hspace{-1.3mm}0.901 $\pm$ 0.025& \hspace{-1.3mm}NR13 \\

    B2050+364 & \hspace{-1.3mm} 78.9 & \hspace{-1.3mm} -5.1 & \hspace{-1.3mm} 0.188 $\pm$ 0.034 & \hspace{-1.3mm}9.14 $\pm$ 0.25 & \hspace{-1.3mm}3.55 $\pm$ 0.30& \hspace{-1.3mm}NR13 \\
    B2050+364 & \hspace{-1.3mm} 78.9 & \hspace{-1.3mm} -5.1 & \hspace{-1.3mm} 0.0428 $\pm$ 0.0067 & \hspace{-1.3mm}14.17 $\pm$ 0.14 & \hspace{-1.3mm}1.16 $\pm$ 0.17& \hspace{-1.3mm}NR13 \\
    B2050+364 & \hspace{-1.3mm} 78.9 & \hspace{-1.3mm} -5.1 & \hspace{-1.3mm} 0.0532 $\pm$ 0.0064 & \hspace{-1.3mm}15.823 $\pm$ 0.089 & \hspace{-1.3mm}0.961 $\pm$ 0.080& \hspace{-1.3mm}NR13 \\
    B2050+364 & \hspace{-1.3mm} 78.9 & \hspace{-1.3mm} -5.1 & \hspace{-1.3mm} 0.032 $\pm$ 0.022 & \hspace{-1.3mm}14.5 $\pm$ 3.9 & \hspace{-1.3mm} 5.3 $\pm$ 2.0& \hspace{-1.3mm}NR13 \\
    B2050+364 & \hspace{-1.3mm} 78.9 & \hspace{-1.3mm} -5.1 & \hspace{-1.3mm} 0.00731 $\pm$ 0.00069 & \hspace{-1.3mm}25.24 $\pm$ 0.17 & \hspace{-1.3mm} 1.97 $\pm$ 0.26& \hspace{-1.3mm}NR13 \\
    B2050+364 & \hspace{-1.3mm} 78.9 & \hspace{-1.3mm} -5.1 & \hspace{-1.3mm} 0.0414 $\pm$ 0.0031  & \hspace{-1.3mm}-0.46 $\pm$ 0.62 & \hspace{-1.3mm} 7.59 $\pm$ 0.52& \hspace{-1.3mm}NR13 \\
    
    \hline   
    \end{tabular}
    \vspace{3mm}
    
    \end{table*}

\begin{table*}[hbt!]
    \centering
   
    \begin{tabular}{ c c c c c c c }
    \textbf{Continued.}\\
    \hline  
    Source  & \hspace{-1.3mm} Longitude & \hspace{-1.3mm} Latitude & \hspace{-1.3mm} Peak optical depth & \hspace{-1.3mm} Center velocity & \hspace{-1.3mm} Broadening parameter & \hspace{-1.3mm} References  \\ [0.5 ex]
     name  & \hspace{-1.3mm} ($l$) (deg.)  & \hspace{-1.3mm} ($b$) (deg.) &  \hspace{-1.3mm} ($\tau_{\text{peak}}$) & \hspace{-1.3mm}($v_{\text{c}})$ (km s$^{-1}$)  & \hspace{-1.3mm} ($b_\text{broad}$) (km s$^{-1}$)&  \hspace{-1.3mm} NR13/NN24 \\ [0.5 ex]
     \hline 
    B2200+420 & \hspace{-1.3mm} 92.6 & \hspace{-1.3mm} -10.4 & \hspace{-1.3mm} 0.00181 $\pm$ 0.00056  & \hspace{-1.3mm}-29.3 $\pm$ 2.3 & \hspace{-1.3mm} 7.7 $\pm$ 3.6& \hspace{-1.3mm}NR13 \\
    B2200+420 & \hspace{-1.3mm} 92.6 & \hspace{-1.3mm} -10.4 & \hspace{-1.3mm} 0.0051 $\pm$ 0.0018  & \hspace{-1.3mm}-17.68 $\pm$ 0.24  & \hspace{-1.3mm} 0.83 $\pm$ 0.35 & \hspace{-1.3mm}NR13 \\
    B2200+420 & \hspace{-1.3mm} 92.6 & \hspace{-1.3mm} -10.4  & \hspace{-1.3mm} 0.0041 $\pm$ 0.0011 & \hspace{-1.3mm}-10.30 $\pm$ 0.53  & \hspace{-1.3mm}2.42 $\pm$ 0.77 & \hspace{-1.3mm}NR13 \\
    B2200+420 & \hspace{-1.3mm} 92.6 & \hspace{-1.3mm} -10.4 & \hspace{-1.3mm} 0.0182 $\pm$ 0.0016 & \hspace{-1.3mm}-20.592 $\pm$ 0.080  & \hspace{-1.3mm}1.22 $\pm$ 0.13 & \hspace{-1.3mm}NR13 \\
    B2200+420 & \hspace{-1.3mm} 92.6 & \hspace{-1.3mm}  -10.4 & \hspace{-1.3mm} 0.192 $\pm$ 0.020 & \hspace{-1.3mm}-1.622 $\pm$ 0.022  & \hspace{-1.3mm}0.557 $\pm$ 0.043 & \hspace{-1.3mm}NR13 \\
    B2200+420 & \hspace{-1.3mm} 92.6 & \hspace{-1.3mm} -10.4 & \hspace{-1.3mm} 0.289 $\pm$ 0.020 & \hspace{-1.3mm}-0.918 $\pm$ 0.094  & \hspace{-1.3mm}1.449 $\pm$ 0.084 & \hspace{-1.3mm}NR13 \\
    B2200+420 & \hspace{-1.3mm} 92.6 & \hspace{-1.3mm} -10.4  & \hspace{-1.3mm} 0.218 $\pm$ 0.012 & \hspace{-1.3mm}1.041 $\pm$ 0.097 & \hspace{-1.3mm}4.18 $\pm$ 0.12 & \hspace{-1.3mm}NR13 \\
    B2200+420 & \hspace{-1.3mm} 92.6 & \hspace{-1.3mm} -10.4  & \hspace{-1.3mm} 0.315 $\pm$ 0.014 & \hspace{-1.3mm}1.354 $\pm$ 0.037 & \hspace{-1.3mm}0.981 $\pm$ 0.040 & \hspace{-1.3mm}NR13 \\
    B2200+420 & \hspace{-1.3mm} 92.6 & \hspace{-1.3mm} -10.4  & \hspace{-1.3mm} 0.0757 $\pm$ 0.0043 & \hspace{-1.3mm}6.428 $\pm$ 0.035 & \hspace{-1.3mm}1.216 $\pm$ 0.072 & \hspace{-1.3mm}NR13 \\
    B2200+420 & \hspace{-1.3mm} 92.6 & \hspace{-1.3mm} -10.4 & \hspace{-1.3mm} 0.0228 $\pm$ 0.0013  & \hspace{-1.3mm}13.808 $\pm$ 0.069 & \hspace{-1.3mm}1.439 $\pm$ 0.098 & \hspace{-1.3mm}NR13 \\
    B2200+420 & \hspace{-1.3mm} 92.6 & \hspace{-1.3mm} -10.4 & \hspace{-1.3mm} 0.1031 $\pm$ 0.0017 & \hspace{-1.3mm}18.149 $\pm$ 0.012 & \hspace{-1.3mm}0.874 $\pm$ 0.017 & \hspace{-1.3mm}NR13 \\
    
       \hline
    B2203-188 & \hspace{-1.3mm} 36.7 & \hspace{-1.3mm} -51.2 & \hspace{-1.3mm} 0.06515 $\pm$ 0.00062 & \hspace{-1.3mm}7.802 $\pm$ 0.010 & \hspace{-1.3mm}1.053 $\pm$ 0.018 & \hspace{-1.3mm}NR13 \\ 
    B2203-188 & \hspace{-1.3mm} 36.7 & \hspace{-1.3mm} -51.2 & \hspace{-1.3mm} 0.02882 $\pm$ 0.00054 & \hspace{-1.3mm}4.948 $\pm$ 0.027 & \hspace{-1.3mm}1.437 $\pm$ 0.043 & \hspace{-1.3mm}NR13 \\ 
    B2203-188 & \hspace{-1.3mm} 36.7& \hspace{-1.3mm} -51.2 & \hspace{-1.3mm} 0.00119 $\pm$ 0.00034 & \hspace{-1.3mm}-21.91 $\pm$ 0.76 & \hspace{-1.3mm}3.6 $\pm$ 1.3 & \hspace{-1.3mm}NR13 \\ 
    B2203-188 & \hspace{-1.3mm} 36.7& \hspace{-1.3mm} -51.2 & \hspace{-1.3mm} 0.00140 $\pm$ 0.00018 & \hspace{-1.3mm}-3.6 $\pm$ 2.1 & \hspace{-1.3mm}15.5 $\pm$ 2.8 & \hspace{-1.3mm}NR13 \\ 
    B2203-188 & \hspace{-1.3mm} 36.7 & \hspace{-1.3mm} -51.2 & \hspace{-1.3mm} 0.00377 $\pm$ 0.00063 & \hspace{-1.3mm}10.23 $\pm$ 0.15 & \hspace{-1.3mm}0.93 $\pm$ 0.23 & \hspace{-1.3mm}NR13 \\ 
    B2203-188 & \hspace{-1.3mm} 36.7 & \hspace{-1.3mm} -51.2 & \hspace{-1.3mm} 0.00324 $\pm$ 0.00096 & \hspace{-1.3mm}-6.138 $\pm$ 0.092 & \hspace{-1.3mm}0.39 $\pm$ 0.13 & \hspace{-1.3mm}NR13 \\ 
          
    \hline
     B2223-052 & \hspace{-1.3mm} 59.0 & \hspace{-1.3mm} -48.8 & \hspace{-1.3mm} 0.0976 $\pm$ 0.0025 & \hspace{-1.3mm}-6.919 $\pm$ 0.013 & \hspace{-1.3mm}1.662 $\pm$ 0.025 & \hspace{-1.3mm}NR13 \\ 
    B2223-052 & \hspace{-1.3mm} 59.0 & \hspace{-1.3mm} -48.8 & \hspace{-1.3mm} 0.0419 $\pm$ 0.0031 & \hspace{-1.3mm}-4.41 $\pm$ 0.13 & \hspace{-1.3mm}4.81 $\pm$ 0.18 & \hspace{-1.3mm}NR13 \\ 
    B2223-052 & \hspace{-1.3mm} 59.0 & \hspace{-1.3mm} -48.8 & \hspace{-1.3mm} 0.1039 $\pm$ 0.0024 & \hspace{-1.3mm}-3.943 $\pm$ 0.010 & \hspace{-1.3mm}1.236 $\pm$ 0.019 & \hspace{-1.3mm}NR13 \\ 
    B2223-052 & \hspace{-1.3mm} 59.0 & \hspace{-1.3mm} -48.8  & \hspace{-1.3mm} 0.0085 $\pm$ 0.0012 & \hspace{-1.3mm}2.34 $\pm$ 0.14 & \hspace{-1.3mm}1.330 $\pm$ 0.212 & \hspace{-1.3mm}NR13 \\ 
    B2223-052 & \hspace{-1.3mm} 59.0 & \hspace{-1.3mm} -48.8 & \hspace{-1.3mm} 0.0232 $\pm$ 0.0024 & \hspace{-1.3mm}5.93 $\pm$ 0.24 & \hspace{-1.3mm}2.23 $\pm$ 0.16 & \hspace{-1.3mm}NR13 \\ 
    B2223-052 & \hspace{-1.3mm} 59.0 & \hspace{-1.3mm} -48.8 & \hspace{-1.3mm} 0.0232 $\pm$ 0.0036 & \hspace{-1.3mm}7.238 $\pm$ 0.046 & \hspace{-1.3mm}1.059 $\pm$ 0.097 & \hspace{-1.3mm}NR13 \\ 
    B2223-052 & \hspace{-1.3mm} 59.0 & \hspace{-1.3mm} -48.8  & \hspace{-1.3mm} 0.00210 $\pm$ 0.00044 & \hspace{-1.3mm}11.27 $\pm$ 0.40 & \hspace{-1.3mm}1.57 $\pm$ 0.57 & \hspace{-1.3mm}NR13 \\ 
    \hline
    1352+314 & \hspace{-1.3mm} 54.6 & \hspace{-1.3mm} 76.1 & \hspace{-1.3mm} 0.01446 $\pm$ 0.00064 & \hspace{-1.3mm}-12.01 $\pm$ 0.04 & \hspace{-1.3mm}1.16 $\pm$ 0.06& \hspace{-1.3mm}NN24 \\
    1352+314 & \hspace{-1.3mm} 54.6 & \hspace{-1.3mm} 76.1 & \hspace{-1.3mm} 0.00807 $\pm$ 0.00067 & \hspace{-1.3mm}-5.55 $\pm$ 0.07 & \hspace{-1.3mm}1.04 $\pm$ 0.11& \hspace{-1.3mm}NN24 \\
    1352+314 & \hspace{-1.3mm} 54.6 & \hspace{-1.3mm} 76.1 & \hspace{-1.3mm} 0.00178 $\pm$ 0.00028 & \hspace{-1.3mm}-28.88 $\pm$ 0.66 & \hspace{-1.3mm}5.10 $\pm$ 0.94& \hspace{-1.3mm}NN24 \\
    1352+314 & \hspace{-1.3mm} 54.6 & \hspace{-1.3mm} 76.1 & \hspace{-1.3mm} 0.00151 $\pm$ 0.00036 & \hspace{-1.3mm}-8.68 $\pm$ 1.24 & \hspace{-1.3mm}7.16 $\pm$ 1.38& \hspace{-1.3mm}NN24 \\
   
     \hline
     1400+621 & \hspace{-1.3mm} 109.6 & \hspace{-1.3mm} 53.1  & \hspace{-1.3mm} 0.00455 $\pm$ 0.00085 & \hspace{-1.3mm}0.02 $\pm$ 0.17 & \hspace{-1.3mm}1.09 $\pm$ 0.23& \hspace{-1.3mm}NN24 \\
     \hline
     1609+266 & \hspace{-1.3mm} 44.2 & \hspace{-1.3mm} 46.2 & \hspace{-1.3mm} 0.09226 $\pm$ 0.00253  & \hspace{-1.3mm}-1.87 $\pm$ 0.03 & \hspace{-1.3mm}2.08 $\pm$ 0.03& \hspace{-1.3mm}NN24 \\
    1609+266 & \hspace{-1.3mm} 44.2 & \hspace{-1.3mm} 46.2 & \hspace{-1.3mm} 0.05487 $\pm$ 0.00548 & \hspace{-1.3mm}-2.01 $\pm$ 0.05 & \hspace{-1.3mm}0.65 $\pm$ 0.04& \hspace{-1.3mm}NN24 \\
    1609+266 & \hspace{-1.3mm} 44.2 & \hspace{-1.3mm} 46.2 & \hspace{-1.3mm} 0.00553 $\pm$ 0.00067 & \hspace{-1.3mm}2.44 $\pm$ 0.10 & \hspace{-1.3mm}1.08 $\pm$ 0.17& \hspace{-1.3mm}NN24 \\
    1609+266 & \hspace{-1.3mm} 44.2 & \hspace{-1.3mm} 46.2 & \hspace{-1.3mm} 0.01123 $\pm$ 0.00057 & \hspace{-1.3mm}-5.67 $\pm$ 0.37 & \hspace{-1.3mm}7.21 $\pm$ 0.24& \hspace{-1.3mm}NN24 \\
    1609+266 & \hspace{-1.3mm} 44.2 & \hspace{-1.3mm} 46.2 & \hspace{-1.3mm} 0.04731 $\pm$ 0.00802 & \hspace{-1.3mm}-10.52 $\pm$ 0.03 & \hspace{-1.3mm}1.85 $\pm$ 0.11& \hspace{-1.3mm}NN24 \\
    1609+266 & \hspace{-1.3mm} 44.2 & \hspace{-1.3mm} 46.2 & \hspace{-1.3mm} 0.06728 $\pm$ 0.00380 & \hspace{-1.3mm}-3.01 $\pm$ 0.05 & \hspace{-1.3mm}0.73 $\pm$ 0.04& \hspace{-1.3mm}NN24 \\
    1609+266 & \hspace{-1.3mm} 44.2 & \hspace{-1.3mm} 46.2 & \hspace{-1.3mm} 0.08992 $\pm$ 0.00813 & \hspace{-1.3mm}-10.61 $\pm$ 0.01 & \hspace{-1.3mm}1.02 $\pm$ 0.03& \hspace{-1.3mm}NN24 \\
    \hline 
    
      1634+627 & \hspace{-1.3mm} 93.6 & \hspace{-1.3mm} 39.4 & \hspace{-1.3mm} 0.00948 $\pm$ 0.00117 & \hspace{-1.3mm}-24.84 $\pm$ 0.06 & \hspace{-1.3mm}1.20 $\pm$ 0.13& \hspace{-1.3mm}NN24 \\          
    1634+627 & \hspace{-1.3mm} 93.6 & \hspace{-1.3mm} 39.4 & \hspace{-1.3mm} 0.00436 $\pm$ 0.00123 & \hspace{-1.3mm}-21.45 $\pm$ 0.13 & \hspace{-1.3mm}1.14 $\pm$ 0.30& \hspace{-1.3mm}NN24 \\   
      1634+627  & \hspace{-1.3mm} 93.6 & \hspace{-1.3mm} 39.4 & \hspace{-1.3mm} 0.00339 $\pm$ 0.00080 & \hspace{-1.3mm}-0.63 $\pm$ 0.08 & \hspace{-1.3mm}0.42 $\pm$ 0.11& \hspace{-1.3mm}NN24 \\          
      1634+627 & \hspace{-1.3mm} 93.6 & \hspace{-1.3mm} 39.4 & \hspace{-1.3mm} 0.00192 $\pm$ 0.00055 & \hspace{-1.3mm}9.78 $\pm$ 0.20 & \hspace{-1.3mm}0.87 $\pm$ 0.29& \hspace{-1.3mm}NN24 \\          
      1634+627 & \hspace{-1.3mm} 93.6 & \hspace{-1.3mm} 39.4 & \hspace{-1.3mm} 0.00674 $\pm$ 0.00147 & \hspace{-1.3mm}-22.84 $\pm$ 0.21 & \hspace{-1.3mm}4.93 $\pm$ 0.47& \hspace{-1.3mm}NN24 \\          
      1634+627 & \hspace{-1.3mm} 93.6 & \hspace{-1.3mm} 39.4 & \hspace{-1.3mm} 0.00114 $\pm$ 0.00038 & \hspace{-1.3mm}-34.68 $\pm$ 0.48 & \hspace{-1.3mm}1.75 $\pm$ 0.69& \hspace{-1.3mm}NN24 \\          
       \hline 
      1638+625 & \hspace{-1.3mm} 93.2 & \hspace{-1.3mm} 39.0  & \hspace{-1.3mm} 0.04821 $\pm$ 0.00347 & \hspace{-1.3mm}-19.99 $\pm$ 0.05 & \hspace{-1.3mm}1.34 $\pm$ 0.04& \hspace{-1.3mm}NN24 \\    
      1638+625 & \hspace{-1.3mm} 93.2 & \hspace{-1.3mm} 39.0 & \hspace{-1.3mm} 0.02119 $\pm$ 0.00164  & \hspace{-1.3mm}-22.06 $\pm$ 0.23 & \hspace{-1.3mm}1.85 $\pm$ 0.17& \hspace{-1.3mm}NN24 \\    
       \hline 
      1927+739  & \hspace{-1.3mm} 105.6 & \hspace{-1.3mm} 23.5 & \hspace{-1.3mm} 0.14169 $\pm$ 0.00396 & \hspace{-1.3mm}-2.73 $\pm$ 0.01 & \hspace{-1.3mm}1.51 $\pm$ 0.02& \hspace{-1.3mm}NN24 \\                           
      1927+739  & \hspace{-1.3mm} 105.6 & \hspace{-1.3mm} 23.5 & \hspace{-1.3mm} 0.04373 $\pm$ 0.00406 & \hspace{-1.3mm}-2.62 $\pm$ 0.04 & \hspace{-1.3mm}3.27 $\pm$ 0.13& \hspace{-1.3mm}NN24 \\                           
      1927+739 & \hspace{-1.3mm} 105.6 & \hspace{-1.3mm} 23.5 & \hspace{-1.3mm} 0.01091 $\pm$ 0.00067 & \hspace{-1.3mm}-61.67 $\pm$ 0.07 & \hspace{-1.3mm}1.52 $\pm$ 0.12& \hspace{-1.3mm}NN24 \\                           
      1927+739 & \hspace{-1.3mm} 105.6 & \hspace{-1.3mm} 23.5 & \hspace{-1.3mm} 0.00323 $\pm$ 0.00044 & \hspace{-1.3mm}-59.05 $\pm$ 0.61 & \hspace{-1.3mm}6.38 $\pm$ 0.67& \hspace{-1.3mm}NN24 \\                           
      1927+739 & \hspace{-1.3mm} 105.6 & \hspace{-1.3mm} 23.5 & \hspace{-1.3mm} 0.00253 $\pm$ 0.00038 & \hspace{-1.3mm}-10.94 $\pm$ 1.33 & \hspace{-1.3mm}18.63 $\pm$ 1.82& \hspace{-1.3mm}NN24 \\                           
      1927+739 & \hspace{-1.3mm} 105.6 & \hspace{-1.3mm} 23.5 & \hspace{-1.3mm} 0.00873 $\pm$ 0.00056 & \hspace{-1.3mm}-11.00 $\pm$ 0.13 & \hspace{-1.3mm}2.74 $\pm$ 0.22& \hspace{-1.3mm}NN24 \\                           
      1927+739 & \hspace{-1.3mm} 105.6 & \hspace{-1.3mm} 23.5 & \hspace{-1.3mm} 0.00237 $\pm$ 0.00062 & \hspace{-1.3mm}-76.28 $\pm$ 0.13 & \hspace{-1.3mm}1.13 $\pm$ 0.34& \hspace{-1.3mm}NN24 \\                           
       \hline                                
      2137-207 & \hspace{-1.3mm} 30.3 & \hspace{-1.3mm} -45.6  & \hspace{-1.3mm} 0.01183 $\pm$ 0.00127  & \hspace{-1.3mm}-0.72 $\pm$ 0.09 & \hspace{-1.3mm}1.08 $\pm$ 0.15& \hspace{-1.3mm}NN24 \\                                 
      2137-207 & \hspace{-1.3mm} 30.3 & \hspace{-1.3mm} -45.6  & \hspace{-1.3mm} 0.00735 $\pm$ 0.00116 & \hspace{-1.3mm}5.55 $\pm$ 0.14 & \hspace{-1.3mm}1.19 $\pm$ 0.24& \hspace{-1.3mm}NN24 \\                 2137-207 & \hspace{-1.3mm} 30.3 & \hspace{-1.3mm} -45.6  & \hspace{-1.3mm} 0.00761 $\pm$ 0.00063 & \hspace{-1.3mm}-2.20 $\pm$ 0.47 & \hspace{-1.3mm}7.44 $\pm$ 0.62& \hspace{-1.3mm}NN24 \\

    \hline

     \hline
    \end{tabular}
    \vspace{3mm}\\
    \textbf{Notes.} Col. 1: Source name of the background continuum objects to which absorption spectra of H{\sc i} are observed in two different\\
    \hspace{-2.8mm}surveys. Col. 2 and Col. 3: Longitude and latitude of these background continuum sources, respectively. Cols. 4, 5, and 6:\\
    \hspace{-2.9mm}Peak optical depth ($\tau_{\text{peak}}$), center velocity ($v_{\text{c}}$) and broadening parameter ($b$) of the gas components observed towards the\\
    \hspace{-0.2mm}background continuum objects. This parameter $b$ is equal to $\sqrt{2}\sigma_{\text{tot}}$, where $\sigma_{\text{tot}}$ is equal to the total velocity dispersion of the\\ 
   \hspace{-3.9mm}line. Col. 7: References of these surveys from where the components are taken for analysis. NR13 denotes the work of\\
   \hspace{-75mm}\citet{2013MNRAS.436.2366R}, NN24 denotes the work of \citet{2024MNRAS.529.4037P}.\\
    \label {tab:table2}
    \end{table*}

\section{Physical parameters of the H{\sc i} spectra from the \texttt{Millennium survey}}\label{Appendix: appendix3}
Physical parameters of the H{\sc i} spectra from the Millennium survey \citep{2003ApJ...586.1067H}, detected only in emission and exhibiting full width half maximum (FWHM) is between 20 and 30 km s$^{-1}$, are listed in Table \ref{Table: tab3}.

\begin{table*}[hbt!]
    \centering
    \caption{Physical parameters of the H{\sc i} spectra from the \texttt{Millennium survey} \citep{2003ApJS..145..329H}, detected only in emission and exhibiting full width half maximum (FWHM) is between 20 and 30 km s$^{-1}$.}
    \begin{tabular}{  c c c c c c c c }
    
    \hline  
  Source &  FWHM$^{\dagger}$   & \hspace{-3.0mm} Column density$^{\dagger}$ & \hspace{-3.0mm} Length scale &\hspace{0.0mm} Kinetic temp.  &\hspace{0.0mm} Spin temp. & \hspace{-3.0mm} Peak optical depth & \hspace{-3.0mm} Peak optical depth  \\ [0.5 ex]
    
  name   & $\Delta V$(km s$^{-1}$) & \hspace{-3.0mm} $N$(H{\sc i})$_{20}$ (cm$^{-2}$) & \hspace{-3.0mm} $L_\text{los}$ (pc) & \hspace{-3.0mm} $T_{\text{k}} (\text{K})$ & \hspace{-3.0mm} $T_{\text{s}} (\text{K})$  & \hspace{-3.0mm} $\tau_{\text{peak}}~(T_{\text{s}} < T_{\text{k}})$ & \hspace{-3.0mm} $\tau_{\text{peak}}~(T_{\text{s}} = T_{\text{k}})$  \\ [0.5 ex]
    \hline 
  3C 33-1  &  26.21  & \hspace{-3.0mm} 0.86 & \hspace{-3.0mm} 80.5 & \hspace{-3.0mm} 10981  & \hspace{-3.0mm} 6230  & \hspace{-3.0mm}  0.00027  & \hspace{-3.0mm} 0.00015   \\ [0.5 ex]
  3C 33-2  &  22.25  & \hspace{-3.0mm} 1.02 & \hspace{-3.0mm} 63.6 & \hspace{-3.0mm} 7316  & \hspace{-3.0mm} 3111  & \hspace{-3.0mm}   0.00076  & \hspace{-3.0mm} 0.00032    \\ [0.5 ex]
  3C 33    &  22.23  & \hspace{-3.0mm} 0.16 & \hspace{-3.0mm} 13.3 & \hspace{-3.0mm} 9731  & \hspace{-3.0mm} 4378  & \hspace{-3.0mm}   0.00008  & \hspace{-3.0mm} 0.00004    \\ [0.5 ex]
  3C 33    &  22.46  & \hspace{-3.0mm} 0.98 & \hspace{-3.0mm} 63.1 & \hspace{-3.0mm} 7548  & \hspace{-3.0mm} 3144  & \hspace{-3.0mm}   0.00072  & \hspace{-3.0mm} 0.00030    \\ [0.5 ex]
  3C 105   & 24.41  & \hspace{-3.0mm} 3.26 & \hspace{-3.0mm} 170.2 & \hspace{-3.0mm} 6122  & \hspace{-3.0mm} 2941   & \hspace{-3.0mm} 0.00234  & \hspace{-3.0mm} 0.00112  \\ [0.5 ex] 
  3C 109   &  26.59  & \hspace{-3.0mm} 4.40 & \hspace{-3.0mm} 247.3 & \hspace{-3.0mm} 6591  & \hspace{-3.0mm} 3008  & \hspace{-3.0mm}  0.00284  & \hspace{-3.0mm} 0.00129  \\ [0.5 ex] 
  3C 120   &  20.35  & \hspace{-3.0mm} 2.27 & \hspace{-3.0mm} 88.7 & \hspace{-3.0mm}  4579 & \hspace{-3.0mm}  2723 & \hspace{-3.0mm}   0.00211  & \hspace{-3.0mm} 0.00126  \\ [0.5 ex]
  3C 123   &  25.36  & \hspace{-3.0mm} 1.89 & \hspace{-3.0mm} 133.2 & \hspace{-3.0mm} 8262  & \hspace{-3.0mm} 3385  & \hspace{-3.0mm}  0.00113  & \hspace{-3.0mm} 0.00046  \\ [0.5 ex]
  3C 167   &  20.60  & \hspace{-3.0mm} 13.00 & \hspace{-3.0mm}190.6  & \hspace{-3.0mm} 1719   & \hspace{-3.0mm} 1484  & \hspace{-3.0mm}0.02192    & \hspace{-3.0mm} 0.01892  \\ [0.5 ex]
  3C 225b  &  20.22  & \hspace{-3.0mm} 0.70 & \hspace{-3.0mm} 38.5 & \hspace{-3.0mm}  6450 & \hspace{-3.0mm} 2988  & \hspace{-3.0mm}   0.00060 & \hspace{-3.0mm}  0.00028 \\ [0.5 ex] 
  3C 228.0 &  21.38  & \hspace{-3.0mm} 0.41 & \hspace{-3.0mm} 28.1 & \hspace{-3.0mm} 8040  & \hspace{-3.0mm} 3235  & \hspace{-3.0mm}  0.00031  & \hspace{-3.0mm} 0.00012  \\ [0.5 ex]
  3C 245   &  21.46  & \hspace{-3.0mm} 1.55 & \hspace{-3.0mm} 78.9 & \hspace{-3.0mm} 5968  & \hspace{-3.0mm} 2919  & \hspace{-3.0mm}  0.00127  & \hspace{-3.0mm} 0.00062  \\ [0.5 ex] 
  3C 272.1 &  20.88  & \hspace{-3.0mm} 1.30 & \hspace{-3.0mm} 65.6 & \hspace{-3.0mm} 5916  & \hspace{-3.0mm} 2912  & \hspace{-3.0mm}  0.00110  & \hspace{-3.0mm} 0.00054  \\ [0.5 ex] 
  3C 273   & 20.99  & \hspace{-3.0mm} 0.26 & \hspace{-3.0mm} 18.2 & \hspace{-3.0mm} 8218  & \hspace{-3.0mm} 3355  & \hspace{-3.0mm}  0.00019  & \hspace{-3.0mm} 0.00008  \\ [0.5 ex] 
  3C 315   &  26.13  & \hspace{-3.0mm} 1.03 & \hspace{-3.0mm} 92.1 & \hspace{-3.0mm} 10480  & \hspace{-3.0mm} 5377  & \hspace{-3.0mm} 0.00038   & \hspace{-3.0mm} 0.00019   \\ [0.5 ex] 
  3C 333   &  27.06  & \hspace{-3.0mm} 0.38 & \hspace{-3.0mm} 43.6 & \hspace{-3.0mm}  13464 & \hspace{-3.0mm} 10455  & \hspace{-3.0mm} 0.00007   & \hspace{-3.0mm} 0.00005  \\ [0.5 ex] 
  3C 333   &  24.47  & \hspace{-3.0mm} 1.83 & \hspace{-3.0mm} 119.9 & \hspace{-3.0mm} 7681  & \hspace{-3.0mm} 3162  & \hspace{-3.0mm}  0.00122  & \hspace{-3.0mm}  0.00050 \\ [0.5 ex] 
  3C 353   &  26.88  & \hspace{-3.0mm} 3.84 & \hspace{-3.0mm} 236.3 & \hspace{-3.0mm} 7214  & \hspace{-3.0mm} 3096  & \hspace{-3.0mm} 0.00238   & \hspace{-3.0mm} 0.00100  \\ [0.5 ex] 
  3C 409   &  24.05  & \hspace{-3.0mm} 1.34 & \hspace{-3.0mm} 92.8 & \hspace{-3.0mm}  8117 & \hspace{-3.0mm} 3287  & \hspace{-3.0mm}  0.00087  & \hspace{-3.0mm}  0.00035 \\ [0.5 ex] 
  3C 410   &  21.77  & \hspace{-3.0mm} 25.37 & \hspace{-3.0mm} 255.1 & \hspace{-3.0mm} 1179  & \hspace{-3.0mm} 1072  & \hspace{-3.0mm} 0.05601   & \hspace{-3.0mm} 0.05095  \\ [0.5 ex] 
  3C 433   &  25.31  & \hspace{-3.0mm} 5.06 & \hspace{-3.0mm} 234.3 & \hspace{-3.0mm} 5428  & \hspace{-3.0mm} 2843  & \hspace{-3.0mm}  0.00362  & \hspace{-3.0mm} 0.00190  \\ [0.5 ex] 
  3C 454.0  &  28.82  & \hspace{-3.0mm} 0.95 & \hspace{-3.0mm} 107.5 & \hspace{-3.0mm} 13272  & \hspace{-3.0mm} 10129  & \hspace{-3.0mm} 0.00017    & \hspace{-3.0mm} 0.00013  \\ [0.5 ex] 
  3C 454.0  &  26.54  & \hspace{-3.0mm} 1.42 & \hspace{-3.0mm} 121.2 & \hspace{-3.0mm} 10006  & \hspace{-3.0mm} 4570  & \hspace{-3.0mm} 0.00060   & \hspace{-3.0mm} 0.00028  \\ [0.5 ex] 
  4C 07.32 &  24.96  & \hspace{-3.0mm} 1.10 & \hspace{-3.0mm} 87.3 & \hspace{-3.0mm}  9307 & \hspace{-3.0mm} 4091  & \hspace{-3.0mm} 0.00056   & \hspace{-3.0mm}  0.00024 \\ [0.5 ex] 
  P0531+19 &  23.65  & \hspace{-3.0mm} 14.3 & \hspace{-3.0mm} 286.2 & \hspace{-3.0mm} 2347  & \hspace{-3.0mm} 1861  & \hspace{-3.0mm} 0.01674   & \hspace{-3.0mm} 0.01328  \\ [0.5 ex]
  P0820+22 &  21.58  & \hspace{-3.0mm} 0.46 & \hspace{-3.0mm} 31.6 & \hspace{-3.0mm}  8056 & \hspace{-3.0mm} 3245  & \hspace{-3.0mm}  0.00034  & \hspace{-3.0mm}  0.00014 \\ [0.5 ex] 
  P1055+20 & 29.62  & \hspace{-3.0mm} 0.61 & \hspace{-3.0mm} 79.5 & \hspace{-3.0mm}  15284 & \hspace{-3.0mm} 13553  & \hspace{-3.0mm} 0.00008   & \hspace{-3.0mm} 0.00007  \\ [0.5 ex]  
  T0556+19 &  20.06  & \hspace{-3.0mm} 53.6 & \hspace{-3.0mm} 219.0 & \hspace{-3.0mm} 479  & \hspace{-3.0mm}  449 & \hspace{-3.0mm}  0.30690  & \hspace{-3.0mm} 0.28773  \\ [0.5 ex]  
    \hline
    
    \end{tabular}\\
    
\hspace{-2mm}\textbf{Notes.} The first column lists the source name, the second column is the full width at half maximum (FWHM) of the components. The third, fourth, fifth, and sixth columns represent the column density, length scale, kinetic temperature, and spin temperature (minimum WF effect) of the gas components, respectively. The sixth and seventh columns show the peak optical depth of the\\ 
\hspace{-3mm} components for the cases where the spin temperature is less than the kinetic temperature (minimum WF effect) and where the spin \\
\hspace{-69mm}temperature equals the kinetic temperature (maximum WF effect), respectively.\\
\hspace{-0mm} ${\dagger}$denotes the parameters adopted directly form the \texttt{Millennium survey} \citep{2003ApJS..145..329H} and the rest of the parameters \\
\hspace{-70mm}we have calculated using our modeling which is discussed in Subsection \ref{non_detection}.
      \label{Table: tab3}
    \end{table*}

\section{Properties of the background quasars and the achieved RMS noise in \texttt{GWA} survey }\label{Appendix: appendix4}

We list the properties of the background quasars and the achieved RMS noise in the \texttt{GWA} survey in Table \ref{Table: tab4}.

\begin{table*}[hbt!]
    \centering
    \caption{37 lines of sight of the \texttt{GWA survey.}}
    \begin{tabular}{ c c c c }
    
    \hline  
     Source & \hspace{-1.3mm} $S_{1.4}$   & \hspace{-1.3mm} $\tau_{\text{rms}}$ & \hspace{-1.3mm} Reference \\ [0.5 ex]
    
     name   & \hspace{-1.3mm} (Jy)   & \hspace{-1.3mm} (10$^{-3}$) & \hspace{-1.3mm} (NR13/NN24)  \\ [0.5 ex]
     \hline
     B0023$-$263 & \hspace{-1.3mm} 7.5    & \hspace{-1.3mm} 1.04 & \hspace{-1.3mm} NR13   \\ [0.5 ex]
     B0114$-$211 & \hspace{-1.3mm} 3.7    & \hspace{-1.3mm} 1.58 & \hspace{-1.3mm} NR13   \\ [0.5 ex]
     B0117$-$155 & \hspace{-1.3mm} 4.2    & \hspace{-1.3mm} 1.51 & \hspace{-1.3mm} NR13  \\ [0.5 ex]
     B0134$+$329 & \hspace{-1.3mm} 16.5   & \hspace{-1.3mm} 0.53 & \hspace{-1.3mm} NR13   \\ [0.5 ex]
     B0202$+$149 & \hspace{-1.3mm} 3.5    & \hspace{-1.3mm} 0.98 & \hspace{-1.3mm} NR13   \\ [0.5 ex]
     B0237$-$233 & \hspace{-1.3mm} 5.7    & \hspace{-1.3mm} 1.19 & \hspace{-1.3mm} NR13   \\ [0.5 ex]
     B0136$+$162 & \hspace{-1.3mm} 7.8    & \hspace{-1.3mm} 0.75 & \hspace{-1.3mm} NR13   \\ [0.5 ex]
     B0316$+$413 & \hspace{-1.3mm} 23.9   & \hspace{-1.3mm} 0.49 & \hspace{-1.3mm} NR13   \\ [0.5 ex]
     B0404$+$768 & \hspace{-1.3mm} 5.8    & \hspace{-1.3mm} 1.16 & \hspace{-1.3mm} NR13   \\ [0.5 ex]
     B0407$-$658 & \hspace{-1.3mm} 16.2   & \hspace{-1.3mm} 1.00 & \hspace{-1.3mm} NR13   \\ [0.5 ex]
     B0518$+$165 & \hspace{-1.3mm} 8.5    & \hspace{-1.3mm} 1.08 & \hspace{-1.3mm} NR13   \\ [0.5 ex]
     B0531$+$194 & \hspace{-1.3mm} 6.8    & \hspace{-1.3mm} 0.99 & \hspace{-1.3mm} NR13   \\ [0.5 ex]
     B0538$+$498 & \hspace{-1.3mm} 22.5   & \hspace{-1.3mm} 0.49 & \hspace{-1.3mm} NR13   \\ [0.5 ex]
     B0831$+$557 & \hspace{-1.3mm} 8.8   & \hspace{-1.3mm} 1.15 & \hspace{-1.3mm} NR13   \\ [0.5 ex]
     B0834$-$196 & \hspace{-1.3mm} 5.0   & \hspace{-1.3mm} 1.08 & \hspace{-1.3mm} NR13   \\ [0.5 ex]
     B0906$+$430 & \hspace{-1.3mm} 3.9   & \hspace{-1.3mm} 0.82 & \hspace{-1.3mm} NR13   \\ [0.5 ex]
     B1151$-$348 & \hspace{-1.3mm} 5.0   & \hspace{-1.3mm} 1.05 & \hspace{-1.3mm} NR13   \\ [0.5 ex]
     B1245$-$197 & \hspace{-1.3mm} 5.3   & \hspace{-1.3mm} 1.23 & \hspace{-1.3mm} NR13   \\ [0.5 ex]
     B1328$+$254 & \hspace{-1.3mm} 6.8   & \hspace{-1.3mm} 0.33 & \hspace{-1.3mm} NR13   \\ [0.5 ex]
     B1328$+$307 & \hspace{-1.3mm} 14.7  & \hspace{-1.3mm} 0.30 & \hspace{-1.3mm} NR13   \\ [0.5 ex]
     B1345$+$125 & \hspace{-1.3mm} 5.2   & \hspace{-1.3mm} 1.07 & \hspace{-1.3mm} NR13   \\ [0.5 ex]
     B1611$+$343 & \hspace{-1.3mm} 4.8   & \hspace{-1.3mm} 0.72 & \hspace{-1.3mm} NR13   \\ [0.5 ex]
     B1641$+$399 & \hspace{-1.3mm} 8.9   & \hspace{-1.3mm} 0.49 & \hspace{-1.3mm}  NR13  \\ [0.5 ex]
     B1814$-$637 & \hspace{-1.3mm} 14.4  & \hspace{-1.3mm} 1.00 & \hspace{-1.3mm}  NR13  \\ [0.5 ex]
     B1827$-$360 & \hspace{-1.3mm} 6.9   & \hspace{-1.3mm} 0.89 & \hspace{-1.3mm}  NR13  \\ [0.5 ex]
     B1921$-$293 & \hspace{-1.3mm} 6.0   & \hspace{-1.3mm} 1.02 & \hspace{-1.3mm}  NR13  \\ [0.5 ex]
     B2050$+$364 & \hspace{-1.3mm} 5.2   & \hspace{-1.3mm} 1.46 & \hspace{-1.3mm}  NR13  \\ [0.5 ex]
     B2200$+$420 & \hspace{-1.3mm} 6.1   & \hspace{-1.3mm} 2.68 & \hspace{-1.3mm}  NR13 \\ [0.5 ex]
     B2203$-$188 & \hspace{-1.3mm} 6.0   & \hspace{-1.3mm} 1.00 & \hspace{-1.3mm} NR13   \\ [0.5 ex]
     B2223$-$052 & \hspace{-1.3mm} 5.7   & \hspace{-1.3mm} 0.79 & \hspace{-1.3mm} NR13   \\ [0.5 ex]
     1352$+$314  & \hspace{-1.3mm} 3.5   & \hspace{-1.3mm} 1.12 & \hspace{-1.3mm} NN24  \\ [0.5 ex]
     1400$+$621  & \hspace{-1.3mm} 4.4   & \hspace{-1.3mm} 1.40 & \hspace{-1.3mm} NN24  \\ [0.5 ex]
     1609$+$266  & \hspace{-1.3mm} 4.8   & \hspace{-1.3mm} 0.87 & \hspace{-1.3mm}  NN24  \\ [0.5 ex]
     1634$+$627  & \hspace{-1.3mm} 4.8   & \hspace{-1.3mm} 0.93  & \hspace{-1.3mm} NN24  \\ [0.5 ex]
     1638$+$625  & \hspace{-1.3mm} 4.5   & \hspace{-1.3mm} 0.89 & \hspace{-1.3mm} NN24  \\ [0.5 ex]
     1927$+$739  & \hspace{-1.3mm} 3.1   & \hspace{-1.3mm} 1.17  & \hspace{-1.3mm} NN24  \\ [0.5 ex]
     2137$-$207  & \hspace{-1.3mm} 3.6   & \hspace{-1.3mm} 1.95 & \hspace{-1.3mm} NN24   \\ [0.5 ex]
    \hline

    \hline

    \end{tabular}
    \vspace{3mm}
    
    \label{Table: tab4}
    \textbf{Notes.} Column 1: Background quasars toward which the absorption spectra were obtained in the GWA survey \citep{2013MNRAS.436.2366R,2024MNRAS.529.4037P}. Column 2: Flux of the background quasars ($S_{1.4}$) at 1.4 GHz. Column 3: RMS optical depth ($\tau_{\text{rms}}$)  of each\\
    \hspace{-114mm}source. Column 4: References for the sources.\\
    \end{table*}

\section{Required integration times for uGMRT, SKA, and ngVLA}\label{Appendix: appendix4_0}

\begin{figure*}
\centering 
 \includegraphics[width=3.4in,height=2.8in,angle=0]{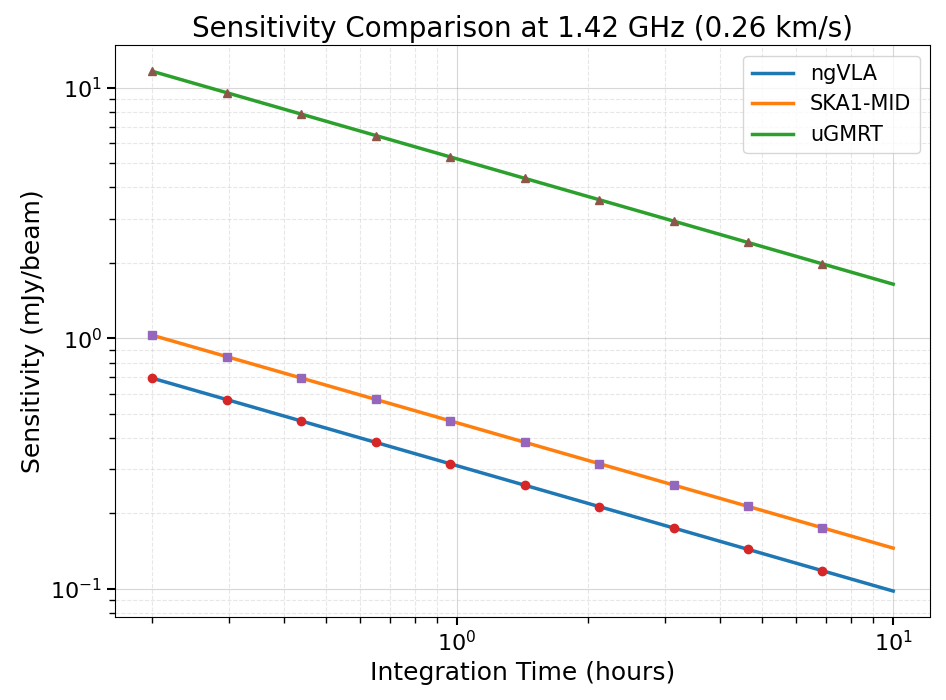}\includegraphics[width=3.4in,height=2.8in,angle=0]{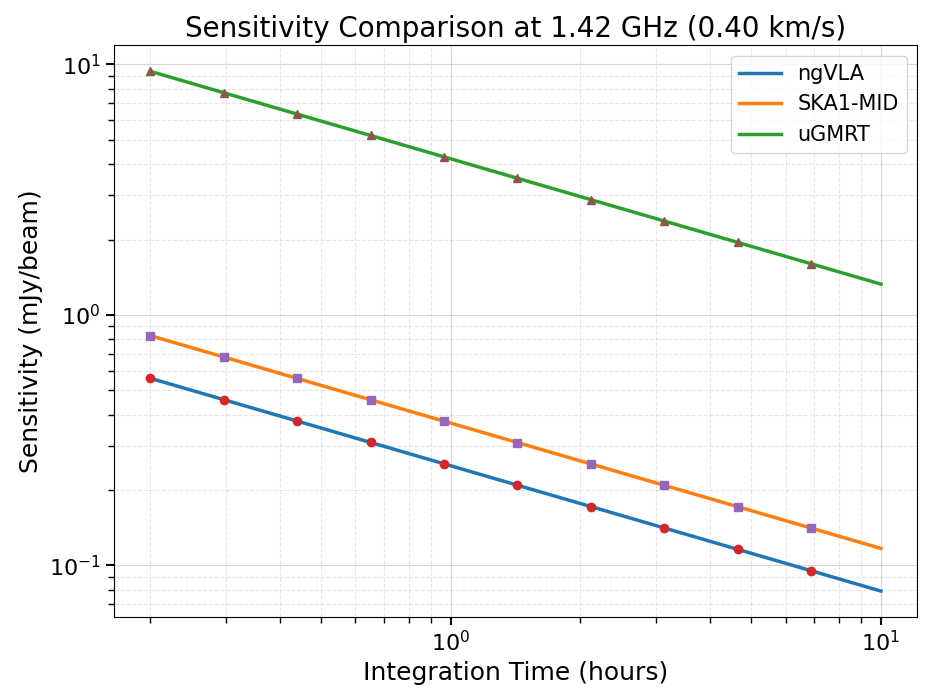}
 \caption{Comparison of sensitivities (mJy/beam) of ngVLA, SKA1-MID, and uGMRT at 1.42 GHz for velocity resolutions of 0.26 km s$^{-1}$ and 0.40 km s$^{-1}$ as a function integration time (hr).}
 \label{fig:fig6}
\end{figure*}

For estimating the spectral line sensitivity we start from the standard radiometer equation for an interferometric array, where the rms noise is given by:

\begin{eqnarray}
  \displaystyle  \sigma_{\text{s}} = \frac{SEFD}{\eta_{\text{s}} \sqrt{N_{\text{ant}}(N_{\text{ant}}-1)N_\text{pol}~\Delta\nu~t_{\text{int}}} },
\end{eqnarray}

with SEFD the system equivalent flux density, $\eta_{\text{s}}$ the system efficiency, $N_{\text{ant}}$ the number of antennas, $N_\text{pol}$ is the number of polarization, $\Delta \nu$ the channel bandwidth, and $t_{\text{int}}$ the on-source integration time. Since the publicly available telescope performance tables provide a reference rms sensitivity  $\sigma_{\text{ref}}$ for a given integration time $t_{\text{ref}}$ and channel width $\Delta v_{\text{ref}}$ we scale the sensitivity using the radiometer dependence on bandwidth and time. The rms noise for an arbitrary integration time is therefore





\begin{eqnarray}
    \sigma(t_{\text{int}},\Delta v) = \sigma_{\text{ref}} \sqrt{\frac{t_{\text{ref}}}{t_{\text{int}}}} \sqrt{\frac{\Delta v_{\text{ref}}}{\Delta v} }.
\end{eqnarray}

For the ngVLA, we adopt a reference rms line sensitivity of $\sigma_{\text{ref,ngVLA}}$ $\sim$ 49.9 $\mu$Jy beam$^{-1}$ for a 1 hr of integration and a 10 km s$^{-1}$ spectral channel under natural weighting, taken from the ngVLA performance specifications\footnote{ngVLA Memo \# 17, ngVLA Reference Design Development \& Performance Estimates, see \citet{ngVLA_memo17_2018}.}. For SKA1-MID, we adopt a representative reference sensitivity of  $\sigma_{\text{ref,SKA}}$ $\sim$ 140 $\mu$Jy beam$^{-1}$ at similar frequencies and channel width $\sim$ 2.83 km s$^{-1}$ based on SKA Phase 1 performance estimates\footnote{Expected Performance of the Square Kilometer Array - Phase 1 (SKA1); see \citet{Braun2019SKA1}.}. For uGMRT. the expected sensitivity improvement is up to a factor of $\sim$ 3 compared to GMRT. We adopt a representative reference rms noise $\sigma_{\text{ref,uGMRT}}$ of $\sim$ 45 $\mu$Jy beam$^{-1}$ in 10 minute and 100 MHz bandwidth \footnote{The upgraded GMRT: Opening new windows on the radio Universe; see \cite{gupta2017ugmrt}.}.\\

\hspace{-5mm}For the spectral resolutions considered here, $\Delta v$ = 0.26 and 0.40 km s$^{-1}$, the spectral scaling factors become 6.20 and 5.00 for the ngVLA telescope. This results in 1-hr rms sensitivities of 0.31 mJy beam$^{-1}$ and 0.25 mJy beam$^{-1}$ for the ngVLA at the respective
spectral resolution. For the SKA1-MID telescope at the corresponding spectral resolutions of 0.26 and 0.40 km s$^{-1}$, the spectral scaling factors become 3.30 and 2.66. This results in 1-hr rms sensitivities of 0.46 mJy beam$^{-1}$ and 0.37 mJy beam$^{-1}$ for the SKA1-MID at the respective spectral resolution. For uGMRT, at the two spectral resolutions of 0.26 and 0.40 km s$^{-1}$, the rms sensitivities achieved in one hour at two different spectral resolutions of 0.26 and 0.40 km s$^{-1}$ are 5.2 and 4.2 mJy beam$^{-1}$, respectively.\\

\begin{table*}[hbt!]
\centering
\caption{Comparison of required sensitivities of ngVLA, SKA1-MID, and uGMRT telescopes.}
\begin{tabular}{c cc cc}
\hline
 & \multicolumn{2}{c}{$\sigma_\text{rms}$ = 0.93 mJy beam$^{-1}$} & \multicolumn{2}{c}{$\sigma_\text{rms}$ = 0.39 mJy beam$^{-1}$} \\
 \cmidrule(r{3pt}){2-3} \cmidrule(l{3pt}){4-5}
Telescope & 0.26 km s$^{-1}$ & 0.40 km s$^{-1}$ & 0.26 km s$^{-1}$ & 0.40 km s$^{-1}$ \\
\cmidrule(r{3pt}){1-1} \cmidrule(r{3pt}){2-3} \cmidrule(l{3pt}){4-5}
\vspace{2mm}
ngVLA & $\sim$6.7 min & $\sim$4.3 min & $\sim$37.9 min & $\sim$24.6 min \\
\vspace{2mm}
SKA1-MID & $\sim$14.7 min & $\sim$9.5 min & $\sim$1.39 hr & $\sim$54.0 min \\
\vspace{2mm}
uGMRT & $\sim$31.2 hr & $\sim$20.4 hr & $\sim$177.6 hr & $\sim$116.1 hr  \\
\hline
\end{tabular}\\
\textbf{Notes:} The first column lists the telescope name. The second and third columns show the required integration times to achieve an rms noise level of $\sigma_\text{rms}$ = 0.93 mJy beam$^{-1}$ at two spectral resolutions, 0.26, and 0.40 km s$^{-1}$. The remaining columns provide the\\
\hspace{-36mm}corresponding integration times for achieving  $\sigma_\text{rms}$ = 0.39 mJy beam$^{-1}$ at the same spectral resolutions.
\label{Table: tab22}
\vspace{2mm}
\end{table*}

The full sensitivity–time curves (for ngVLA, SKA1-MID, and uGMRT) shown in Fig. \ref{fig:fig6} are then obtained by applying the temporal scaling 

\begin{eqnarray}
  \sigma (t) = \frac{\sigma_{\text{1hr}}}{\sqrt{t_{int}}},    
\end{eqnarray}

\hspace{-5mm}across the chosen integration time range. These curves provide an approximate but physically motivated comparison of the expected noise performance of the three facilities under consistent assumptions of natural weighting, matched observing frequency, and similar imaging setups.\\

\hspace{-5mm}From the GWA survey the median background quasar flux ($S_{1.4}$) is  5.7 Jy which we have obtained from Table \ref{Table: tab4},and the median $\tau_\text{peak}$ of the WNM components is 0.00082 when $T_\text{s} <$ $T_\text{k}$  and  0.00034 when $T_\text{s} =$ $T_\text{k}$ obtained from Table \ref{Table: tab3}. We adopt the median rather than the mean, as a few data points exhibit extreme deviations from the rest of the sample and can disproportionately influence the mean.\\

\hspace{-5mm}For optically thin lines, which are mainly applicable to WNM components, the depth of the absorption feature, $\Delta I$, is proportional to the continuum flux density $I_{0}$. This is expressed as follows: 

\begin{eqnarray}
 \Delta I = \tau I_{0}.   
\end{eqnarray}

\hspace{-5mm}For $I_{0}$ = 5.7 Jy and optical depths $\tau$ = 0.00082 and  0.00034, the corresponding absorption depths are $\Delta I$ = 4.67 and 1.94 mJy beam$^{-1}$, respectively. For a 5$\sigma_\text{rms}$ detection, the required rms noise levels are therefore 0.93 and 0.39 mJy beam$^{-1}$, respectively. For the ngVLA and SKA-MID, the required integration times range from a few minutes to approximately 1.5 hours. In contrast, the uGMRT requires substantially longer integration times, ranging from about 20 to nearly 180 hours, depending on the desired sensitivity and spectral resolution. Such long integration times make the uGMRT observations challenging. Therefore, a more practical approach for the uGMRT would be to identify sufficiently bright background quasars, for which the required sensitivity level can be achieved within a few hours of integration. The detailed results are presented in Table \ref{Table: tab22}.\\

\section{Other physical properties of the CNM and UNM gas clouds}\label{other_clouds_properties}

We also plot the other physical properties of the H{\sc i} clouds assuming $T_\text{s}=T_\text{k}$. Since the results do not differ significantly from those obtained for the $T_\text{s}<T_\text{k}$ case, we present these results here in Fig. \ref{fig:fig100_11}.\\

\begin{figure*}
\includegraphics[width=2.4in,height=2.1in,angle=0]{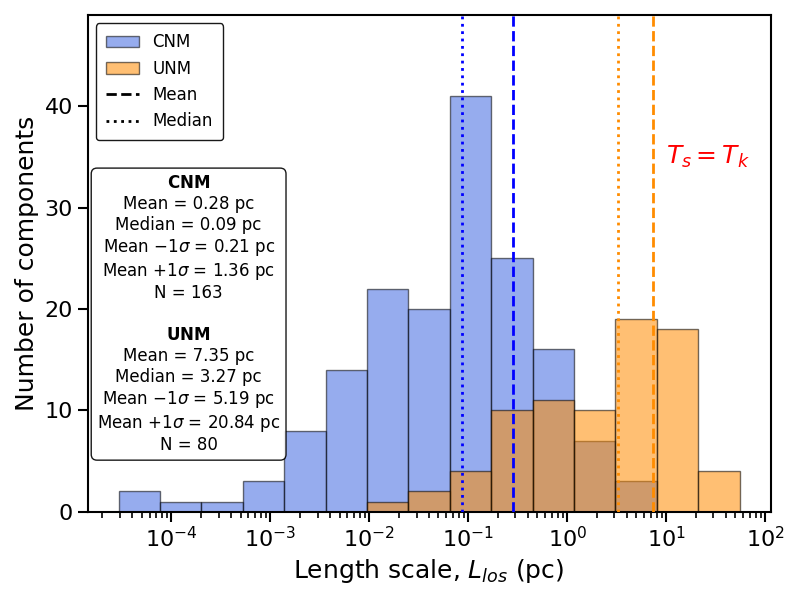}
\includegraphics[width=2.4in,height=2.1in,angle=0]{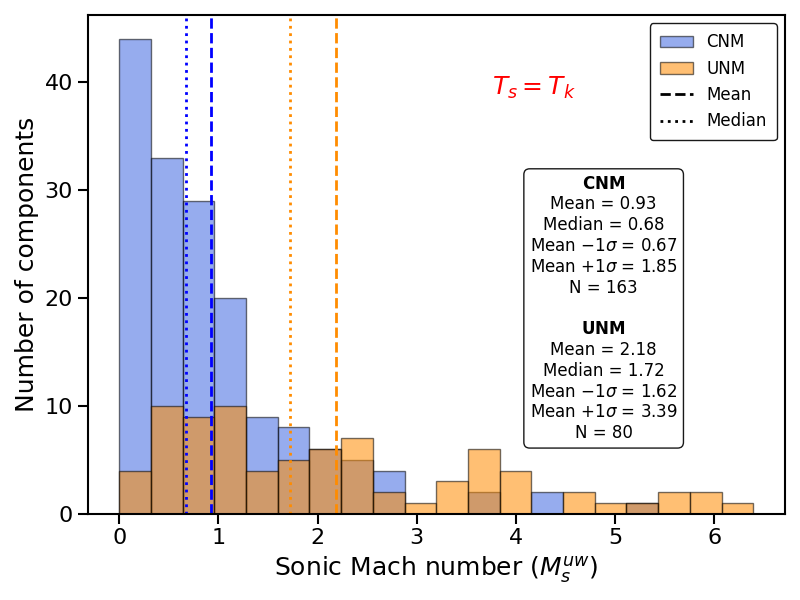}
\includegraphics[width=2.4in,height=2.1in,angle=0]{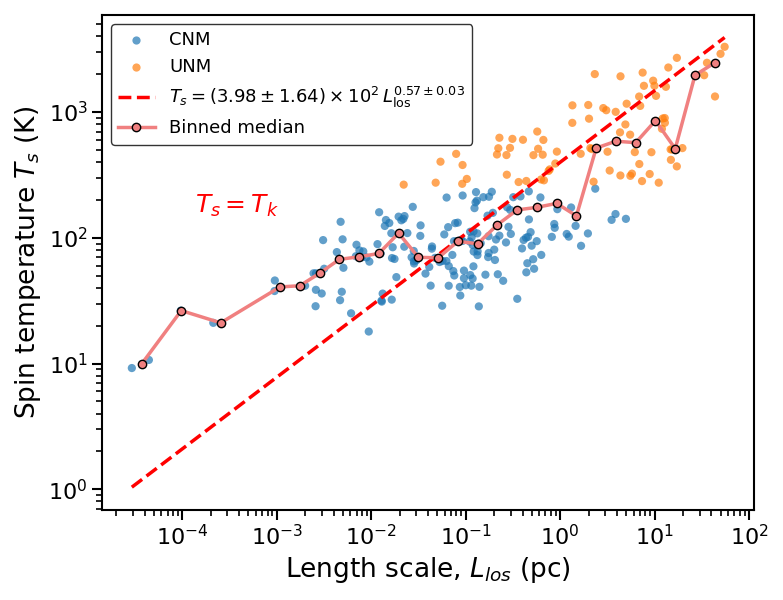}\\
\includegraphics[width=2.4in,height=2.1in,angle=0]{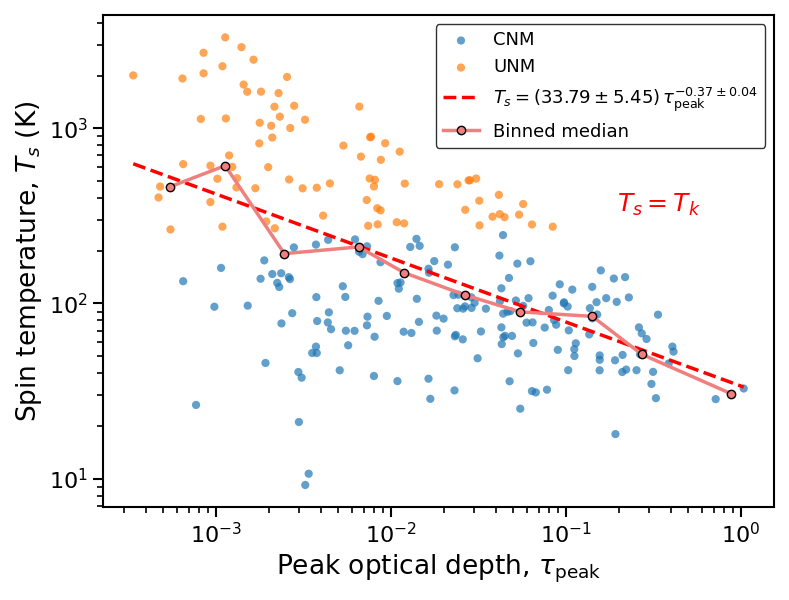}
\includegraphics[width=2.4in,height=2.1in,angle=0]{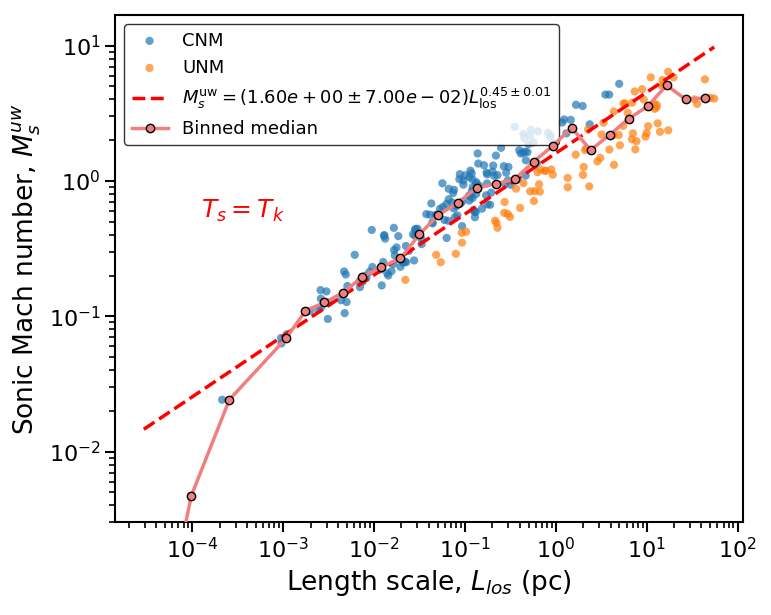}
\includegraphics[width=2.4in,height=2.1in,angle=0]{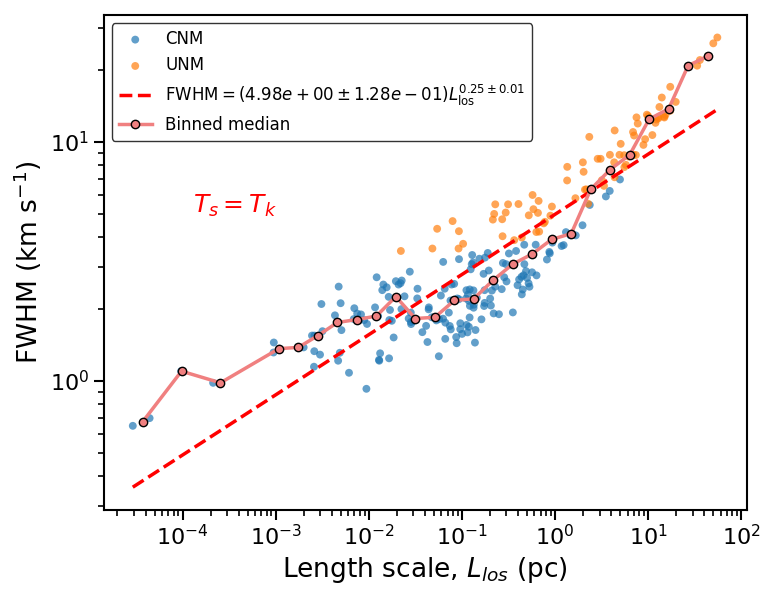}\\
\includegraphics[width=2.4in,height=2.1in,angle=0]{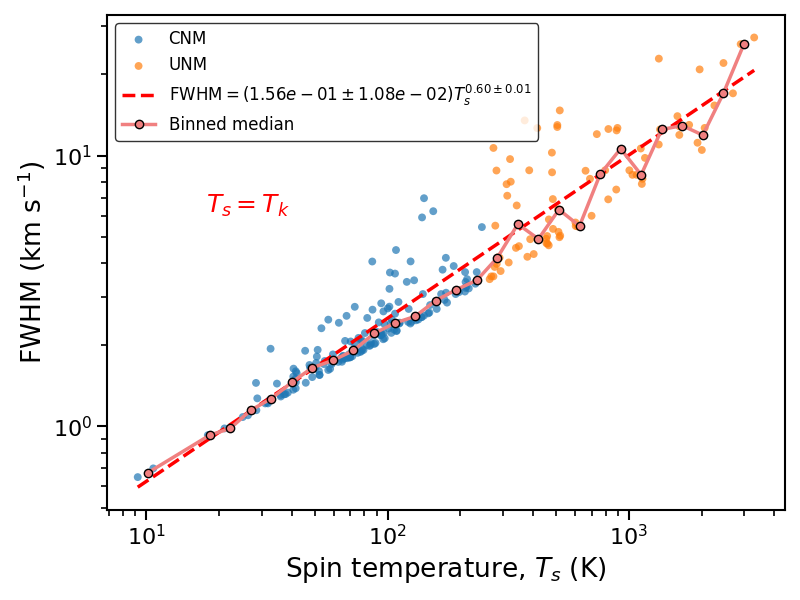}
\includegraphics[width=2.4in,height=2.1in,angle=0]{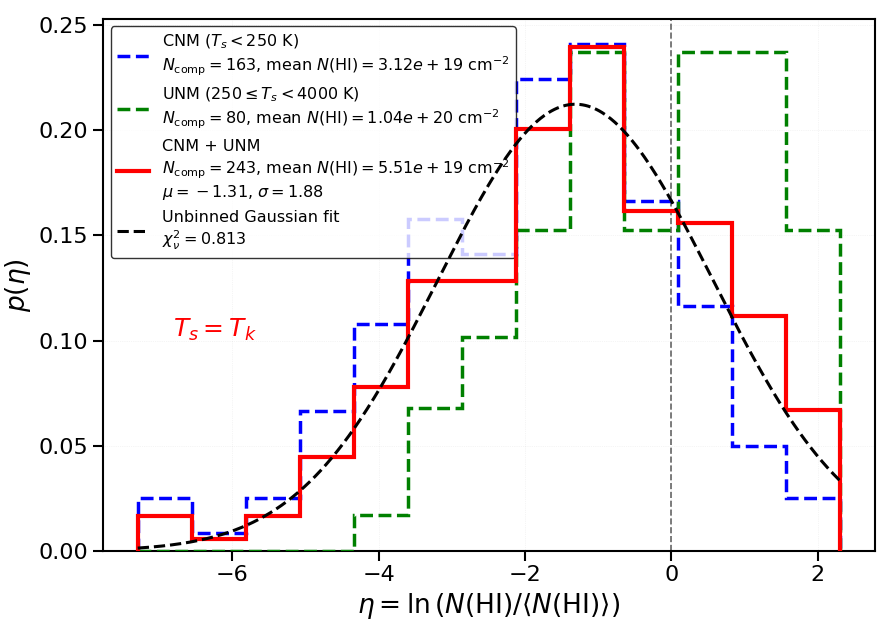}
 \includegraphics[width=2.4in,height=2.1in,angle=0]{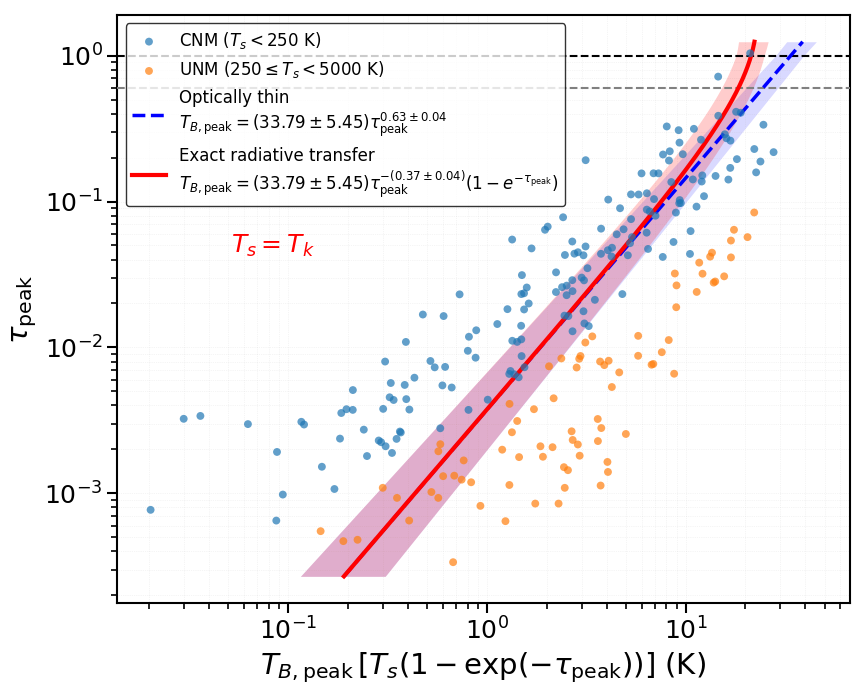}\\
\caption{Upper: Probability distribution functions (PDFs) of the length scale ($L_\text{los}$) and unweighted sonic Mach number ($M_\text{s}^\text{uw}$) for the CNM and UNM gas components, together with the correlation between spin temperature ($T_\text{s}$) and $L_\text{los}$, which follows a power-law relation with an index of 0.57. Middle: Correlations between $T_\text{s}$ and peak optical depth ($\tau_\text{peak}$), $M_\text{s}^\text{uw}$ and $L_\text{los}$, and FWHM and $L_\text{los}$, with power-law indices of $-$0.37, 0.45, and 0.25, respectively. Lower: Correlation between FWHM and $T_\text{s}$ with power law index of 0.60, along with the PDFs of the H{\sc i} column density, $N$(H{\sc i}), for the CNM, UNM, and combined CNM+UNM components, shown by the dashed blue, green, and solid red curves, respectively. The combined PDF is fitted with a Gaussian distribution, yielding a mean of $\mu=-1.31$ and a dispersion of $\sigma=1.88$. The final panel presents the correlation between the peak brightness temperature ($T_{\text{B,peak}}$) and $\tau_\text{peak}$, comparing the optically thin approximation and the full radiative-transfer model, shown by the blue dashed and red solid curves, respectively, with the corresponding $1\sigma$ uncertainties indicated by the shaded regions.}
 \label{fig:fig100_11}
 \end{figure*}

}

\begin{appendix}

\onecolumn

\appendix

\end{appendix}

\vspace{20 mm}


\end{document}